%% file: ms.tex
\documentclass{mn2e}

\usepackage{graphicx,lscape,amssymb,amsmath,url,placeins}

\title[S2CLS: properties of the galaxies unveiled in the deep 850${\mathbf{\rm \mu m}}$ survey]{The SCUBA-2 Cosmology Legacy Survey: galaxies in the 
deep 850$\bmath{{\rm \mu m}}$ survey, and the star-forming `main sequence'}
\author[M. Koprowski et al.]{M.P. Koprowski$^{1}$\thanks{E-mail: mpk@roe.ac.uk},
J.S.~Dunlop$^{1}$,
M.J.~Micha{\l}owski$^{1}$,
I.~Roseboom$^{1}$,
J.E.~Geach$^{2}$, 
\and M.~Cirasuolo$^{1,3}$, I. Aretxaga$^4$,  
R.A.A. Bowler$^{1}$, M. Banerji$^{5}$, N. Bourne$^{1}$, 
\and K.E.K. Coppin$^{2}$, S. Chapman$^{6}$, D.H. Hughes$^{4}$,  T. Jenness$^{7}$, 
R.J. McLure$^{1}$, 
\and M. Symeonidis$^8$, P. van der Werf$^{9}$\\
$^{1}$SUPA\thanks{Scottish Universities Physics Alliance}, Institute for Astronomy, University of Edinburgh, Royal Observatory, Edinburgh, EH9 3HJ\\
$^{2}$Centre for Astrophysics Research, Science \& Technology Research Institute, University of Hertfordshire, College Lane, Hatfield AL10 9AB\\
$^3$UK Astronomy Technology Centre, Royal Observatory, Edinburgh EH9 3HJ\\
$^4$Instituto Nacional de Astrofisica, Optica y Electronica (INAOE), Aptdo. Postal 51 y 216, 72000 Puebla, Pue., Mexico\\
$^5$Institute of Astronomy, University of Cambridge, Madingley Road, Cambridge
CB3 0HA\\
$^6$Department of Physics and Atmospheric Science, Dalhousie University, Coburg Road, Halifax, NS B3H 1A6, Canada\\
$^7$LSST Project Office, 933 N. Cherry Ave, Tucson, AZ 85719 USA\\
$^8$Mullard Space Science Laboratory, University College London, Holmbury St. Mary, Dorking, Surrey RH5 6NT\\ 
$^9$Leiden Observatory, Leiden University, PO Box 9513, 2300 RA Leiden, the Netherlands}
\begin{document}

\date{Accepted ???. Received ???; in original form ???}

\pagerange{\pageref{firstpage}--\pageref{lastpage}} \pubyear{???}

\maketitle

\label{firstpage}

\vspace{-1.5cm}

\begin{abstract}
We investigate the properties of the galaxies selected from the deepest 850-$\mu$m 
survey undertaken to date with SCUBA-2 on the JCMT as part of the SCUBA-2 Cosmology Legacy 
Survey. A total of 106 sources ($>$5$\sigma$) were uncovered at 850\,$\mu$m from an  
area of $\simeq 150$\,arcmin$^2$ in the centre of the COSMOS/UltraVISTA/CANDELS field, 
imaged to a typical depth of $\sigma_{850} \simeq 0.25$\,mJy. We utilise the available multi-frequency 
data to identify galaxy counterparts for 80 of these sources (75\%), 
and to establish the complete redshift distribution for this sample, yielding $\bar{z} = 2.38\pm 0.09$.
We have also been able to determine the stellar masses of the majority of the galaxy identifications, 
enabling us to explore their location on the star-formation rate~:~stellar-mass (SFR:$M^*$)
plane. Crucially, our new deep 850-$\mu$m selected sample reaches flux densities equivalent to ${\rm SFR} \simeq 100\,{\rm M_{\odot} yr^{-1}}$, 
enabling us to confirm that sub-mm galaxies form the high-mass 
end of the `main sequence' (MS) of star-forming galaxies at $z > 1.5$ (with a mean specific SFR of 
${\rm sSFR}=2.25 \pm 0.19\,{\rm Gyr^{-1}}$ at $z \simeq 2.5$). Our results are consistent with no significant 
flattening of the MS towards high masses at these redshifts. However, our results add to the growing evidence that average sSFR rises only slowly at high redshift, 
resulting in $\log_{10}$sSFR being an apparently simple linear function of the age of the Universe.

\end{abstract}


\section{Introduction}
\label{sec:intro}

It is now well known that approximately half of the starlight in the Universe is re-processed by cosmic dust and re-emitted 
at far-infrared wavelengths (Dole et al. 2006). However, due to a combination of the inescapable physics of diffraction, the molecular content 
of our atmosphere, and the technical difficulties of sensitive high-background imaging, it has proved difficult to connect the UV/optical and 
far-infrared/sub-mm views of the Universe into a consistent and complete picture of galaxy formation/evolution. Thus, while the advent of SCUBA
on the 15-m James Clerk Maxwell Telescope (JCMT) in the late 1990s (Holland et al. 1999) enabled the first discovery of distant dusty galaxies with star-formation rates 
${\rm SFR} \simeq 1000\,{\rm M_{\odot} yr^{-1}}$ (e.g. Smail et al. 1997; Hughes et al. 1998; Barger et al. 1998; Eales et al. 1999), 
such objects initially seemed too extreme and unusual to be easily related to the more 
numerous, `normal' star-forming galaxies being uncovered at UV/optical wavelengths at comparable redshifts ($z \simeq 2 - 4$) 
by Keck (e.g. Steidel et al. 1996) and the {\it Hubble Space Telescope} (HST) (e.g. Madau et al. 1996). 
In recent years the study of rest-frame UV-selected galaxies has been extended out beyond $z \simeq 10$ (see Dunlop 2013 for a review, and Coe et al. 2013;
Ellis et al. 2013; McLure et al. 2013; Bouwens et al. 2015; Bowler et al. 2012, 2014; Oesch et al. 2014;  Finkelstein et al. 2015; Ishigaki et al. 2015; McLeod et al. 2015), 
while a number of sub-mm selected galaxies 
have now been confirmed at $z>4$ (Capak et al. 2008; Coppin et al. 2009; Daddi et al. 2009a,b; Knudsen et al. 2010; Riechers et al. 2010; Cox et al. 2011; Combes et al. 2012; Weiss et al. 2013) 
with the current redshift record holder at $z=6.34$ (Riechers et al. 2013).
However, while such progress is exciting, at present there is still relatively little meaningful 
intersection between these UV/optical and far-infrared/sub-mm studies of the high-redshift Universe (although see Walter et al. 2012).

At more moderate redshifts, however, recent years have seen increasingly successful efforts 
to bridge the gap between the unobscured and dust-enshrouded views of 
the evolving galaxy population. Of particular importance in this endeavour has been the power 
of deep $24\,\mu$m imaging with the MIPS instrument on board {\it Spitzer}, which has proved capable of providing 
a useful estimate of the dust-obscured star-formation activity in a significant fraction of optically-selected 
galaxies out to $z \simeq 1.5 - 2$ (e.g. Caputi et al. 2006; Elbaz et al. 2010). Indeed, MIPS imaging of the GOODS survey fields played a key role in 
establishing what has proved to be a fruitful framework for the study of 
galaxy evolution, namely the existence of a so-called ``main sequence'' (MS) for star-forming galaxies, in which star-formation 
rate is found to be roughly proportional to stellar mass (${\rm SFR} \propto M_*$; Noeske et al. 2007; Daddi et al. 2007; Renzini \& Peng 2015), with a normalisation 
that rises with increasing redshift (e.g. Santini et al. 2009; 
Oliver et al. 2010; Elbaz et al. 2011; Karim et al 2011; Rodighiero et al. 2011, 2014; Tasca et al. 2015; Salmon et al. 2015; Schreiber et al. 2015; Johnston et al. 2015).

Interest in the MS of star-forming galaxies has continued to grow (see Speagle et al. 2014 for a useful and comprehensive 
overview), not least 
because of the difficulty encountered by most current models of galaxy formation in reproducing its apparently 
rapid evolution between $z \simeq 0$ and $z \simeq 2$ (e.g. Mitchell et al. 2014).
However, it has, until now, proved very difficult to extend the robust study of the MS beyond $z \simeq 2$ and to the highest 
masses (e.g. Steinhardt et al. 2014; Salmon et al. 2015;
Leja et al. 2015). This is because an increasing 
fraction of star formation is enshrouded in dust in high-mass galaxies, and {\it Spitzer} 
MIPS and {\it Herschel} become increasingly ineffective in the study of dust-enshrouded SF with increasing redshift (due 
to a mix of wavelength and resolution limitations), as the far-infrared emission from dust is redshifted into the sub-mm/mm regime.

A complete picture of star-formation in more massive galaxies at high-redshift can therefore only be achieved with 
ground-based sub-mm/mm observations, which provide image quality 
at sub-mm wavelengths that is vastly superior to what can currently be achieved from space. The challenge, then, is to connect the population 
of dusty, rapidly star-forming high-redshift galaxies revealed by ground-based sub-mm/mm surveys to the population 
of more moderate star-forming galaxies now being revealed by optical/near-infrared observations 
out to the highest redshifts. On a source-by-source basis this can now be achieved by targeted follow-up of 
known optical/infrared-selected galaxies with ALMA (e.g. Ono et al. 2014). However, this will inevitably produce a biased perspective 
which can only be re-balanced by also continuing to undertake ever deeper and wider sub-mm/mm surveys capable of 
detecting highly-obscured objects (again, potentially, for ALMA follow-up; Karim et al. 2013; Hodge et al. 2013), 
and thus completing our census of star-forming galaxies in the young Universe.

This is one of the primary science drivers for the SCUBA-2 Cosmology Legacy Survey (S2CLS). The S2CLS is advancing the 
field in two directions. First, building on previous efforts with SCUBA (e.g. Scott et al. 2002; Scott, Dunlop \& Serjeant 2006; Coppin et al. 2006), MAMBO
(e.g. Bertoldi et al. 2000; Greve et al. 2004), LABOCA (e.g. Weiss et al. 2009) 
and AzTEC (e.g. Austermann et al. 2010; Scott et al. 2012), the S2CLS is using the improved mapping capabilities of SCUBA-2 (Holland et al. 2013) 
to extend surveys for bright ($S_{850} > 5$\,mJy) sub-mm sources to areas of several square degrees, yielding large 
statistical samples of such sources ($>1000$). Second, the S2CLS is exploiting the very dryest (Grade-1) conditions 
at the JCMT on Mauna Kea, Hawaii to obtain very deep 450\,$\mu$m imaging of small areas of sky centred on the HST CANDELS 
fields (Grogin et al. 2011), which provide the very best multi-wavelength supporting data to facilitate galaxy counterpart identification and study.
The first such deep 450\,$\mu$m image has been completed in the centre of the COSMOS-CANDELS/UltraVISTA field, with the 
results reported by Geach et al. (2013) and Roseboom et al. (2013). Here we utilise the ultra-deep
850\,$\mu$m image of the same region, which was automatically obtained in parallel with the 450\,$\mu$m imaging.
While the dryest weather is more essential for the shorter-wavelength imaging at the JCMT, such excellent conditions 
(and long integrations) inevitably also benefit the parallel 850\,$\mu$m imaging. Consequently, the 850\,$\mu$m 
data studied here constitute the deepest ever 850\,$\mu$m survey ever undertaken over an area $\simeq 150$\,arcmin$^2$.

The depth of the new S2CLS 850\,$\mu$m imaging is typically $\sigma_{850} \simeq 0.25\,{\rm mJy}$. 
This is important because it means that galaxies detected near the limit of this survey 
have ${\rm SFR} \simeq 100\,{\rm M_{\odot} yr^{-1}}$, 
which is much more comparable to the highest SFR values derived from 
UV/optical/near-infrared studies than the typical SFR sensitivity achieved with previous single-dish sub-mm/mm 
imaging (i.e. ${\rm SFR} \simeq 1000 \,{\rm M_{\odot} yr^{-1}}$ as a result of $\sigma_{850} \simeq 2\,{\rm mJy}$).
Ultimately, of course, ALMA will provide even deeper sub-mm surveys with the resolution required to overcome 
the confusion limit of the single-dish surveys. However, because of its modest field of view ($\sim 20$\,arcsec at 
850\,$\mu$m) it is observationally expensive to survey large areas of blank sky with ALMA, and contiguous mosaic surveys 
are hard to justify at depths where the source surface density is significantly less than one per pointing. Thus,
at the intermediate depths probed here, the S2CLS continues to occupy a unique and powerful niche in the 
search for dust-enshrouded star-forming galaxies.

\begin{figure*}
\begin{center}
\includegraphics[scale=0.8]{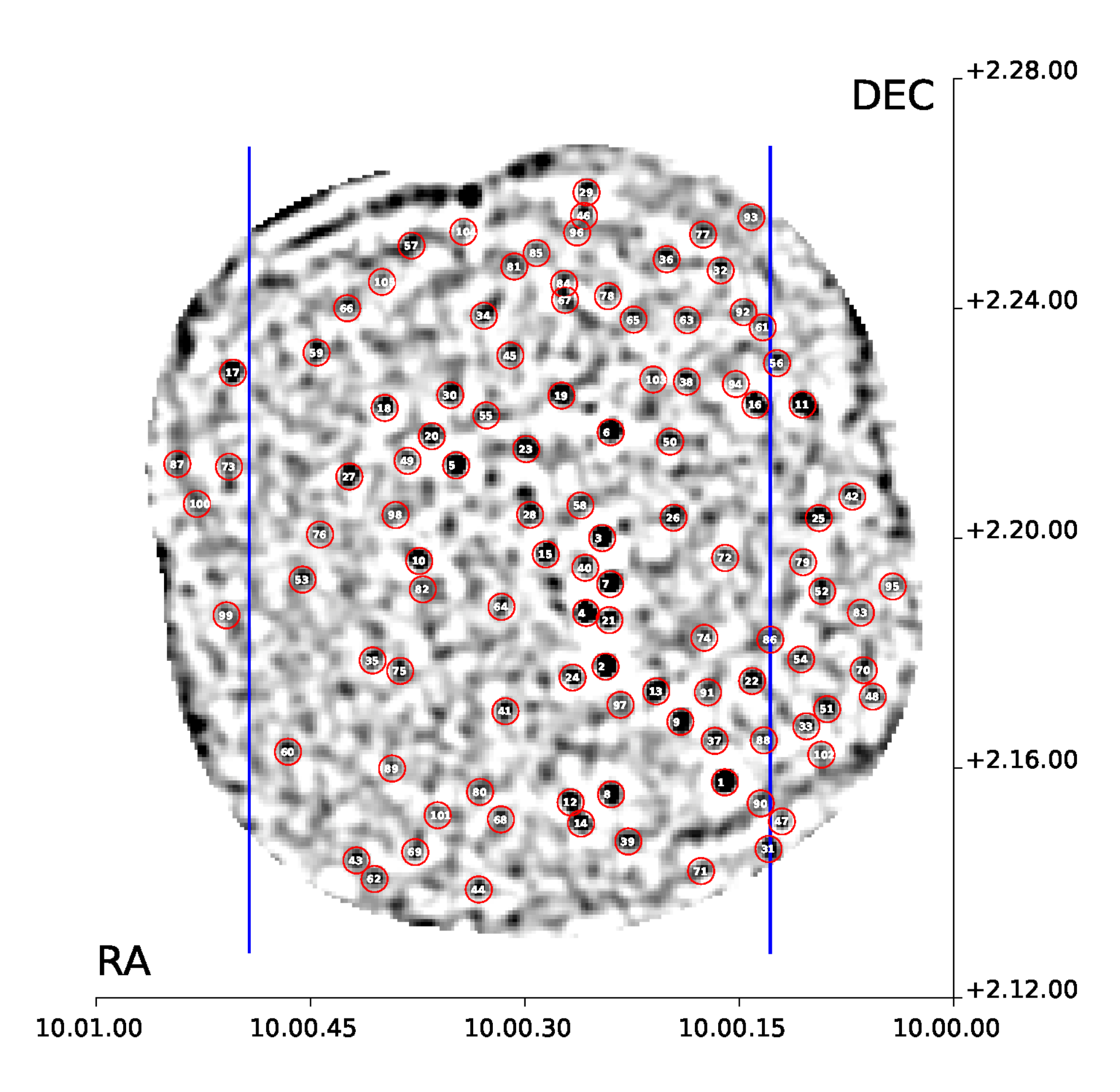}
\end{center}
\caption{The SCUBA-2 850\,${\rm \mu m}$ map of the central sub-region of the COSMOS/UltraVISTA field. 
All 106 sources extracted with ${\rm SNR}>5\sigma$ are highlighted with red circles, and marked with the 
ID number by which the sources are tabulated in Table \ref{tab:sample}. As explained in Section \ref{sec:auxdata} 
two optical--infrared catalogues were utilised in this work. The catalogue with the {\it HST} 
CANDELS and deconfused IRAC SEDS data, which contains sources extracted from the map covering 
the CANDELS area (enclosed by the two blue vertical lines), and the catalogue without the {\it HST} data 
(outside the blue lines) which was used only for 20 sources.}
\label{fig:scuba2_image}
\end{figure*}

The fact that previous sub-mm/mm surveys were only generally capable of detecting very extreme objects has undoubtedly 
contributed to some of the confusion/controversy over the nature of galaxies selected at sub-mm/mm wavelengths; 
while Micha{\l}owski et al. (2012b) and Roseboom et al. (2013) have presented evidence that sub-mm selected galaxies 
lie on the high mass end of the MS at $z = 2 - 3$, others have continued to argue that, like many local ULIRGs, they 
are extreme pathological objects driven by recent major mergers (e.g. Hainline et al. 2011). Some of this debate reflects   
disagreements over the stellar masses of the objects rather than their star-formation rates (e.g. Micha{\l}owski et al. 2014). 
Nevertheless, the fact that even high-mass galaxies on the MS lay right at the detection limits of previous sub-mm surveys inevitably 
resulted in many sub-mm selected objects apparently lying above the MS, fueling arguments about whether they were indeed 
significant outliers, or whether we have simply been uncovering the positive tail in SFR around the MS (see Roseboom et al.
2013).

The much deeper 850\,$\mu$m survey studied here is capable of settling this issue, provided of course we can overcome the 
now customary challenge of identifying the galaxy counterparts of most of the sub-mm sources, and determining their redshifts, SFRs and 
stellar masses ($M_*$) (e.g. Ivison et al. 2007; Dunlop et al. 2010; Biggs et al. 2011; Wardlow et al. 2011; Micha{\l}owski et al. 2012a; 
Koprowski et al. 2014). However, in this effort, we are also aided by the depth of the SCUBA-2 data, and by the additional positional 
information provided by the (unusual) availability of 450\,$\mu$m detections (with FWHM $\simeq$ 8\,arcsec) for 50\% of the sample.
We also benefit hugely from the unparalleled multi-frequency supporting data available in the CANDELS fields, provided by 
{\it HST}, Subaru, CFHT, Vista, {\it Spitzer}, {\it Herschel} and the VLA.

The paper is structured as follows.
In Section 2 we present the SCUBA-2 and other multi-wavelength data 
utilized in this work. Then, in Section 3 we explain how optical/infrared galaxy counterparts were 
established for the SCUBA-2 sources, and summarize the resulting identification statistics.
Next, in Section 4 we explain the calculation of the photometric redshifts, both from the optical-infrared data for the 
galaxy identifications, and from the long-wavelength data for the unidentified or spuriously identified sources. The resulting 
redshift distribution for the complete 106-source S2CLS sample is presented here, and compared with the redshift distributions 
derived from other recent sub-mm/mm surveys. In Section 5 we move on to derive and discuss the 
physical properties of the sources (such as dust temperature, bolometric luminosity, SFR, stellar mass), culminating in the 
calculation of specific SFR, and the exploration of the star-forming MS. Our conclusions are summarized in 
Section 6. All magnitudes are quoted in the AB system (Oke 1974; Oke \& Gunn
1983) and all cosmological calculations 
assume $\Omega_{M}=0.3, \Omega_{\Lambda}=0.7$ and $H_{0}=70$\,kms$^{-1}$Mpc$^{-1}$.

\section{Data}
\label{sec:data}

\subsection{SCUBA-2 imaging \& source extraction}
\label{sec:scuba}

We used the deep 850\,${\rm \mu m}$ and 450\,${\rm \mu m}$ S2CLS imaging of the central $\simeq 150\,{\rm arcmin^2}$ of the COSMOS/UltraVISTA 
field, coincident with the {\it Spitzer} SEDS (Ashby et al. 2013) and {\it HST} CANDELS (Grogin et al. 2011) imaging.
The observations were taken with SCUBA-2 mounted on the JCMT between October 2011 and March 2013, reaching depths of $\sigma_{850}\simeq 0.25$\,mJy 
and $\sigma_{450}\simeq 1.5$\,mJy (Geach et al. 2013, Roseboom et al. 2013, Geach et al. 2016 in preparation). In order to enable effective 450\,$\mu m$ observations, only the very best/dryest conditions were used
(i.e. $\tau_{225{\rm GHz}}<0.05$), and to maximise depth the imaging was undertaken with a ``daisy'' mapping pattern 
(Bintley et al. 2014). 

The details of the reduction process are described in Roseboom et al. (2013), and so 
only a brief description is given here.

The data were reduced with the {\sc smurf} package\footnote{http://www.starlink.ac.uk/docs/sun258.htx/sun258.html} V1.4.0 
(Chapin et al. 2013) with flux calibration factors (FCFs) of 606\,Jy\,pW$^{-1}$\,Beam$^{-1}$ for 
450\,$\mu$m and 556\,Jy\,pW$^{-1}$\,Beam$^{-1}$ for 850\,$\mu$m (Dempsey et al. 2013).
 
The noise-only maps were constructed by inverting an odd half of the $\sim 30$ min 
scans and stacking them all together. In the science maps the large-scale background 
was removed by applying a high-pass filter above 1.3\,Hz to the data (equivalent to 
120\,arcsec given the SCUBA-2 scan rate). Then a ``whitening filter'' was applied to 
suppress the noise in the map whereby the Fourier Transform of the map is divided by the noise-only 
map power spectrum, normalised by the white-noise level and transformed back into  
real space. The effective point-source response function (PRF) was constructed from a 
Gaussian with a full-width-half-maximum (FWHM) of 14.6\,arcsec following the same procedure.  
Finally, the real sources with a signal-to-noise ratio (SNR) of better than five were 
extracted by convolving the whitened map with the above PRF (see \S4.2 of Chapin et al. 2013).

\begin{figure}
\begin{center}
\includegraphics[scale=0.45,trim={1.7cm 2cm 0cm 5.5cm}]{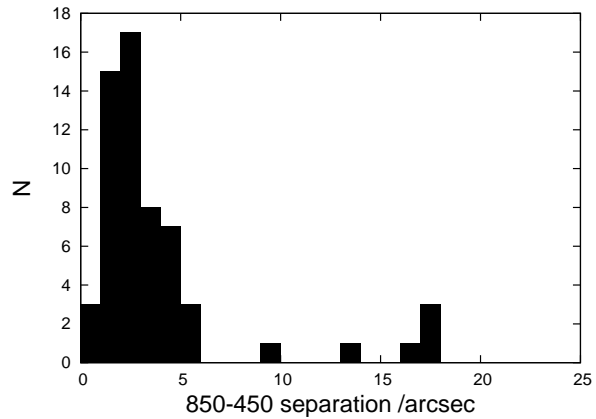}
\end{center}
\caption{The distribution of the separation between 850\,${\rm \mu m}$ and 450\,${\rm \mu m}$ positions, where the maximum search radius was set to 25\,arcsec. 
Based on this distribution, it was decided that a 450\,${\rm \mu m}$ source and an 850\,${\rm \mu m}$ source lying within 6\,arcsec of each other correspond to 
the same galaxy (for such pairs, the mean separation is $2.7\pm 0.2$\,arcsec). 
As detailed in Table \ref{tab:sample}, there are 53 850\,${\rm \mu m}$ sources with 450\,${\rm \mu m}$ counterparts (50\%).}
\label{fig:850_450_sep}
\end{figure}

The 850\,$\mu$m image and the sources extracted from it are shown in Fig. \ref{fig:scuba2_image}, while the positions and sub-mm photometry 
for the sources are listed in Appendix \ref{app_tables}, Table \ref{tab:sample}.

A total of 106 850\,$\mu$m sources were found within the map with a SNR $>5$. The photometry 
at 450\,$\mu$m was performed in the same manner,but assuming the PRF at 450\,$\mu$m to be a 
Gaussian of FWHM = 8\,arcsec. The 450\,$\mu$m counterparts to the 850\,$\mu$m sources were 
adopted if a 450\,$\mu$m-selected source  was found within 6\,arcsec 
of the 850\,$\mu$m centroid. As seen in Fig. \ref{fig:850_450_sep}, 53 850\,$\mu$m sources have 
450\,$\mu$m counterparts with the mean separation of $2.7\pm 0.2$\,arcsec. Otherwise, for the purpose of SED 
fitting, the 450\,$\mu$m flux density was measured at the 
850\,$\mu$m position (flags 1 and 0 in Table \ref{tab:sample} respectively).

\begin{figure}
\begin{center}
\includegraphics[angle=270, scale=0.32]{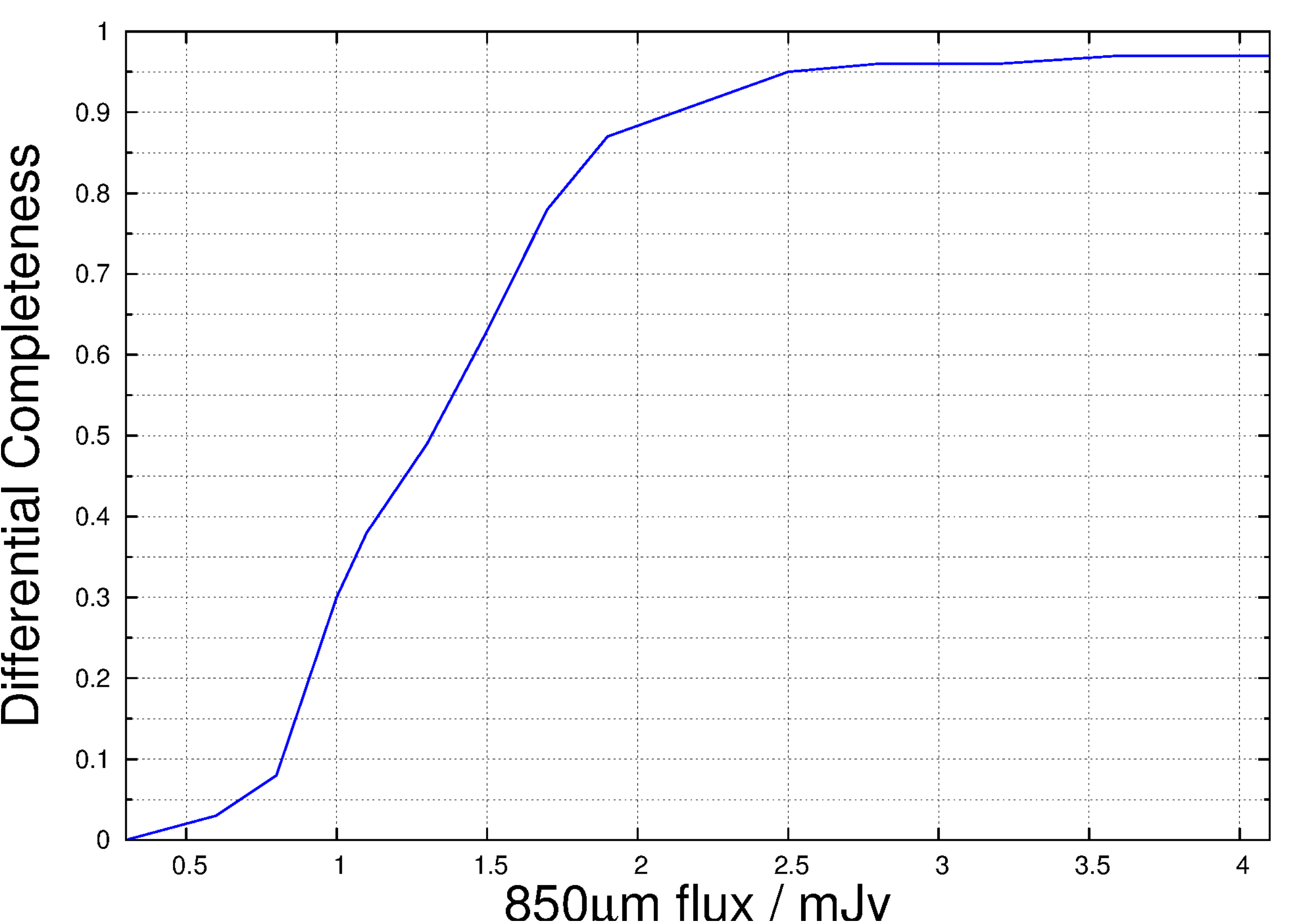}
\end{center}
\caption{Completeness of the 850\,${\rm \mu m}$ source sample as a function of flux density 
based on 2000 simulated maps with source extraction performed on $10^4$ artificially-created objects.}
\label{fig:completeness}
\end{figure}

The completeness of the 850\,$\mu$m catalogue was assessed by injecting sources of known flux density 
into the noise-only maps. Overall $10^4$ objects were used, split into 10 logarithmically-spaced flux-density 
bins between 1 and 60\,mJy. In total 2000 simulated maps were created and the source extraction was 
performed in the same way as with the real maps. The completeness was then assessed by dividing the 
number of extracted sources by the number of sources inserted into the 
noise-only maps, and the results are shown in Fig. \ref{fig:completeness}.

\subsection{Supporting multi-frequency data}
\label{sec:auxdata}

This first deep S2CLS pointing within the COSMOS/UltraVISTA field was chosen 
to maximise the power of the available ancillary multi-wavelength data, 
in particular the {\it HST} Cosmic Assembly Near-infrared Deep Extragalactic Legacy Survey 
(CANDELS)\footnote{http://candels.ucolick.org} imaging (Grogin et al. 2011). In addition, the 
optical Canada-France-Hawaii Telescope Legacy Survey (CFHTLS; Gwyn 2012), 
the Subaru/Suprime-Cam {\it z}'-band (Taniguchi et al. 2007; Furusawa et al., in preparation) 
and UltraVISTA near-infrared data (McCracken et al. 2012) 
were used. The catalogues were made by smoothing all the ground-based and 
{\it HST} data to the seeing of the UltraVISTA {\it Y}-band image with the 
Gaussian of FWHM = 0.82 arcsec (for details, see Bowler et al. 2012, 2014, 2015).  
The catalogue was selected in the smoothed CANDELS {\it H}-band image and photometry 
was measured in 3 arcsec apertures using the dual-mode function in SE{\sc xtractor} 
(Bertin \& Arnouts 1996) on all other PSF homogenised images.

The {\it Spitzer} IRAC flux densities at 3.6\,$\mu$m and 4.5\,$\mu$m were measured from the 
S-COSMOS survey (Sanders et al. 2007), after image deconfusion based on the UltraVISTA {\it K}$_s$-band 
image; using {\sc galfit} (Peng et al. 2002) the $K_S$-band images were modelled, and 
the corresponding structural parameters were then applied to both the 3.6\,$\mu$m and 4.5\,$\mu$m data 
and the flux-densities allowed to vary until the optimum fit to the IRAC image of each object was achieved
(after convolution with the appropriate PSFs). The infinite-resolution scaled model IRAC images created 
in this way were than smoothed again to match the seeing of the UltraVISTA {\it Y}-band image, after which 
the IRAC flux densities were measured within 3 arcsec apertures. 
For the small number of objects selected from the SCUBA-2 map which lay outside the 
area with CANDELS {\it HST} imaging (see Fig. \ref{fig:scuba2_image}) the {\it K}$_s$-band UltraVISTA image was used as the primary 
image for near-infrared candidate counterpart selection.

The 24\,$\mu$m catalogue was constructed using the MIPS 24\,$\mu$m imaging from the 
S-COSMOS survey (LeFloch et al. 2009). The source extraction was performed on the 
publicly-available imaging using the {\sc starfinder idl} package (Diolaiti et al. 2000). 
The resulting catalogue covers $\sim 2.1$ deg$^2$ and reaches the depth of 
$\sigma \simeq 13\,\mu$Jy (for details, see Roseboom et al. 2013). 

For the extraction of far-infrared flux densities and limits 
we used the {\it Herschel} (Pilbratt et al. 2010) Multi-tiered Extragalactic 
Survey (HerMES; Oliver et al. 2012) and the Photodetector Array Camera and Spectrometer 
(PACS; Poglitsch et al. 2010) Evolutionary Probe (PEP; Lutz et al. 2011) data obtained 
with the Spectral and Photometric Imaging Receiver (SPIRE; Griffin et al. 2010)
and PACS instruments, covering the entire COSMOS field. We utilised {\it Herschel} maps at 
$100$, $160$, $250$, $350$ and $500\,\mu$m with beam sizes of $7.39$, $11.29$, $18.2$, 
$24.9$, and $36.3$ arcsec, and $5\sigma$ sensitivities of $7.7$,  $14.7$,  
$24.0$,  $27.5$, and $30.5$\,mJy, respectively.
We obtained the fluxes of each SCUBA-2 source in the following way. We extracted  
120-arcsec wide stamps from each {\it Herschel} map around each SCUBA-2 source and used 
the PACS ($100$, $160\,\mu$m) maps to simultaneously fit Gaussians with the FWHM of 
the respective map, centred at all radio and 24\,${\rm \mu m}$ sources located 
within these cut-outs, and at the positions of the SCUBA-2 optical identifications 
(IDs, or just sub-mm positions if no IDs were selected). Then, to deconfuse the 
SPIRE ($250$, $350$ and $500\,\mu$m) maps in a similar way, we used the positions 
of the $24\,\mu$m sources detected with PACS (at $>3\sigma$), the positions of 
all radio sources, and the SCUBA-2 ID positions.

Finally the Very Large Array (VLA) COSMOS Deep catalogue was used where 
the additional VLA A-array observations at 1.4\,GHz were obtained and combined 
with the existing data from the VLA-COSMOS Large project (for details, see Schinnerer et al. 2010). 
This catalogue covers $\simeq 250$\,arcmin$^2$ and reaches 
a sensitivity of $\sigma = 12\,\mu$Jy\,beam$^{-1}$.

\begin{table*}
\begin{normalsize}
\begin{center}
\caption{The radio/infrared/optical identification statistics for the 850\,${\rm \mu m}$ S2CLS 
COSMOS sample used in this work. The number of reliably (with a probability of chance association, $p < 0.05$), 
tentatively (with $0.05<p\leq 0.1$) and all ($p\leq 0.1$) identified sources are given 
(with the percentage, out of 106 sources, in brackets). The columns give the ID success rate 
at a given band followed by the overall radio/mid-infrared ID success rate, 
the raw optical ID success rate, and revised optical 
ID success rate after checking for consistency with the long-wavelength photometric redshifts
(see Fig. \ref{fig:zcorrection} and Section \ref{sec:zcorrect}).}
\label{tab:IDrate}
\setlength{\tabcolsep}{1.7 mm} 
\begin{tabular}{lcccccc}
\hline
                             & 1.4\,GHz      & 24\,$\mu$m    & 8\,$\mu$m   & radio/IR     & optical         & optical         \\
                             &             &             &             & overall      & before corr.    & after corr.     \\
\hline
reliable  ($p\leq 0.05$)     & 15 (14$\%$) & 62 (58$\%$) & 37 (35$\%$) & 67 (63$\%$)  & 67 (63$\%$)     & 54 (51$\%$)     \\
tentative ($0.05<p\leq 0.1$) & 0  (0$\%$)  & 11 (10$\%$) & 20 (19$\%$) & 13 (12$\%$)  & 13 (12$\%$)     & 8  (8$\%$)      \\
all       ($p\leq 0.1$)      & 15 (14$\%$) & 73 (69$\%$) & 57 (54$\%$) & 80 (75$\%$)  & 80 (75$\%$)     & 62 (58$\%$)     \\
\hline
\end{tabular}
\end{center}
\end{normalsize}
\end{table*}

\section{SCUBA-2 source identifications}
\label{sec:id}

In order to find the optical counterparts for sub-mm sources, for which positions are measured with 
relatively large beams, a simple closest-match approach is not sufficiently accurate. 
We therefore use the method outlined in Dunlop et al. (1989) and Ivison et al. (2007) where we adopt the $2.5\sigma$ search radius 
around the SCUBA-2 position based on the signal-to-noise ratio (SNR): $r_s=2.5\times 0.6\times \rm{FWHM}/\rm{SNR}$, 
where $\rm{FWHM}\simeq 15$\,arcsec. In order to account for systematic astrometry shifts (caused by pointing 
inaccuracies and/or source blending; e.g. Dunlop et al. 2010) we enforce a minimum search radius of 4.5\,arcsec. 
Within this radius we calculate the corrected Poisson probability, $p$, that a given counterpart could have been selected by chance.

For reasons explained below, the VLA 1.4\,GHz and {\it Spitzer} MIPS 24\,$\mu$m and IRAC 8\,$\mu$m (with addition of 3.6\,$\mu$m) 
bands were chosen for searching for galaxy counterparts. In the case of the MIPS 24\,$\mu$m band, 
the minimum search radius was increased to 5\,arcsec to account for the significant MIPS beam size 
($\simeq 6$\,arcsec). The optical/near-infrared catalogues were then matched with these coordinates using 
a search radius of $r=1.5$\,arcsec and the closest match taken to be the optical counterpart. 
In addition, we utilised the {\it Herschel}, SCUBA-2 and VLA photometry to help isolate
likely incorrect identifications (Section~\ref{sec:zcorrect}).

The results of the identification process are summarized in Table \ref{tab:pstats}, where the most reliable 
IDs ($p\leq 0.05$) are marked in bold, the tentative IDs ($0.05<p\leq 0.1$) are marked in italics, 
and incorrectly identified sources (as discussed in Section~\ref{sec:zcorrect}) are marked with asterisks. 

\begin{figure}
\begin{center}
\includegraphics[angle=270, scale=0.46]{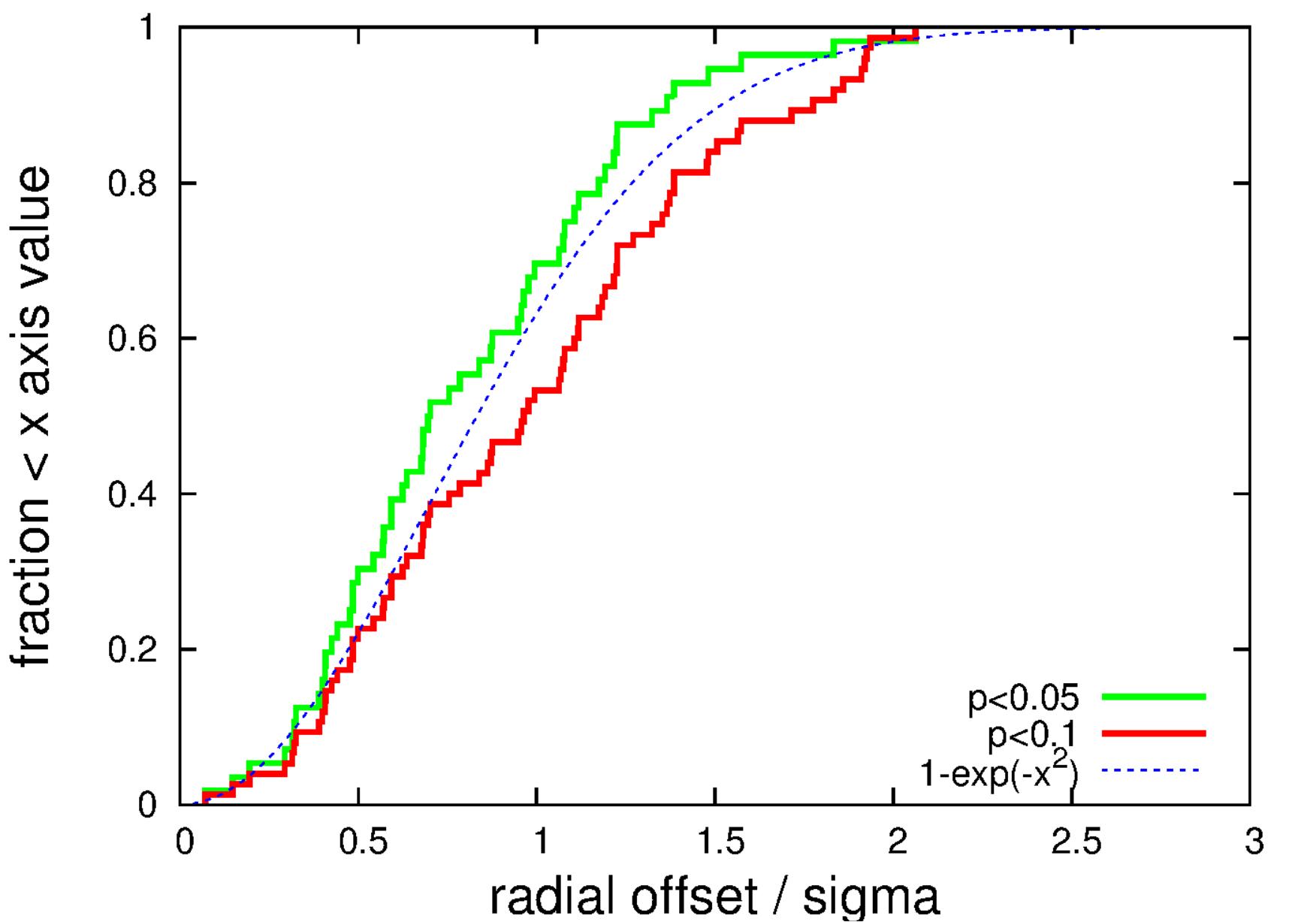}
\end{center}
\caption{A demonstration that the distribution of positional offsets between the 850\,$\mu$m sources and 
identified galaxy counterparts is consistent with statistical expectations. The histograms show the cumulative distribution
of positional offset divided by positional uncertainty, where the positional uncertainty for each SCUBA-2--counterpart 
association is derived by calculating the uncertainty in the position of both the 850\,$\mu$m source and its counterpart
(based on the standard formula $\sigma = 0.6 \times {\rm FWHM / SNR}$; see Section \ref{sec:id}), and adding these in 
quadrature, with an additional 2\,arcsec added in quadrature to account for JCMT pointing uncertainties. The histograms 
show the distributions for the most secure $p < 0.05$, and less secure $p < 0.1$ identifications, while the curve shows 
the prediction assuming a Gaussian distribution. This plot provides reassurance that, at the flux level of the sources considered
here, source confusion has not significantly distorted the source positions, and source blending is not a significant issue.}
\label{fig:radial}
\end{figure}

Given the depth of the 850\,$\mu$m imaging utilised here, it is important to check that source positions have not been significantly distorted 
by source confusion. We have therefore checked that the distribution of positional offsets between the 850\,$\mu$m 
sources and their adopted multi-frequency frequency counterparts is as expected, assuming the 
standard formula for positional uncertainty (i.e. $\sigma = 0.6 \times {\rm FWHM / SNR}$). The results, shown 
in Fig. \ref{fig:radial}, provide reassurance that the vast majority of source positions have not been significantly distorted by 
confusion/blending, and that our association process is statistically valid.

\subsection{Radio and 24\,$\bmath{\mu}$m counterparts}
\label{sec:radioid}

The 850\,$\mu$m band is sensitive to the cool dust re-radiating energy absorbed from hot, young stars. 
The radio band also traces recent star formation via synchrotron radiation from 
relativistic electrons produced within supernovae (SNe; Condon 1992). The 24\,$\mu$m waveband is in turn 
sensitive to the emission from warm dust, and since sub-mm selected galaxies are dusty star-forming galaxies, 
they are also expected to be reasonably luminous in this band. There is thus a good physical motivation for 
searching for the counterparts of SCUBA-2 sources in the VLA and MIPS imaging. In addition, the surface density of sources 
in these wavebands is low enough for chance positional coincidences to be rare (given a sufficiently small 
search radius).

As seen in Table \ref{tab:IDrate} (before the corrections of Section~\ref{sec:zcorrect}) at 1.4\,GHz 
the ID success rate is only 14$\%$ (15 out of 106 sources, all with $p\leq 0.05$) but at 
24\,$\mu$m the success rate is 69$\%$ (73 out of 106, 62 of which have $p\leq 0.05$). Combining both methods, 
the successful identification rate is 70$\%$ (74 out of 106, 63 of which have $p\leq 0.05$). The striking difference in 
these statistics is due to the fact that the S-COSMOS 24\,${\rm \mu m}$ imaging utilised here is 
relatively deeper than the radio data currently available in the COSMOS field.

\subsection{8\,$\bmath{\mu}$m counterparts}
\label{sec:iracid}

In order to maximise the identification success rate we also searched for counterparts in the 
the S-COSMOS IRAC 8\,$\mu$m imaging. At the redshifts of interest, 
this waveband traces the rest-frame near-infrared light coming from the older, mass-dominant 
stellar populations in galaxies. Given the growing evidence that sub-mm galaxies are masssive, 
it is expected they will be more luminous than average in this waveband (e.g. Dye et al. 2008; 
Michalowski et al. 2010; Biggs et al. 2011; Wardlow et al. 2011). 
We found that 57 of the 106 SCUBA-2 sources (54\%) had 8\,$\mu$m counterparts, 37 of which have $p\leq 0.05$. 
However, unsurprisingly, several of these identifications simply confirmed 
the identifications already secured via the radio and/or 24\,$\mu$m cross matching, 
and the search for 8\,$\mu$m counterparts only added 5 
new identifications (2 of which have $p\leq 0.05$) 
to the results described in the previous sub-section.

\subsection{Optical counterparts}
\label{sec:optid}

In total, therefore, we identified radio/mid-infrared counterparts for 80 of the 106 SCUBA-2 850\,$\mu$m sources 
(67 of which have $p\leq 0.05$; see Table \ref{tab:pstats}), and hence achieved an identification success rate of 75$\%$. The 
identification success rate achieved in each individual waveband is given in Table \ref{tab:IDrate}. In addition, we 
present postage-stamp images for all the sources in the online version (Fig. B1), with 
all the identifications marked with the appropriate symbols.

To complete the connection between the SCUBA-2 sources and their host galaxies, within the area covered by the 
CANDELS {\it HST} WFC3/IR imaging (Fig.\,\ref{fig:scuba2_image}) we matched the statistically-significant 
mid-infrared and radio counterparts to the galaxies in the CANDELS 
$H_{160}$-band imaging using a maximum matching radius of 1.5\,arcsec. This yielded accurate positions for 
the optical identifications of 60 of the SCUBA-2 sources (Table \ref{tab:photo1}). For those few SCUBA-2 sources which lie 
outside the CANDELS {\it HST} imaging, we matched the statistically-significant 
mid-infrared and radio counterparts to the galaxies in the $K_S$-band UltraVISTA imaging (using the same 
maximum matching radius). This yielded accurate positions for the optical identifications 
of the remaining 20 sources (Table \ref{tab:photo2}). We note that galaxies SC850-37, 46 and 61, even 
though successfully identified in the optical/near-infrared, turned out to be too close to a foreground star 
for reliable photometry (Fig. B1 in the online version) and therefore no optical redshifts or stellar masses 
were derived and utilised in the subsequent analysis.

\section{Redshifts}
\label{sec:redshifts}

\subsection{Photometric redshifts}
\label{sec:photoz}

For all the identified sources, the multi-band photometry given in 
Tables \ref{tab:photo1} and \ref{tab:photo2} was used to derive 
optical-infrared photometric redshifts using a $\chi^2$-minimization method (Cirasuolo et al. 2007, 2010) 
with a template-fitting code based on the H{\small YPER}Z package (Bolzonella et al. 2000). To create template 
galaxy SEDs, the stellar population synthesis models of Bruzual \& Charlot (2003) were applied, using 
the Chabrier (2003) stellar initial mass function (IMF) with a lower and upper mass cut-off of 
$0.1$ and $100\,{\rm M_{\odot}}$ respectively. A range of single-component star-formation 
histories were explored, as well as double-burst models. Metallicity was fixed at solar, 
but dust reddening was allowed to vary over the range $0 \leq A_V\leq 6$, assuming the law of 
Calzetti (2000). The HI absorption along the line-of-sight was applied according to Madau (1995).
The optical-infrared photometric redshifts for the 77 optically-identified sources for which 
photometry could be reliably extracted (i.e. the 80 identified sources excluding 
SC850-37, 46 and 61) are given in Table \ref{tab:z}. Also given in this table are the optical spectroscopic 
redshifts where available. We note that, in general, $z_{spec}$ and $z_p$ are 
in excellent agreement, except for the two SCUBA-2 sources which are associated with active galactic 
nuclei (AGN; sources 65 and 72), presumably because 
no AGN template was included in the photometric redshift fitting procedure.

In addition, for {\it every} SCUBA-2 source we used the 450 and 850\,$\mu$m photometry as well as the 
{\it Herschel} $100$, $160$, $250$, $350$, $500\,\mu$m and VLA 1.4\,GHz flux densities (or limits)  
to obtain `long-wavelength' photometric redshifts ($z_{\rm LW}$). This was achieved by fitting the average 
SED template of sub-mm galaxies from Michalowski et al. (2010) to the measured flux densities and errors in 
all 8 of these long-wavelength bands (including flux-density measurements corresponding to non detections). 
The resulting `long-wavelength' redshift estimates for all 106 sources are also given in Table \ref{tab:z}.

\subsection{Redshift/identification refinement}
\label{sec:zcorrect}

Given the statistical nature of the identification process described above, there is always a possibility that some identifications are incorrect
(as revealed by interferometric follow-up -- e.g. Hodge et al. 2013; Koprowski et al. 2014), and indeed, 
even when the probability of chance coincidence is extremely small, it can transpire that the optical counterpart is not, in fact,  
the correct galaxy identification, but is actually an intervening galaxy, gravitationally lensing a more distant sub-mm source (e.g. Dunlop et al. 2004). In either case, 
a mis-identification will lead to an under-estimate of the true redshift of the sub-mm source, and indeed dramatic discrepancies between 
$z_p$ and $z_{LW}$ can potentially be used to isolate mis-identified sources. 

\begin{figure*}
\begin{center}
\begin{tabular}{cc}
\includegraphics[scale=0.47,angle=270]{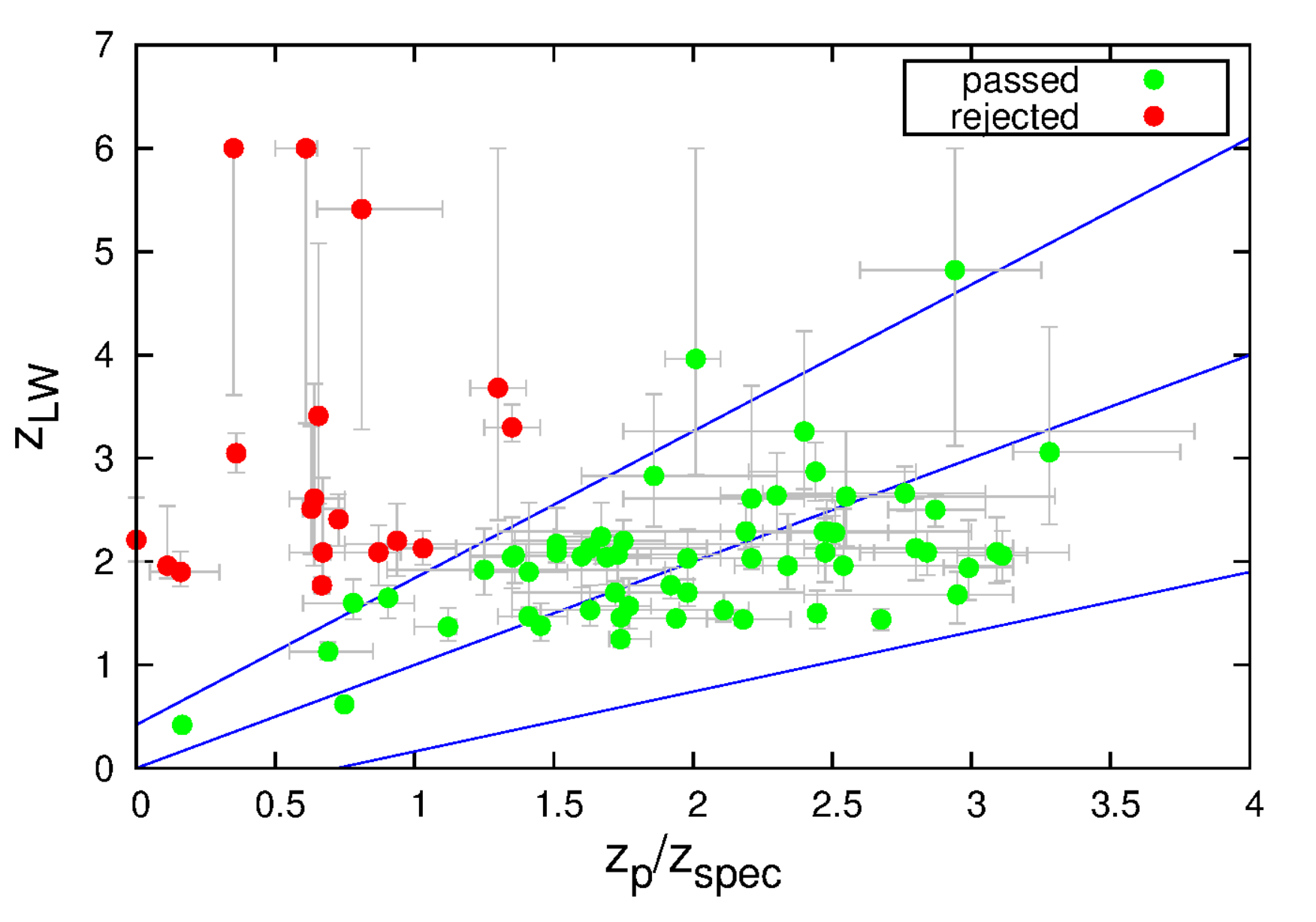}
\includegraphics[scale=0.47,angle=270]{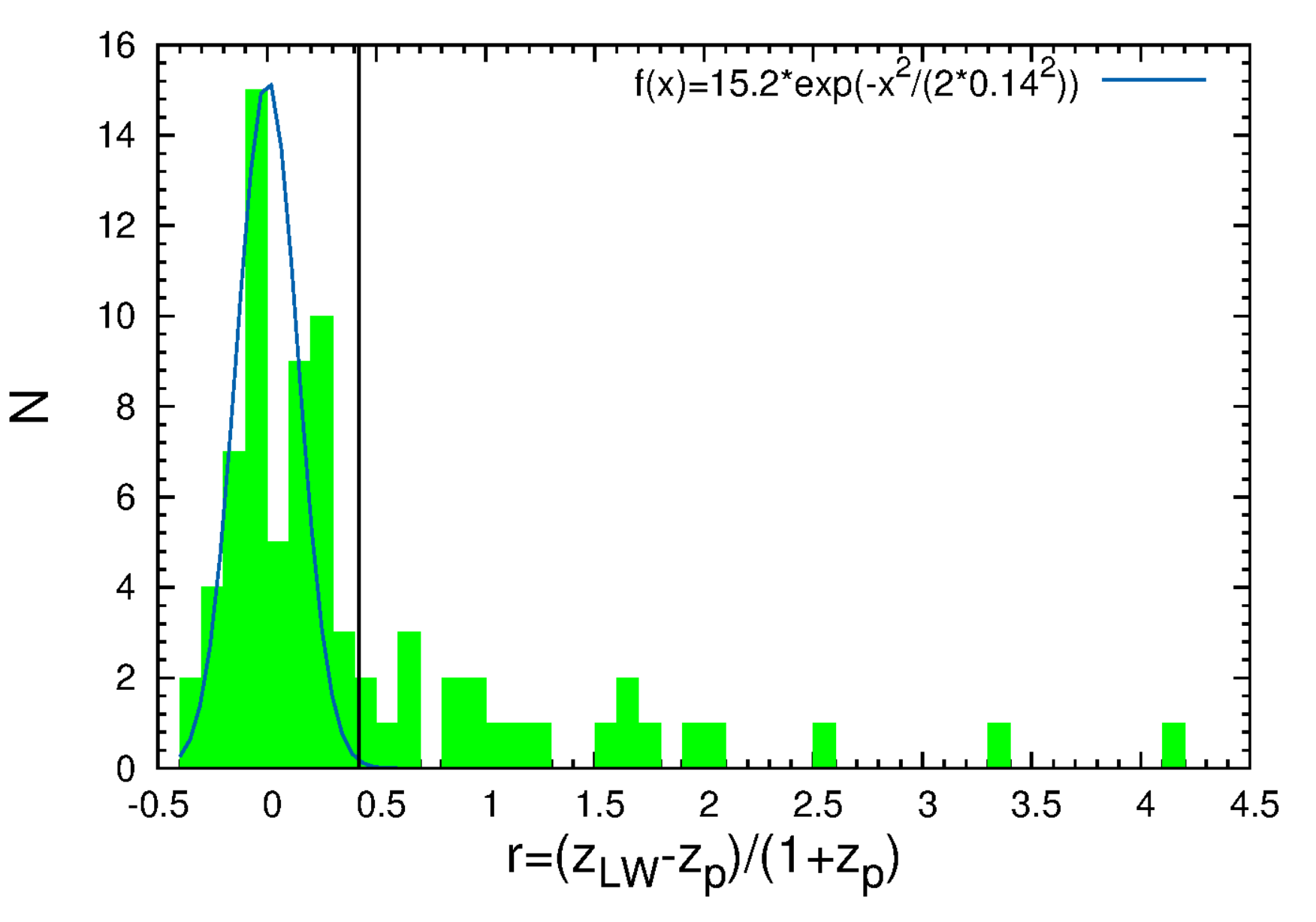}
\end{tabular}
\end{center}
\caption{The left-hand panel shows the `long-wavelength' photometric redshifts ($z_{LW}$) 
derived for the SCUBA-2 sources plotted against the optical-infrared photometric redshifts ($z_{p}$) 
of the optical identifications (see Section \ref{sec:zcorrect}). The central blue solid line shows the 1:1 relation. 
As illustrated in the right-hand panel, the sources lying below the 1:1 relation 
display a distribution of normalized redshift offsets (i.e. $r=({\rm z_{LW}}-{\rm z_p})/(1+{\rm z_p})$) 
which is approximately Gaussian with $\sigma=0.14$. The positive side of this distribution is also 
reasonably well fitted by this same Gaussian, but there is a long positive tail, indicative of the fact 
that a significant subset of the identifications have a value of $z_p$ which is much smaller than the (identification independent)
`long-wavelength' photometric redshift of the SCUBA-2 source ($z_{LW}$). Given the potential for mis-identification (e.g. 
through galaxy-galaxy gravitional lensing) we view such discrepancies as evidence that $z_{p}$, or more likely 
the galaxy identification itself, is in error. 
The upper and lower blue solid lines in the left-hand panel show the $\pm3\sigma$ limits of the Gaussian distribution, and so we 
choose to reject the optical identifications (and hence also $z_p$) for the sources that lie above the $3\sigma$ limit (red dots).
This same $3\sigma$ limit is shown by the black vertical line in the right-hand panel.}
\label{fig:zcorrection}
\end{figure*}

In Fig. \ref{fig:zcorrection} we have therefore plotted $z_{LW}$ versus $z_{p}$ in an attempt to test the consistency of these two independent redshift estimators. From this 
plot it can be seen that, for the majority of sources, the two redshift estimates are indeed consistent, with the normalized offset in $z_{LW}$ 
($r= (z_{\rm LW}-z_{\rm p})/(1+z_{\rm p})$) displaying a Gaussian distribution with $\sigma=0.14$. However, 
there is an extended {\it positive} tail to this distribution, indicative of the fact 
that a significant subset of the identifications have a value of $z_p$ which is much smaller than the (identification independent)
`long-wavelength' photometric redshift of the SCUBA-2 source, $z_{LW}$. Given the aforementioned potential for mis-identification (and 
concomitant redshift under-estimation) we have chosen to reject the optical identifications (and hence also $z_p$) for the sources that lie 
more than $3\sigma$ above the 1:1 redshift relation (see Fig. \ref{fig:zcorrection} and caption for details). This may lead to the rejection of a few correct identifications, but this is less important than the key aim of removing any significant redshift biases due to mis-identifications, and also the value 
of retaining only the most reliable set of identified sources for further study.

\begin{figure*}
\begin{center}
\begin{tabular}{cc}
\includegraphics[scale=0.47,angle=270]{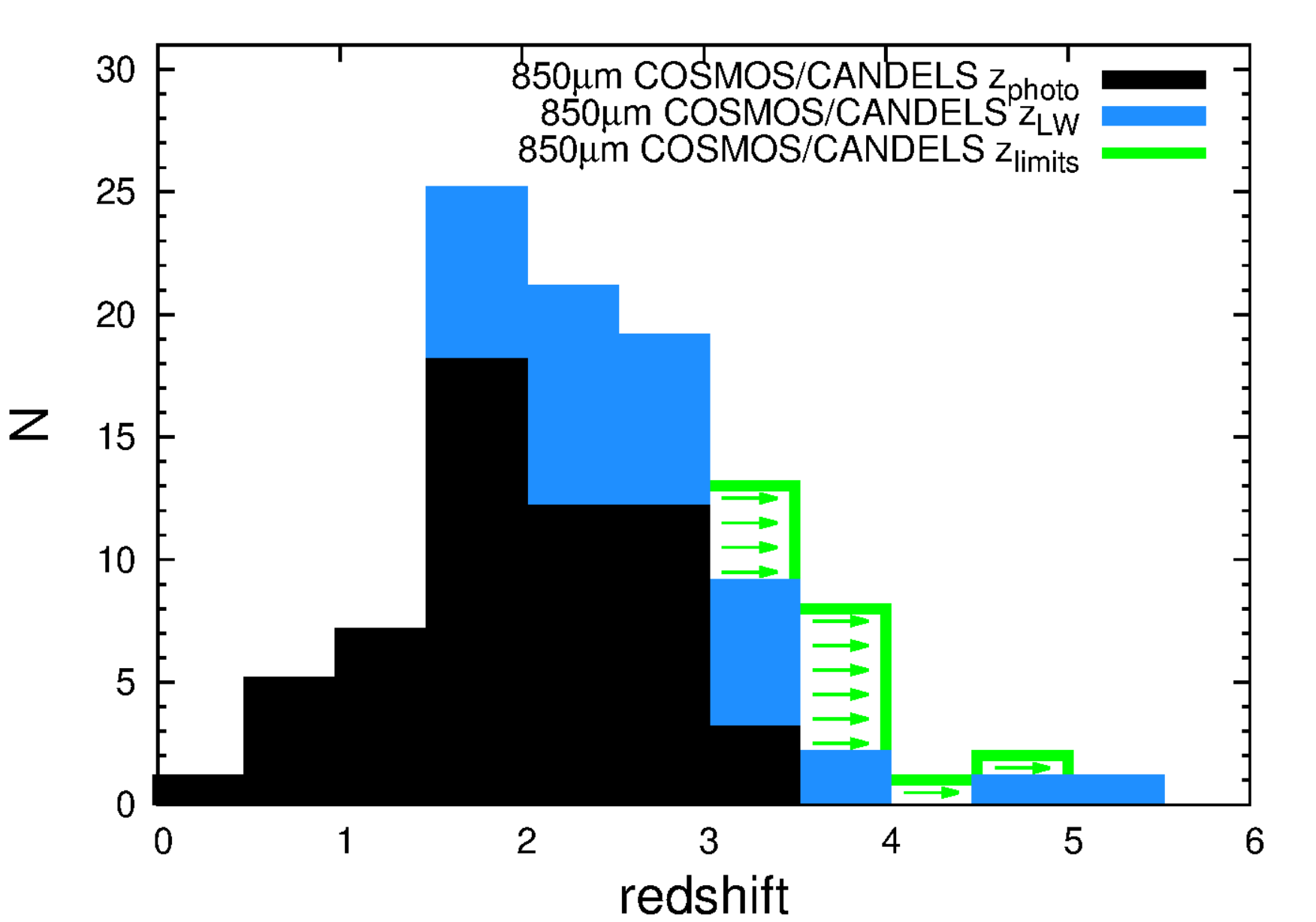}
\includegraphics[scale=0.47,angle=270]{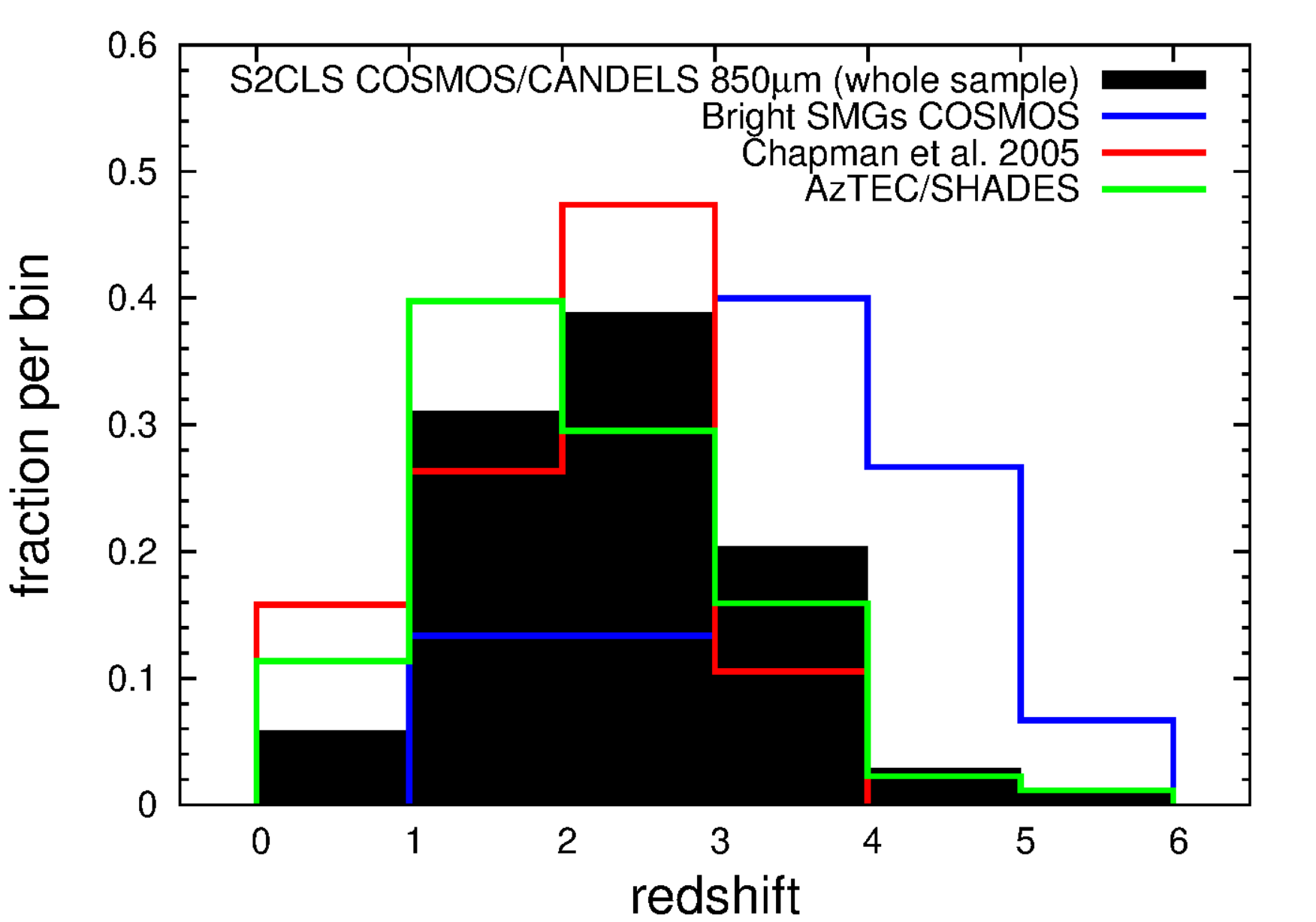}
\end{tabular}
\end{center}
\caption{\textbf{Left-hand panel:} The redshift distribution of our full 106-source S2CLS 850\,${\rm \mu m}$ 
sample in the COSMOS field (Table \ref{tab:z}). The black area shows the distribution for the robust optical 
identifications with spectroscopic or optical-infrared photometric redshifts, which has a 
mean redshift of $\bar{z}=1.97 \pm 0.09$. 
The blue line depicts the redshift distribution of the enlarged sample 
which results from adding the SCUBA-2 sources which lack robust optical identifications, but which 
have reliable `long-wavelength' redshifts ($z_{LW}$). Finally, the additional histrogram maked by 
the green arrows indicates the objects for which only lower-limits on redshift could be derived 
from the long-wavelength photometry. Adopting these lower limits, 
the mean redshift for the whole sample is $\bar{z}=2.38 \pm 0.09$. 
\textbf{Right-hand panel:} The redshift distribution for the whole S2CLS COSMOS sample overlaid with the redshift 
distributions derived by Chapman et al. (2005) ($\bar{z} = 2.00\pm 0.09$), and for the robust galaxy 
identifications in the AzTEC/SHADES survey presented by Michalowski et al. (2012a) ($\bar{z} = 2.00\pm 0.10$). 
In addition we plot the redshift distribution of the sample of luminous (sub-)mm sources in the COSMOS field
presented by Koprowski et al. (2014) ($\bar{z}=3.53 \pm 0.19$).}
\label{fig:zhistogram}
\end{figure*}

\begin{figure*}
\begin{center}
\includegraphics[scale=0.7,trim={1.5cm 2cm 0cm 0cm}]{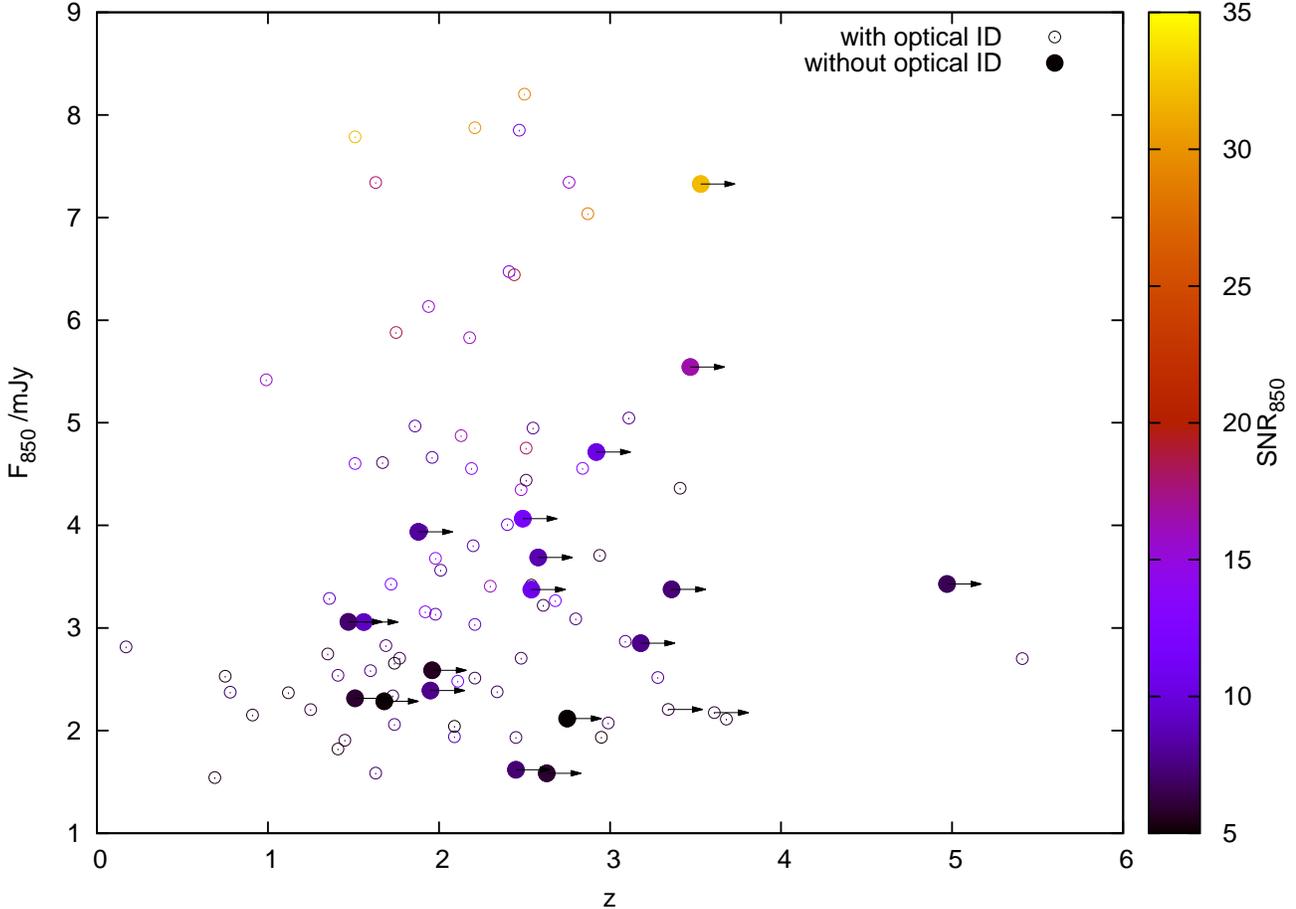}
\end{center}
\caption{850\,$\mu$m flux density as a function of redshift for the whole SCUBA-2 850-$\mu$m sample used in this study. 
The open symbols show the 80 sources which possess 
optical IDs and the filled circles show the 26 sources that lack optical IDs, with arrows signifying lower 
limits on `long-wavelength' redshifts. All the objects are colour-coded according to their 850\,$\mu$m signal-to-noise-ratio. 
It can be seen that the galaxies that lack optical associations span a similar range in SNR to the optically identified sources. 
This suggests that the failure to uncover optical counterparts for these 26 sources is not due to them being 
spurious sub-mm detections, but instead suggests that either they lie at somewhat higher redshifts than the 
typical identified source, and/or that these unidentified SCUBA-2 sources are blends of individual fainter sources whose counterparts
lie below the radio/24\,$\mu$m detection threshold.}
\label{fig:flux_z}
\end{figure*}

\begin{table*}
\begin{normalsize}
\begin{center}
\caption{Five sources in our sample that have been the subject of previous 
detailed study. Four of them (the ones with 
$z_{prev}$) were previously followed up with the mm/sub-mm interferometry. The columns show respectively 
our ID, the source name from previous work (the full previous ID name for the AzTEC source is AzTEC J100025.23+022608.0), 
interferometric RA and Dec (or, in the case of the AzTEC source, single-dish coordinates), the separation 
between the interferometric position (except AzTEC) and the optical ID found in this work, 
our final redshifts (from Table \ref{tab:z}), the redshift estimates from previous studies, and finally references. 
Where two references are given, the first one refers to the coordinates and the second to 
the previous redshift estimate (i.e. the source of $z_{prev}$).}
\label{tab:other_assoc}
\setlength{\tabcolsep}{1.7 mm} 
\begin{tabular}{lccccccl}
\hline
ID     & ${\rm ID_{other}}$ & RA & Dec & separation & $z$                & $z_{prev}$         & References                \\
SC850- &                    & /deg & /deg & /arcsec &                  &                         &                          \\
\hline
1      & MM1      & 150.0650 & 2.2636 & 2.62 & $ 3.30^{+0.22}_{-0.14}$ & $ 3.10^{+0.50}_{-0.60}$ & Aravena et al. (2010)   \\
6      & COSLA-35 & 150.0985 & 2.3653 & 0.13 & $ 2.50^{+0.20}_{-0.15}$ & $ 3.16^{+0.24}_{-0.26}$ & Smolcic et al. (2012); Koprowski et al. (2014) \\
14     & COSLA-8  & 150.1064 & 2.2523 & 2.76 & $ 2.18^{+0.17}_{-0.13}$ & $ 1.90^{+0.11}_{-0.22}$ & Smolcic et al. (2012); Koprowski et al. (2014) \\
29     & AzTEC    & 150.1051 & 2.4356 & 2.03 & $ 2.41^{+0.24}_{-0.26}$ & ...                  & Scott et al. (2008)\\
31     & COSLA-38 & 150.0525 & 2.2456 & 0.27 & $ 2.47^{+0.08}_{-0.12}$ & $ 2.44^{+0.12}_{-0.11}$ & Smolcic et al. (2012)\\
\hline
\end{tabular}
\end{center}
\end{normalsize}
\end{table*}

The effect of this cut is the rejection of 18 of the 80 optical identifications derived in Section \ref{sec:id}. These rejected optical IDs are 
flagged with asterisks in Table \ref{tab:pstats} and zeros in Table \ref{tab:z}. We emphasize that the rejection of these low-redshift identifications
does not impact significantly on the investigation of the physical properties of the sub-mm sources at $z \simeq 1 - 4$ pursued further below,
because if the low-redshift IDs were retained they would not feature in the relevant redshift bins, while adoption of the long-wavelength redshift 
for these sources means that we do not include these sources in the sample of objects with reliable stellar masses. We also stress that only a small subset 
of these objects are likely lenses (5 possible examples are highlighted in Fig.\,B2 available in the online version), but while a revised search for 
galaxy counterparts for the other sources might yield alternative counterparts with $z_p$ consistent with $z_{LW}$, we prefer not to confuse subsequent 
analysis by the inclusion of what would be inevitably less reliable galaxy identifications.

As tabulated in Table \ref{tab:IDrate}, with this redshift refinement, the effective optical ID success rate for the most reliable ($p\leq 0.05$) IDs 
drops from 63$\%$ to 51$\%$, while the overall ($p\leq 0.1$) ID success rate drops from 75$\%$ to 58$\%$. However, while this reduces the 
number of reliably identified SCUBA-2 sources to $\simeq 50$\% of the sample, this has the advantage or removing the most dubious identifications. 
Moreover, we stress that we retain redshift information for {\it every one} of the 106 SCUBA-2 sources, 
in the form of $z_{LW}$ if neither $z_{spec}$ nor a reliable value for $z_p$ are available.

In Fig.\,B2 (available in the online version) we present 12 $\times$ 12\,arcsec near-infrared postage-stamp images for every source, with the positions of all the IDs 
marked. In this figure we give the source name in red if the optical ID was in fact subsequently rejected in the light of $z_{LW}$.
It can be seen from this figure that at least some of these incorrect identifications are indeed due to galaxy-galaxy lensing (see figure caption for details).

\subsection{Redshift distribution}
\label{sec:zdist}

The differential redshift distribution for our SCUBA-2 galaxy sample is presented in Fig.~\ref{fig:zhistogram}. In the 
left-hand panel the black area depicts the redshift distribution for the sources with reliable optical IDs (and hence $z_{spec}$ or $z_p$), 
while the histrogram indicated in blue includes the additional unidentified SCUBA-2 sources with meaningful measurements of $z _{LW}$.
Finally, the green histogram containing the green arrows indicates the impact of also including those sources for which only 
lower limits on their estimated redshifts could be derived from the long-wavelength photometry. 
The mean and median redshifts for the whole sample are $\bar{z} = 2.38\pm 0.09$ (strictly speaking, a lower limit) and $z_{med}=2.21\pm 0.06$ respectively whereas,
for the confirmed optical IDs with optical spectroscopic/photometric redshifts the corresponding numbers are $\bar{z} = 1.97\pm 0.09$ and $z_{med}=1.96\pm 0.07$. 
This shows that, as expected, the radio/infrared identification process biases the mean redshift towards lower redshifts, but in 
this case only by about $\simeq 10\%$ in redshift. In addition, to make sure that our unidentified sources are in fact not spurious, 
which would manifest itself as them having low SNR values, we also plot in Fig. \ref{fig:flux_z} the 850\,$\mu$m flux as a function of redshift for the
whole sample used here, colour-coded according to SNR. It can be clearly seen that the unidentified sources exhibit a wide range of SNRs and hence are most likely
real.

In the right-hand panel of Fig.~\ref{fig:zhistogram} we compare the redshift distribution of the deep 850\,$\mu$m selected sample studied here 
with example redshift distributions from previous studies. Although our sample is somewhat deeper/fainter than the sub-mm samples studied previously by 
Chapman et al. (2005) and Micha{\l}owski et al. (2012a), the redshift distributions displayed by the optically-identified subset of sources from each study  
are remarkably consistent; we find $\bar{z} = 1.97\pm 0.09$, while Chapman et al. (2005) reported $\bar{z} = 2.00\pm 0.09$, and Micha{\l}owski et al. (2012a) reported
$\bar{z} = 2.00\pm 0.10$. 

While inclusion of our adopted values of $z_{LW}$ for our unidentified sources moves the mean redshift up to at least $\bar{z} \simeq 2.4$, it is clear that 
the redshift distribution found here cannot be consistent with that found by Koprowski et al. (2014) for the subset of very bright sub-mm/mm sources in 
the COSMOS field (see also Smolcic et al. 2012), for which $\bar{z} = 3.53\pm 0.19$. This is not due to any obvious inconsistency in redshift 
estimation techniques, as can be seen from Table \ref{tab:other_assoc} (discussed further below), and indeed the analysis methods used here are near identical to those
employed by Koprowski et al. (2014). Rather, as discussed in Koprowski et al. (2014), there must either be a trend for the most luminous 
sub-mm/mm sources (i.e. $\bar{S}_{850\mu m} \ge 8$\,mJy) to lie at significantly higher redshifts than the more typical sources studied here, or the COSMOS bright source sample 
of Scott et al. (2008) imaged by Younger et al. (2007, 2009) and Smolcic et al. (2012) must be unusually dominated by a high-redshift over-density in the COSMOS field.

\begin{figure*}
\begin{center}
\includegraphics[scale=0.55,angle=270]{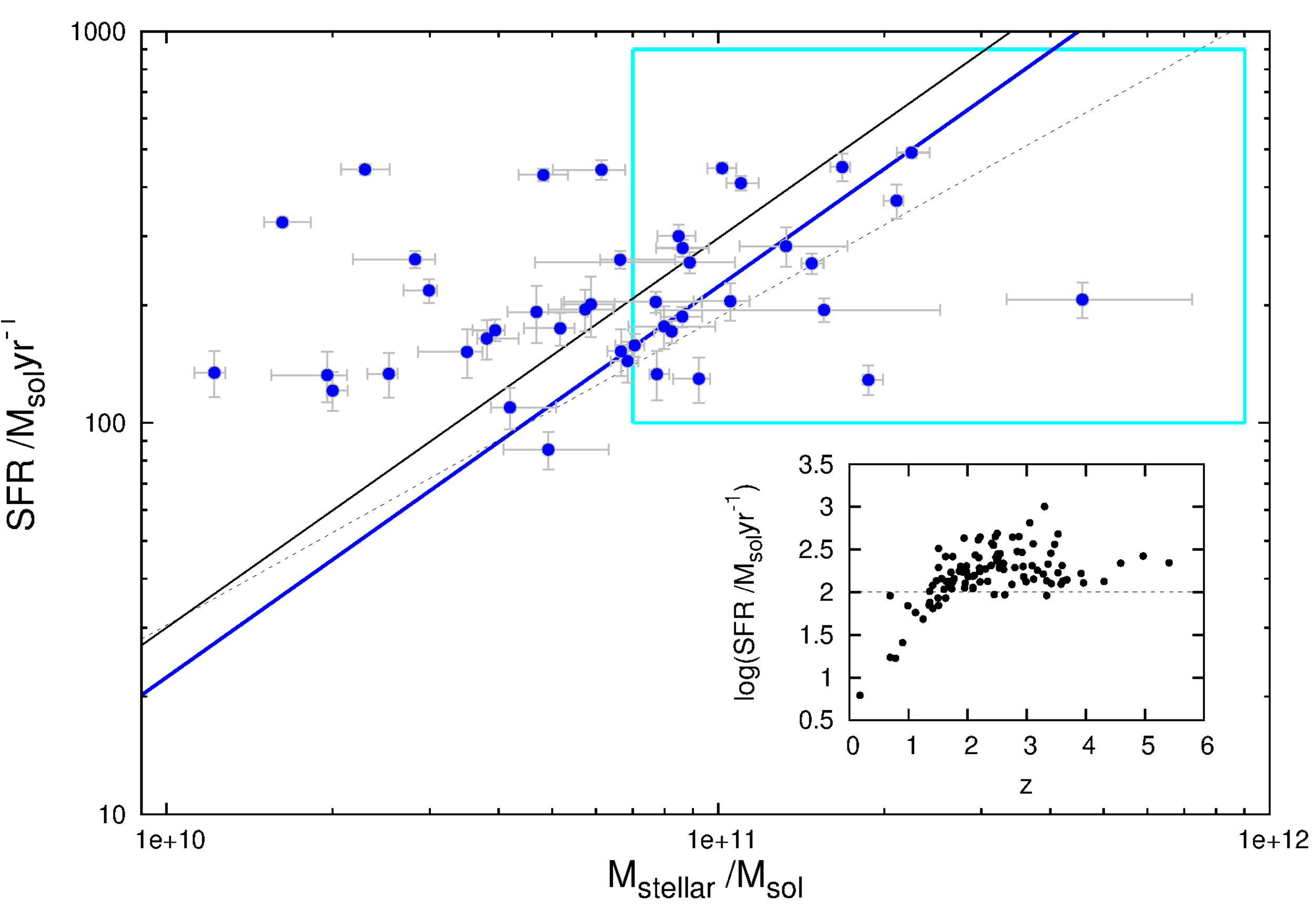}
\end{center}
\caption{The star-formation rate (SFR) as a function of stellar mass ($M_{\star}$) for the robustly identified SCUBA-2 sources with $z > 1.5$. 
As can be seen from the inset plot, due to the impact of the negative K-correction at 850\,$\mu$m, for $z > 1.5$ the flux-density limit 
of the current sample essentially equates to ${\rm SFR \simeq 100\,M_{\odot}yr^{-1}}$. The black solid line in the main plot 
shows the position of the so-called `Main-Sequence' (MS) of star-forming galaxies at $z \simeq 2.5$ as deduced by Elbaz et al. (2011),
while the black dashed line depicts the MS at $z > 1.5$ as given by Rodighiero et al. (2011). 
The sensitivity of our deep SCUBA-2 sample to values of SFR as low as $100\,{\rm M_{\odot}yr^{-1}}$ means that, for objects with stellar masses 
$M_{\star}>7\times 10^{10}\,{\rm M_{\odot}yr^{-1}}$ (i.e., inside the cyan rectangle), we are able for the first time to 
properly compare the positions of sub-mm selected galaxies on the SFR:$M_{\star}$ plane with the MS in an unbiased manner.
As shown in Fig.\,\ref{fig:MSfit}, we find that, confining our attention to $M_{\star}>7\times 10^{10}\,{\rm M_{\odot}yr^{-1}}$, the SCUBA-2 sources display a 
Gaussian distribution in specific SFR peaking at ${\rm sSFR}=2.25\pm 0.19\,{\rm Gyr^{-1}}$ (corresponding to the main sequence shown here by the 
blue solid line), demonstrating that the SCUBA-2 sources 
lie on the high-mass end of the normal star-forming MS at $z \simeq 2$.}
\label{fig:sfr}
\end{figure*}

\subsection{Previous literature associations}
\label{sec:other_assoc}

Five of the sub-mm sources in our SCUBA-2 sample have been previously studied in some detail, and so, in Table \ref{tab:other_assoc}, we compare our ID positions and redshifts 
with the pre-existing information. Four of these bright sources were previously the subject of  
interferometric mm/sub-mm observations, yielding robust optical identifications and photometric redshifts in good agreement with our results. 
The source separation for SC850-29 (2.03\,arcsec) is perfectly plausible since this is the separation between the original AzTEC single-dish coordinate 
and our chosen ID. The small separations between the positions of our adopted IDs for SC850-6 and 31 and their mm/sub-mm interferometric centroids 
confirm the reliability of our ID selection. For SC850-1 the rather large source separation of 2.62\,arcsec supports our rejection of the 
optical ID for this source. Finally, the rather large separation for SC850-14 clearly casts doubt on our adopted ID, but in this case $z_p$ is very similar to $z_{LW}$ (which, of course, is why we did not reject the ID) and so the final redshift distribution is unaffected by whether or not the ID is correct.

\section{Physical Properties}
\label{sec:otherprop}

\subsection{Stellar masses and star-formation rates}
\label{sec:sfr}

For the 58 SCUBA-2 sources for which we have secure optical identifications+redshifts (after the sample refinement discussed in Section~\ref{sec:zcorrect}) 
we were able to use the results of the SED fitting (used to determine $z_p$) to obtain an estimate of the stellar mass, $M_{\star}$, for each galaxy. 
The derived stellar masses were based on the models of Bruzual \& Charlot (2003) assuming double-burst star-formation histories (see Micha{\l}owski 
et al. 2012b), and we assumed a Chabrier (2003) IMF. 

We were also able to estimate the star-formation rate, SFR, for each of these sources by using the average long-wavelength SED of the 
sub-mm galaxies from Micha{\l}owski et al. (2010), applied to the 850\,$\mu$m flux-density of each source at the relevant photometric redshift,
to estimate the far-infrared luminosity of each source.

The resulting SFRs are plotted against $M_{\star}$ in Fig.\,\ref{fig:sfr}. In the main plot, for clarity we have confined attention to the sources with $z_p > 1.5$ because,
as shown in the inset plot, due to the impact of the negative K-correction at 850\,$\mu$m, at $z > 1.5$ the flux-density limit 
of the current sample essentially equates to ${\rm SFR \simeq 100\,M_{\odot}yr^{-1}}$ at all higher redshifts. 
In this plot we also show the position of the `Main-Sequence' (MS) of star-forming 
galaxies, as deduced at $z \simeq 2.5$ by Elbaz et al. (2011), and at $z > 1.5$ by Rodighiero et al. (2011). 
The sensitivity of our deep SCUBA-2 sample to values of SFR as low as $100\,{\rm M_{\odot}yr^{-1}}$ means that, for objects with stellar masses 
$M_{\star}>7\times 10^{10}\,{\rm M_{\odot}}$, we are able for the first time to 
properly compare the positions of sub-mm selected galaxies on the SFR:$M_{\star}$ plane with the MS in an unbiased manner.

\subsection{Specific star-formation rates}
\label{sec:sfr}

\begin{figure}
\begin{center}
\includegraphics[scale=0.47,angle=270]{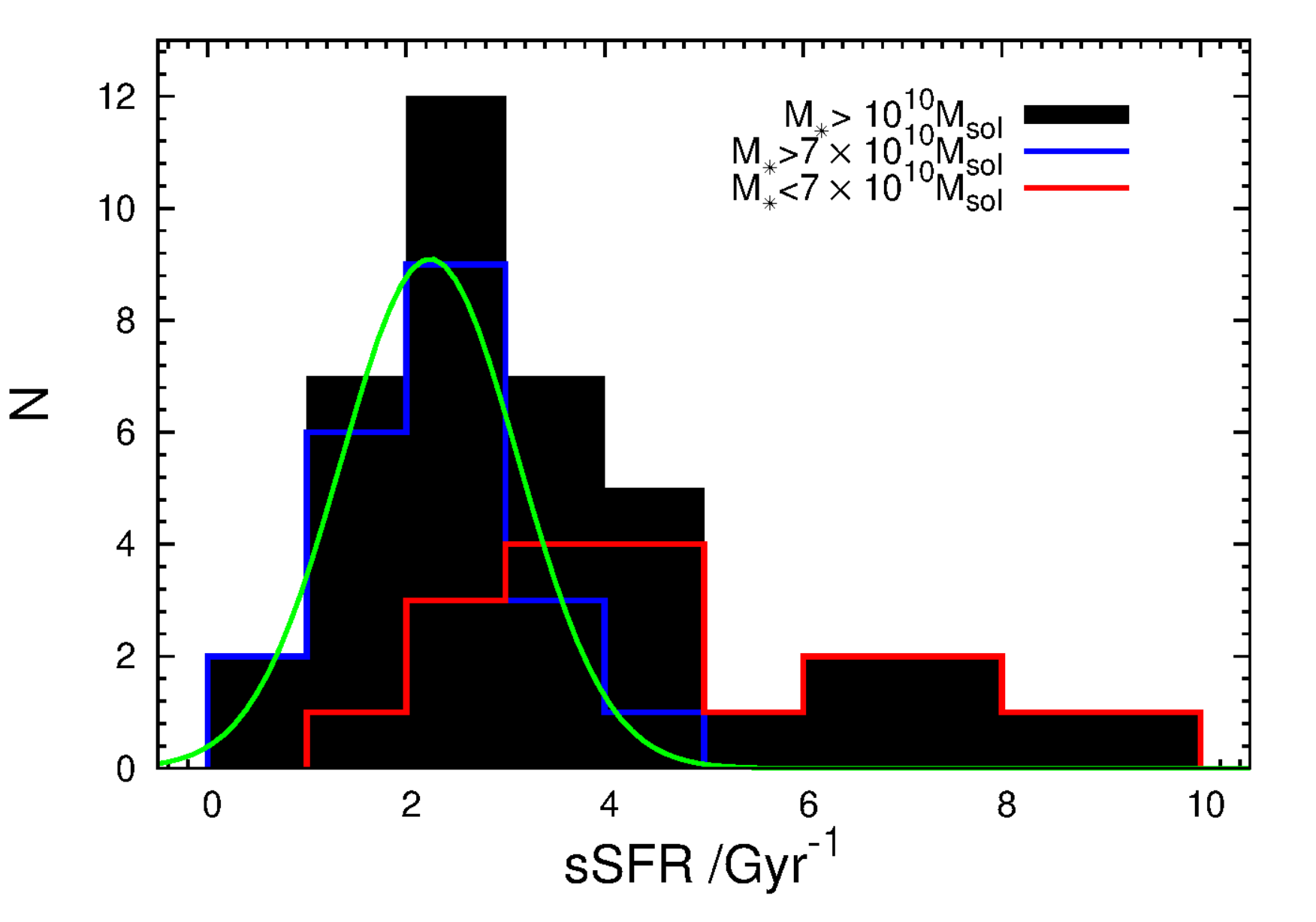}
\end{center}
\caption{The distribution of specific star-formation rate, sSFR, as derived from the 
values of SFR and $M_{\star}$ plotted in Fig. \ref{fig:sfr}. The black histogram shows the distribution for the whole robustly-identified sample of SCUBA-2 
sources at $z > 1.5$, but this can be subdivided by mass into the sub-sample with $M_{\star}>7\times 10^{10}\,{\rm M_{\odot}}$ 
(blue histogram) and the complementary sub-sample of sources with $M_{\star}<7\times 10^{10}\,{\rm M_{\odot}}$ (red histogram). 
It can be seen that, for the high-mass sample, in which SFR is not biased by the effective flux-density limit of the deep SCUBA-2 survey,
the distribution resembles closely a Gaussian peaked at ${\rm sSFR=2.25\,Gyr^{-1}}$ with $\sigma=0.89\,{\rm Gyr^{-1}}$, as shown by the green curve. This demonstrates 
that, where their distribution on the SFR:$M_{\star}$ plane can now finally be probed in an unbiased manner, the SCUBA-2 galaxies lie 
on the MS of star-forming galaxies at $z \simeq 2$.}
\label{fig:MSfit}
\end{figure}

\begin{figure}
\begin{center}
\includegraphics[scale=0.45,angle=270]{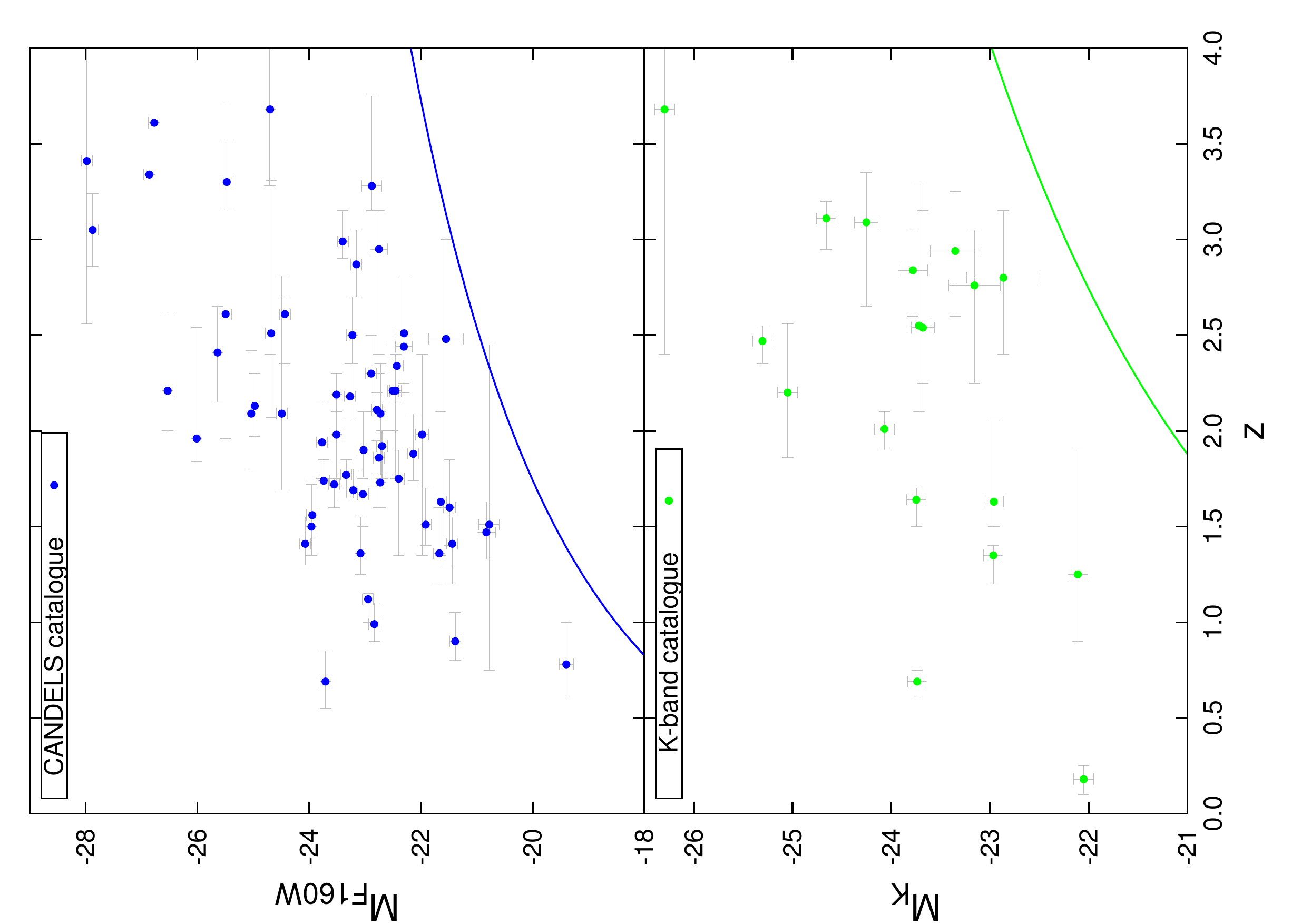}
\end{center}
\caption{Absolute magnitude versus redshift for the secure SCUBA-2 near-infrared IDs as derived for the sources lying within the {\it HST} imaging (i.e. CANDELS $H$-band neasurements; upper panel) and those lying outside the {\it HST} imaging (i.e. UltraVISTA $K_S$-band measurements; lower panel). 
The blue and green solid curves represent the detection limits of our CANDELS $H$-band and the UltraVISTA $K_S$-band selected catalogues respectively. It can be seen that 
virtually all the sources lie well above the detection limits, indicating that this our sample is not vulnerable to serious biases or incompleteness 
in stellar mass.}
\label{fig:M_z}
\end{figure}

In Fig. \ref{fig:MSfit}, we collapse the information shown in Fig. \ref{fig:sfr} into distributions of specific SFR.
The black histogram shows the distribution of sSFR for the whole robustly-identified sample of SCUBA-2 sources at $z > 1.5$,
but this can be subdivided by mass into the sub-sample with $M_{\star}>7\times 10^{10}\,{\rm M_{\odot}}$ 
(blue histogram) and the complementary sub-sample of sources with $M_{\star}<7\times 10^{10}\,{\rm M_{\odot}}$ (red histogram). 
Refering back to Fig. \ref{fig:sfr}, it can be seen that, at lower stellar masses, the measurement of sSFR is inevitably biased high 
by the effective SFR limit $>100\,{\rm M_{\odot} yr^{-1}}$, and so it is difficult to tell if these SCUBA-2 sources genuinely lie
above the MS, or if we are simply sampling the high-sSFR tail of the distribution around the MS. However, at $M_{\star}>7\times 10^{10}\,{\rm M_{\odot}}$
it is clear that the SFR limit would not produce a significantly biased sampling of the distribution of galaxies on the MS. 
In essence, because of the depth of the SCUBA-2 imaging, for sub-mm selected galaxies with $M_{\star}>7\times 10^{10}\,{\rm M_{\odot}}$ we 
should now be able to perform the first unbiased estimate of their sSFR at $z \simeq 1.5 - 3$.

In fact, for the high-mass sub-sample, in which SFR is not biased by the effective flux-density limit of the deep SCUBA-2 survey,
the distribution of sSFR resembles closely a Gaussian peaked at ${\rm sSFR=2.25\,Gyr^{-1}}$ with $\sigma=0.89\,{\rm Gyr^{-1}}$. This Gaussian fit 
is shown by the green curve in Fig. \ref{fig:MSfit}, and is completely consistent with the normalization and scatter ($\simeq 0.25$ dex) in the MS reported by 
Rodighiero et al. (2011).

Finally, to check whether we could be biased towards high-mass (and hence low sSFR) objects at high redshift, as a consequence of the flux-density limits of our optical/near-infrared catalogues, we plot the near-infrared (CANDELS $H$-band and UltraVISTA $K_S$-band) absolute magnitudes of our source IDs against redshift in Fig.\,\ref{fig:M_z}. 
The measured values are generally not close to the detection limits of our catalogues and therefore we conclude that the sample is not biased against high sSFRs at high redshifts on account of an inability to detect low-mass galaxies. 

We conclude, therefore, within the stellar mass range where we are able to sample the distribution of sSFR in an unbiased way, the sub-mm sources 
uncovered from this deep SCUBA-2 850\,$\mu$m image, display exactly the mean sSFR and scatter expected from galaxies lying on the 
high-mass end of the star-forming main-sequence at $z \simeq 2$.

\subsection{The `main sequence' and its evolution}

Given that the SCUBA-2 sources seem to, in effect, define the high-mass end of 
the star-forming main sequence (MS) of galaxies over the redshift 
range probed by our sample (i.e. $1.5 < z < 3$) it is of interest to
explore how the inferred normalization and slope of the MS as derived here 
compares to that derived from other independent studies based on very different
selection techniques over a wide range of redshifts.

\begin{figure*}
\begin{center}
\includegraphics[scale=0.6,angle=270]{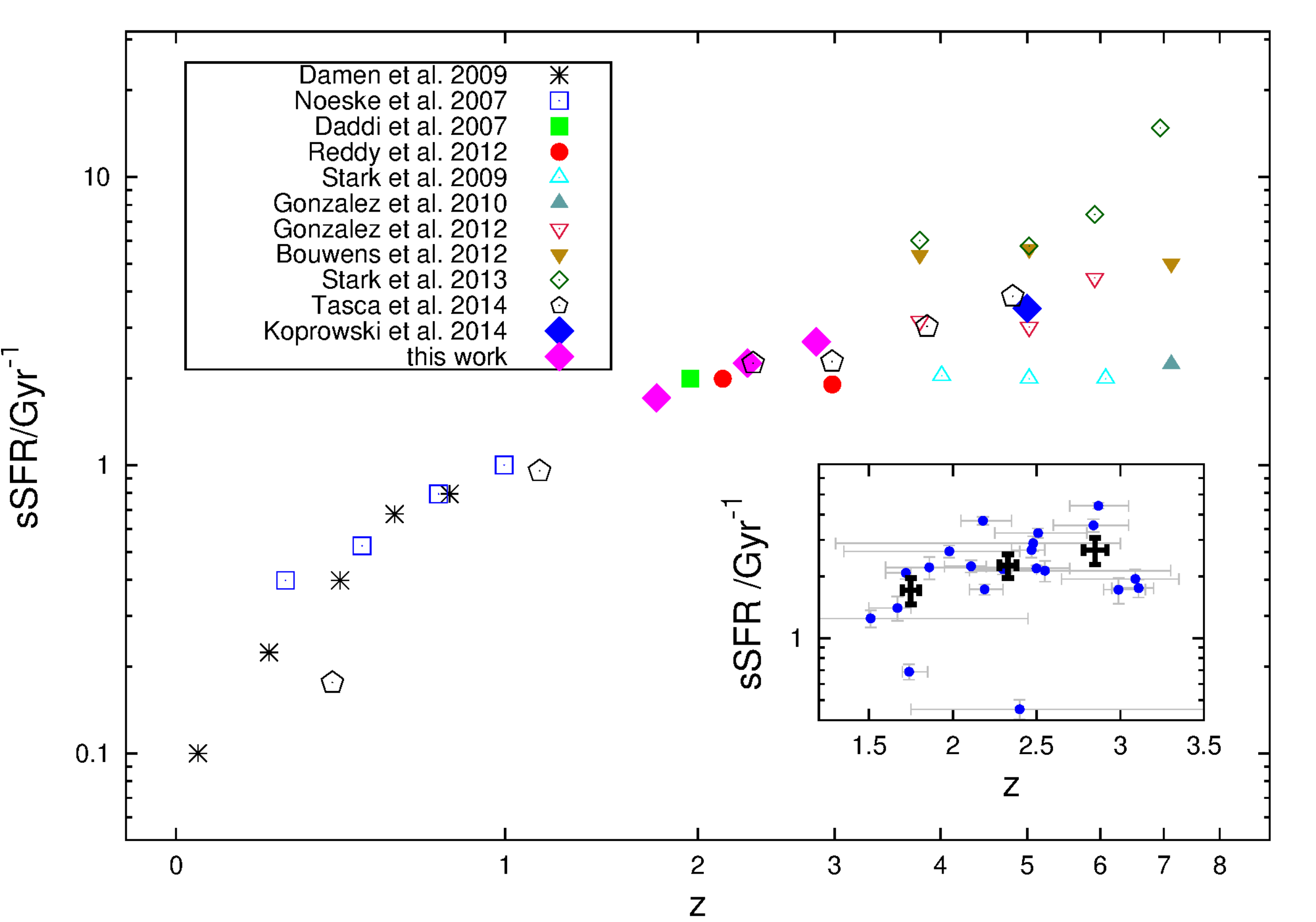}
\end{center}
\caption{Mean sSFR as a function of the redshift. The results of this work (inset plot with the mean values represented by black points with thick error bars) 
calculated using the unbiased sample (from inside the cyan rectangle in Fig.\,\ref{fig:sfr}) are shown by the magenta diamonds. 
It can be seen that the evolution of characteristic sSFR ($\equiv$ to the normalization of the MS) inferred from the SCUBA-2 galaxies 
is in excellent accord with the results from many other studies.} 
\label{fig:ssfr}
\end{figure*}

Thus, in Fig. \ref{fig:ssfr} we divide our (high-mass) sample into three redshift bins
to place the inferred evolution of sSFR within the wider context of studies 
spanning virtually all of cosmic time (i.e. $0 < z < 8$).

The first obvious striking feature of Fig. \ref{fig:ssfr} is that our new determination of 
average sSFR over the redshift range $1.5 < z < 3$ follows very closely the trend defined 
by the original studies of the MS undertaken by Noeske et al. (2007) and Daddi et al. (2007).
Since such studies were based on very different samples, sampling lower stellar masses, this result
also implies that we find no evidence for a high-mass turnover in the MS at these redshifts
(i.e. a decline in sSFR, or change in the slope of the MS above some characteristic mass).
Evidence for a decline in the slope of the MS above a stellar mass $\log (M^*/{\rm M_{\odot}}) \simeq 10.5$ 
has been presented by several authors (e.g. Whitaker et al. 2014; Tasca et al. 2015) 
but these results are based on optical/near-infrared studies, and suffer from two problems.
First, as recently discussed by Johnston et al. (2015), the results of optically-based studies 
depend crucially on how one selects star-forming galaxies, and colour selection can yield an 
apparent turn-over in the MS at high masses simply due to increased contamination from 
passive galaxies/bulges (see also Renzini \& Peng 2015; Whitaker et al. 2015). Second, 
and more important, at the high SFRs of interest here, it is well known that SED fitting 
to optical-infrared data struggles to capture the total star-formation rate because the vast 
majority of the star-formation activity in high-mass galaxies is deeply obscured.
It is therefore interesting that other recent studies of the MS based on far-infrared/sub-mm data 
also find no evidence for a high mass turnover in the MS at high-redshift; for example Schreiber
et al. (2015), from their {\it Herschel} stacking study of the MS, report that any evidence for a
flattening of the MS above $\log (M^*/{\rm M_{\odot}}) \simeq 10.5$ becomes less prominent with 
increasing redshift and vanishes by $z \simeq 2$.

As is clear from Fig. \ref{fig:sfr}, the present study does not provide sufficient dynamic 
range to enable a new measurement of the precise value and redshift evolution of the slope of the 
MS (see Speagle et al. 2014 for results from a compilation of 25 studies). Nevertheless, the advantages
of sub-mm selection for an unbiased study of the high-mass end of the MS are clear (i.e. no contamination 
from passive galaxies, and a complete census of dust-enshrouded star formation), and our results show that
the slope of the MS must remain close to unity up to stellar masses $M^*\simeq 2 \times 10^{11}\,{\rm M_{\odot}}$ at $z \simeq 2 - 3$.
We note that it is sometimes claimed that studies of the MS based on far-IR or sub-mm selected samples yield 
vastly different determinations of the SFR--$M^*$ relation from the MS (e.g. Rodighiero et al. 2014), 
but it needs to be understood that this is because previous studies based on such samples did not reach 
sufficient sensitivity in SFR (for individual objects) to properly
sample the MS at high redshift. As emphasized in Section \ref{sec:sfr}, and in Fig. \ref{fig:sfr}, even the deepest ever 
850\,$\mu$m survey analysed here only enables us to properly explore the MS at the very highest masses, due to
the effective SFR sensitivity limit; clearly the sources detected in the present study at lower masses 
are outliers from the MS, and can only provide indirect information of the scatter in the MS at 
masses of a few $\times 10^{10}\,{\rm M_{\odot}}$, rather than its normalization. 

\begin{figure*}
\begin{center}
\includegraphics[scale=0.6,angle=270]{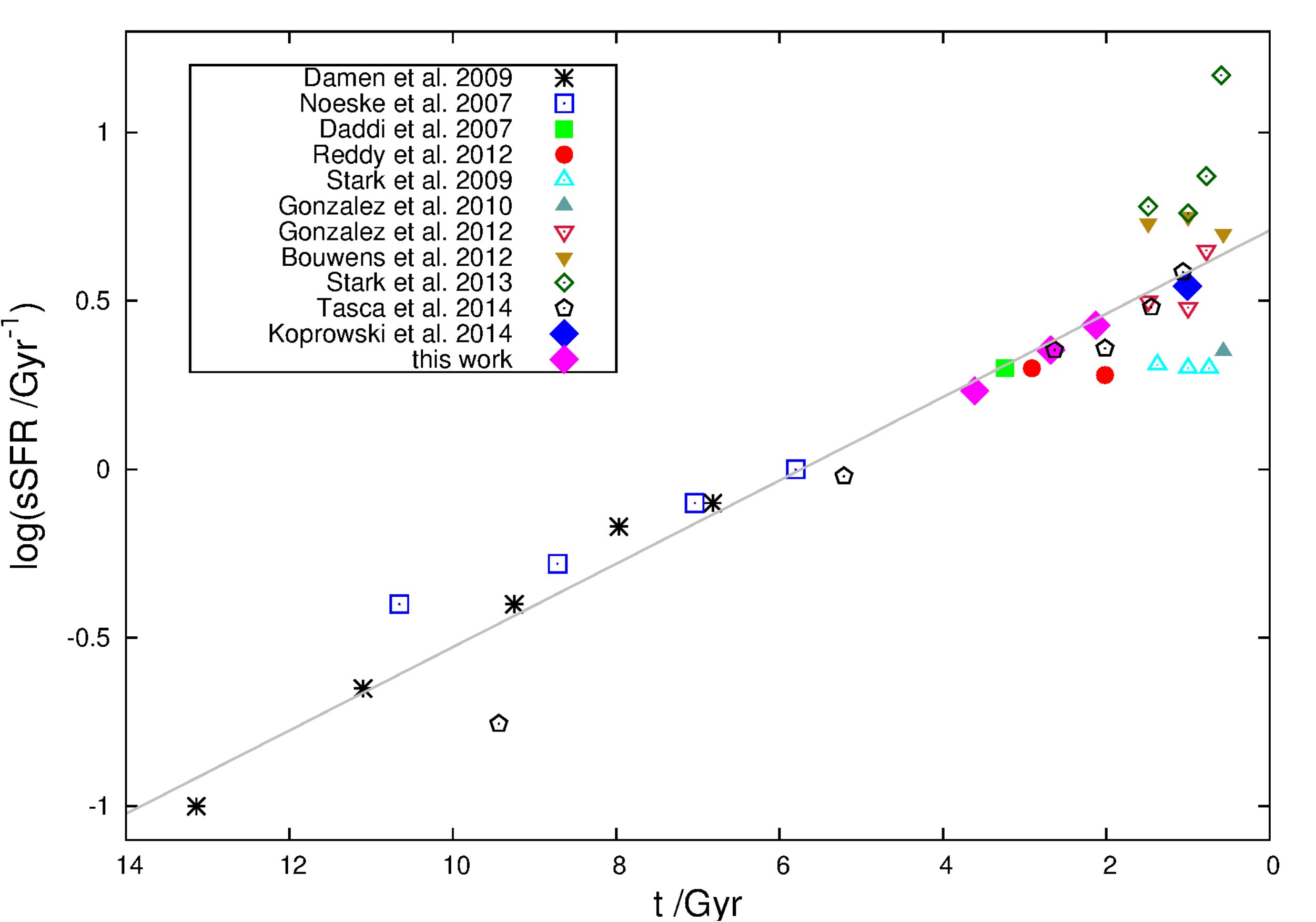}
\end{center}
\caption{Mean sSFR as a function of cosmic time. Data/symbols are as in 
Fig. \ref{fig:ssfr}. The straight-line fit to the data has the form
$\log_{10} {\rm (sSFR/Gyr^{-1})} = -0.12 (t/Gyr)+ 0.71$.}
\label{fig:ssfr_t}
\end{figure*}

Finally, looking to higher redshifts, Fig. \ref{fig:ssfr} shows that the present study does not provide useful information
on characteristic sSFR beyond $z \simeq 3$, but also demonstrates that the trend indicated here over $1.5 < z < 3$ extends naturally out 
to our previous determination of sSFR in very high-redshift sub-mm/mm galaxies at $z \simeq 5$ (Koprowski et al. 2014).
There is currently considerable debate over the normalization of the MS at $z \simeq 4$, due in large part to uncertainty 
over the impact of nebular emission lines on the estimation of stellar masses (see e.g. Stark et al. 2013; Smit et al. 2014).
However, the sub-mm studies of high-mass star-forming galaxies are clearly consistent with the results of several existing studies
(e.g. Steinhardt et al. 2014), and (despite their 
supposedly extreme star-formation rates) sub-mm--selected  galaxies provide additional support for the presence of a `knee' in the evolution of 
sSFR around $z \simeq 2$ (as originally suggested by the results of 
Gonzalez et al. 2010, 2012). The ability of theoretical models of galaxy formation to reproduce this transition remains 
the subject of continued debate, with smooth cold accretion onto dark matter halos leading to expectations 
that sSFR should rise $\propto (1+z)^{2.5}$ (Dekel et al. 2009, 2013; Facher-Gigu\'{e}re et al. 2011; Rodr\'{i}guez-Puebla et al. 2016), and a range of  
hydrodynamical and semi-analytic models of galaxy formation yielding predictions 
of characteristic sSFR at $z \simeq 2$ that fall short of the results shown in Fig. \ref{fig:ssfr} 
by a factor of $2 - 6$ (see discussion in Johnston et al. 2015, and references 
therein). However, in Fig. \ref{fig:ssfr_t} we show that when the redshift axis is re-cast in terms of 
cosmic time, there is really no obvious feature in the evolution of characteristic sSFR. Rather, the challenge for theoretical
models is to reproduce the apparently simple fact that $\log_{10}$sSFR is a linear function of the age of the Universe, 
at least out to the highest redshifts probed to date.

\section{Summary}
\label{sec:summary}

We have investigated the multi-wavelength properties of the galaxies selected from the deepest 850-$\mu$m 
survey undertaken to date with SCUBA-2 on the JCMT. This deep 850-$\mu$m imaging was taken in 
parallel with deep 450\,$\mu$m imaging in the very best observing conditions as part of the SCUBA-2 Cosmology Legacy 
Survey. A total of 106 sources ($>$5$\sigma$) were uncovered at 850\,$\mu$m from an  
area of $\simeq 150$\,arcmin$^2$ in the centre of the COSMOS/UltraVISTA/CANDELS field, 
imaged to a typical depth of $\sigma_{850} \simeq 0.25$\,mJy. Aided by radio, mid-IR, and 450-$\mu$m positional 
information, we established statistically-robust galaxy counterparts 
for 80 of these sources ($\simeq 75$\%). 

By combining the optical-infrared photometric redshifts, $z_p$, of these 
galaxies with independent `long-wavelength' estimates of redshift, $z_{LW}$
(based on {\it Herschel}/SCUBA-2/VLA photometry), we have been able to refine the list of 
robust galaxy identifications. This approach has also enabled us to 
complete the redshift content of the whole sample, yielding $\bar{z} = 2.38\pm 0.09$, a mean redshift 
comparable with that derived from all but the brightest previous sub-mm samples. 

Because our new deep 850-$\mu$m selected galaxy sample reaches flux densities equivalent to star-formation rates 
${\rm SFR} \simeq 100\,{\rm M_{\odot} yr^{-1}}$, we have been able to confirm that sub-mm galaxies form the high-mass 
end of the `main sequence' (MS) of star-forming galaxies at $z > 1.5$ (with a mean specific SFR of 
${\rm sSFR}=2.25 \pm 0.19\,{\rm Gyr^{-1}}$ at $z \simeq 2.5$). Our results are consistent with no significant 
flattening of the MS towards high stellar masses at these redshifts
(i.e. SFR continues $\propto M_*$), suggesting that reports of such flattening are based on contamination by 
passive galaxies/bulges, and/or under-estimates of dust-enshrouded star-formation activity in massive star-forming galaxies. 
However, our findings contribute to the growing evidence that average sSFR rises only slowly at high redshift, from 
${\rm sSFR}\simeq 2 \,{\rm Gyr^{-1}}$ at $z \simeq 2$ to ${\rm sSFR} \simeq 4\,{\rm Gyr^{-1}}$ at $z \simeq 5$.
These results are consistent with a rather simple evolution of global characteristic sSFR, 
in which $\log_{10}$sSFR is a linear function of the age of the Universe, 
at least out to the highest redshifts probed to date.

\section*{Acknowledgments}

MPK acknowledges the support of the UK Science and Technology Facilities Council. JSD and RAAB acknowledge the support of the 
European Research Council via the award of an Advanced Grant (PI J. Dunlop). JSD also acknowledges the contribution of the EC FP7 SPACE project 
ASTRODEEP (Ref.No: 312725). MJM acknowledges the support of the UK Science and Technology Facilities Council, 
and the FWO Pegasus Marie Curie Fellowship. MC acknowledges the support of the UK Science and Technology Facilities Council via an Advanced Fellowship.

The James Clerk Maxwell telescope has historically been operated by the Joint Astronomy
Centre on behalf of the Science and Technology Facilities Council
of the United Kingdom, the National Research Council of Canada, and the Netherlands Organisation for Scien-
tific Research. Additional funds for the construction of SCUBA-2 were provided by
the Canada Foundation for Innovation. 
This work is based in part on data products from observations made with ESO Telescopes at the 
La Silla Paranal Observatories under ESO programme ID 179.A-2005 and on data 
products produced by TERAPIX and the Cambridge Astronomy survey Unit 
on behalf of the UltraVISTA consortium.
This work is based in part on observations obtained with MegaPrime/MegaCam a joint project of CFHT and CEA/DAPNIA, 
at the Canada-France-Hawaii Telescope (CFHT) which is operated by the National Research Council (NRC) of 
Canada, the Institut National des Science de l'Univers of the Centre National de la Recherche Scientifique (CNRS) of France, 
and the University of Hawaii.
This work is based in part on data products produced at TERAPIX and 
the Canadian Astronomy Data Centre as part of the Canada-France-Hawaii Telescope Legacy Survey, a collaborative 
project of NRC and CNRS.
 This work is based in part on observations made with the NASA/ESA {\it Hubble Space Telescope}, which is operated by the Association 
of Universities for Research in Astronomy, Inc, under NASA contract NAS5-26555.
This work is also based in part on observations made with the {\it Spitzer Space Telescope}, which is operated by the Jet Propulsion Laboratory, 
California Institute of Technology under NASA contract 1407, as well as
the observations made with ESO Telescopes at the La Silla or Paranal Observatories under programme ID 175.A-0839.
{\it Herschel} is an ESA space observatory with science instruments provided by European-led
Principal Investigator consortia and with important participation
from NASA. We thank the staff of the Subaru telescope for their 
assistance with the $z'$-band imaging utilised here.
This research has made use of the NASA/IPAC Infrared Science Archive, which is operated 
by the Jet Propulsion Laboratory, California Institute of Technology, under contract with the National Aeronautics and Space Administration.

{}

\clearpage

\appendix

\section{Data Tables}
\label{app_tables}

In this appendix we provide tables detailing: {\bf i)} 
the sub-mm properties of the 
deep 106-source 
850$\mu$m-selected SCUBA-2 sample utilised in this study, {\bf ii)}
the results of the galaxy counterpart identification process, {\bf iii)}
the optical-infrared photometry for the galaxy identifications, and {\bf iv)}
the estimated redshifts and derived physical properties of the sub-mm galaxies.

\begin{table*}
\caption{The basic properties of the 106-source SCUBA-2 850\,$\mu$m-selected sample. The penultimate column gives the SCUBA-2 colour where,  
if the significance of the 450\,${\rm \mu m}$ detection is less than $2\sigma$, the SCUBA-2 colour is based on a 450\,$\mu$m limit with $S_{450}< S_{450}+2\sigma$. 
The flag given in the final column indicates whether the 450\,$\mu$m flux density was taken from 450\,$\mu$m catalogue (1) or simply measured at the 850\,$\mu$m position 
(0); the latter measurement was adopted if no 450\,$\mu$m-selected source with $S_{450}>4\sigma$ was found within 6\,arcsec of the 850\,$\mu$m source position).} 
\setlength{\tabcolsep}{2.7 mm} 
\include{a1}
\label{tab:sample}
\end{table*}

\addtocounter{table}{-1}
\begin{table*}
\caption{(continued).}
\setlength{\tabcolsep}{2.7 mm} 
\include{a2}
\end{table*}

\begin{table*}
\caption{The results of the radio/mid-infrared statistical identification process described in Section \ref{sec:id}. The columns give the SCUBA-2 source number, 
the positions of the adopted optical ID, the VLA 1.4\,GHz coordinates (where a radio ID exists), and the relevant 
mid-infrared and radio flux densities, angular offsets (from the SCUBA-2 850\,$\mu$m position), and corrected probabilities, $p$, that each association 
could have occurred by chance (given the depth of the supporting data, the relevant number counts,  and the counterpart search radius).
If a given ID is listed more than once, the counterpart with the lowest $p$-value was treated as a correct association. 
The robust IDs ($p\leq 0.05$) are shown in bold, the more tentative IDs ($0.05<p\leq 0.1$) in italics, and the sources for which the optical/near-infrared IDs were rejected 
on the basis of the $z_p$ -- $z_{LW}$ comparison (see Section \ref{sec:zcorrect}) are marked with asterisks.}
\setlength{\tabcolsep}{1.0 mm} 
\include{p_stats}
\label{tab:pstats}
\end{table*}

\begin{landscape}
\begin{table}
\caption{The optical--infrared photometry for the SCUBA-2 identifications that lie within the CANDELS {\it HST} imaging. The table gives CFHTLS optical, Subaru {\it z}'-band, UltraVISTA 
near-infrared, IRAC mid-infrared, and {\it HST}
AB magnitudes as measured through 3\,arcsec-diameter apertures. Errors are the $1\sigma$ values, with a minimum adopted error of 0.1\,mag.}
\setlength{\tabcolsep}{0.7 mm} 
\include{photo1}
\label{tab:photo1}
\end{table}
\end{landscape}

\begin{landscape}
\begin{table}
\caption{The optical--infrared photometry for the SCUBA-2 identifications which lie outside the CANDELS {\it HST} imaging.
The table gives CFHTLS optical, Subaru {\it z}'-band, UltraVISTA near-infrared and IRAC mid-infrared AB magnitudes measured through 3\,arcsec-diameter apertures. 
Errors are the $1\sigma$ values with a minimum adopted error of 0.1\,mag.}
\setlength{\tabcolsep}{0.7 mm} 
\include{photo2}
\label{tab:photo2}
\end{table}
\end{landscape}

\begin{table*}
\caption{The derived physical properties of all 106 sources in the SCUBA-2 sample. The columns show, respectively, SCUBA-2 source number, 
optical spectroscopic redshift (should it exist for a robust ID), optical--infrared photometric redshift $z_p$, 
the `long-wavelength' photometric redshift $z_{LW}$, the normalized redshift offset $r={\rm (z_{LW}-z_p)/(1+z_p)}$ (see Section \ref{sec:zcorrect}), 
a flag indicating the status of the redshift information, and our final adopted redshift (with estimated errors), star-formation rate (SFR) and stellar mass ($M_{\star}$). 
If a source's optical/near-infrared ID was rejected on the basis of an excessive value of $r$, it is flagged here with 0, if accepted it is flagged with 1 and if no ID was found the flag is set to 2. 
For objects flagged with 1 the final redshift $z$ is the optical--infrared photometric redshift $z_p$ (or $z_{spec}$ if it exists), 
and therefore a stellar mass can be estimated for the galaxy and is given in the final column. If the flag is 0 or 2, the final adopted redshift $z$ becomes $z_{LW}$, but no stellar mass can be 
calculated due to the absence of any optical--infrared photometry. The last column gives the source for the spectroscopic redshifts. These include 11 redshifts from DR1 of the 
zCOSMOS redshift survey undertaken in the COSMOS field with the VIMOS spectrograph (Lilly et al. 2007), 2 redshifts obtained as a part of the 3D-{\it HST} observations of the COSMOS field (Brammer et al. 2012, Skelton et al. 2014) and 
2 redshifts from the spectroscopic survey undertaken in the COSMOS field with DEIMOS spectrograph (PI: Jehan Kartheltepe).}
\setlength{\tabcolsep}{2.7 mm} 
\include{z1}
\label{tab:z}
\end{table*}

\addtocounter{table}{-1}
\begin{table*}
\caption{(continued).}
\setlength{\tabcolsep}{2.7 mm} 
\include{z2}
\end{table*}

\FloatBarrier

\section{Multi-frequency images and source identifications}

In this second appendix we first provide multi-wavelength postage-stamp 
images of all 106 SCUBA-2 sources to illustrate the results of the identification process,
and then show higher-resolution versions of the 
deepest available near-infrared images of the sources, highlighting both 
the identified and unidentified sources, along with their estimated redshifts.

\begin{figure*}
\begin{center}
\includegraphics[scale=0.8,trim=5cm 2cm 5cm 2cm]{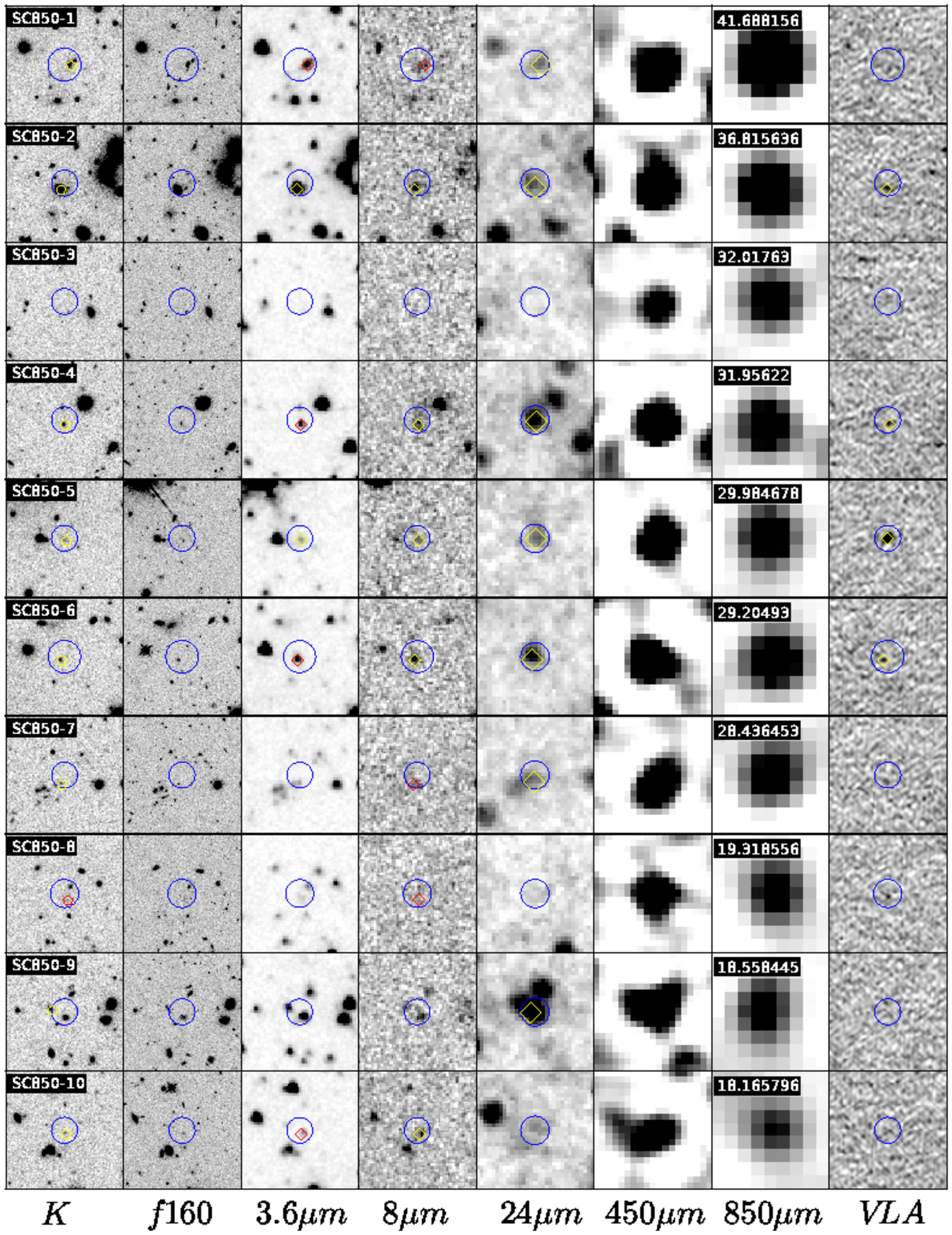}
\end{center}
\caption{Multi-wavelength postage-stamp images of all 106 SCUBA-2 sources. Each image is centred on the 850\,$\mu$m source position, and is 
36 $\times$ 36\,arcsec, with North to the top and East to the left.
The blue circle in each stamp shows the search radius adopted for each source when searching for 
counterparts and calculating the significance of any positional association. The yellow and red symbols depict the most 
probable counterparts with $p<0.05$ and $0.05<p<0.1$ respectively. The number given on the 850\,${\rm \mu m}$ stamp is the 
signal-to-noise ratio of the 850\,${\rm \mu m}$ detection; the sources are ranked by this number.}
\label{fig:8filters}
\end{figure*}

\addtocounter{figure}{-1}
\begin{figure*}
\begin{center}
\includegraphics[scale=0.8,trim=5cm 2cm 5cm 2cm]{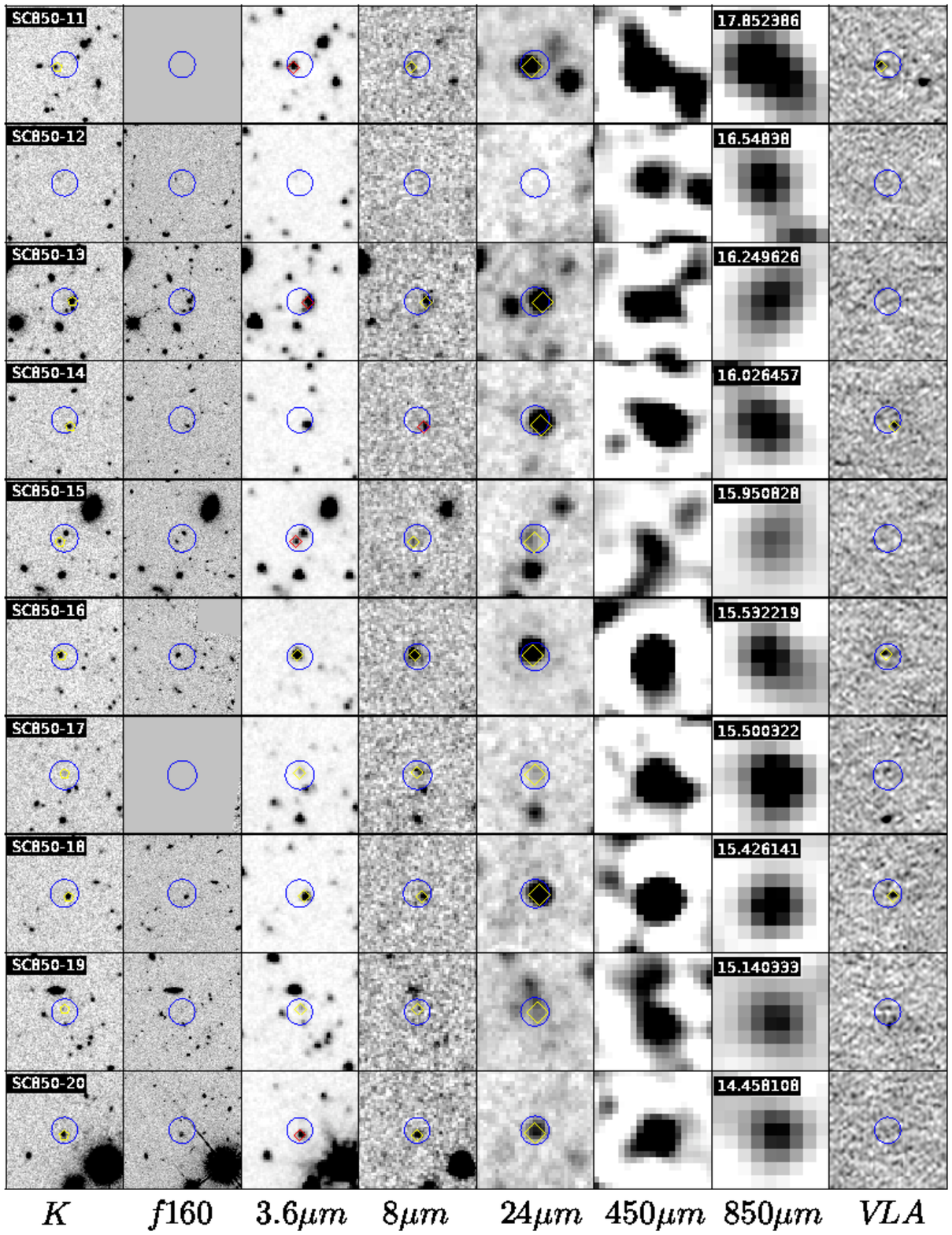}
\end{center}
\caption{(continued).}
\end{figure*}

\addtocounter{figure}{-1}
\begin{figure*}
\begin{center}
\includegraphics[scale=0.8,trim=5cm 2cm 5cm 2cm]{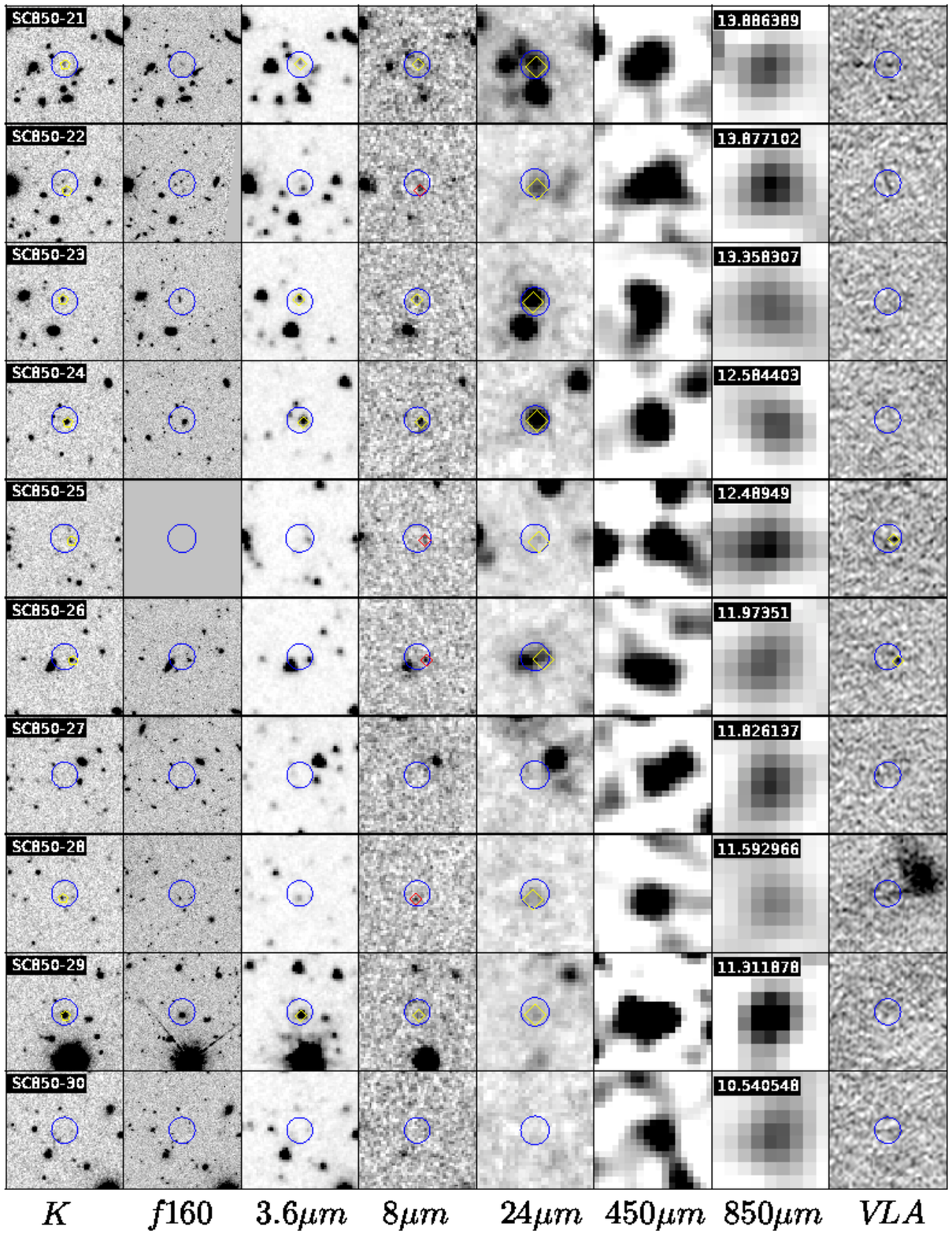}
\end{center}
\caption{(continued).}
\end{figure*}

\addtocounter{figure}{-1}
\begin{figure*}
\begin{center}
\includegraphics[scale=0.8,trim=5cm 2cm 5cm 2cm]{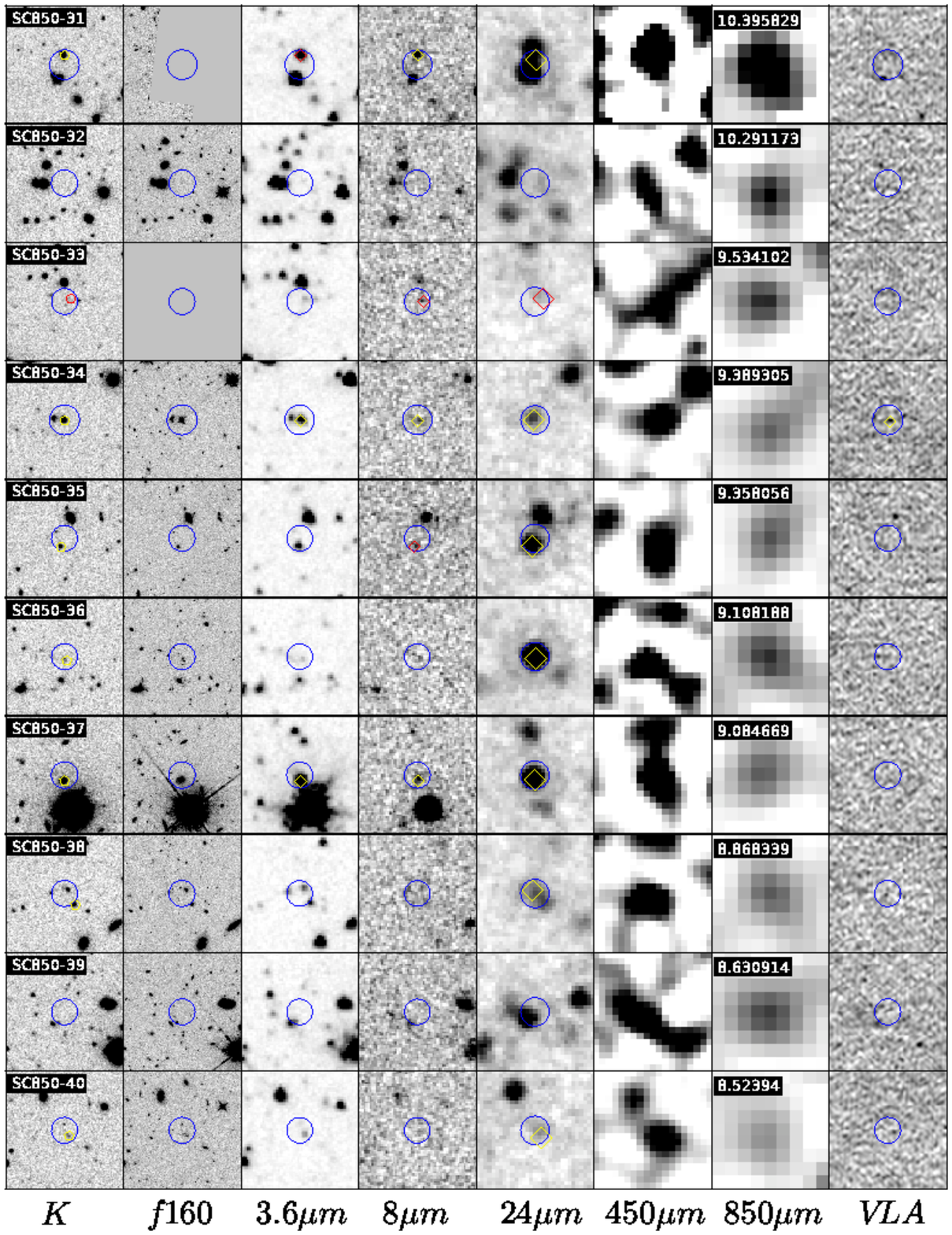}
\end{center}
\caption{(continued).}
\end{figure*}

\addtocounter{figure}{-1}
\begin{figure*}
\begin{center}
\includegraphics[scale=0.8,trim=5cm 2cm 5cm 2cm]{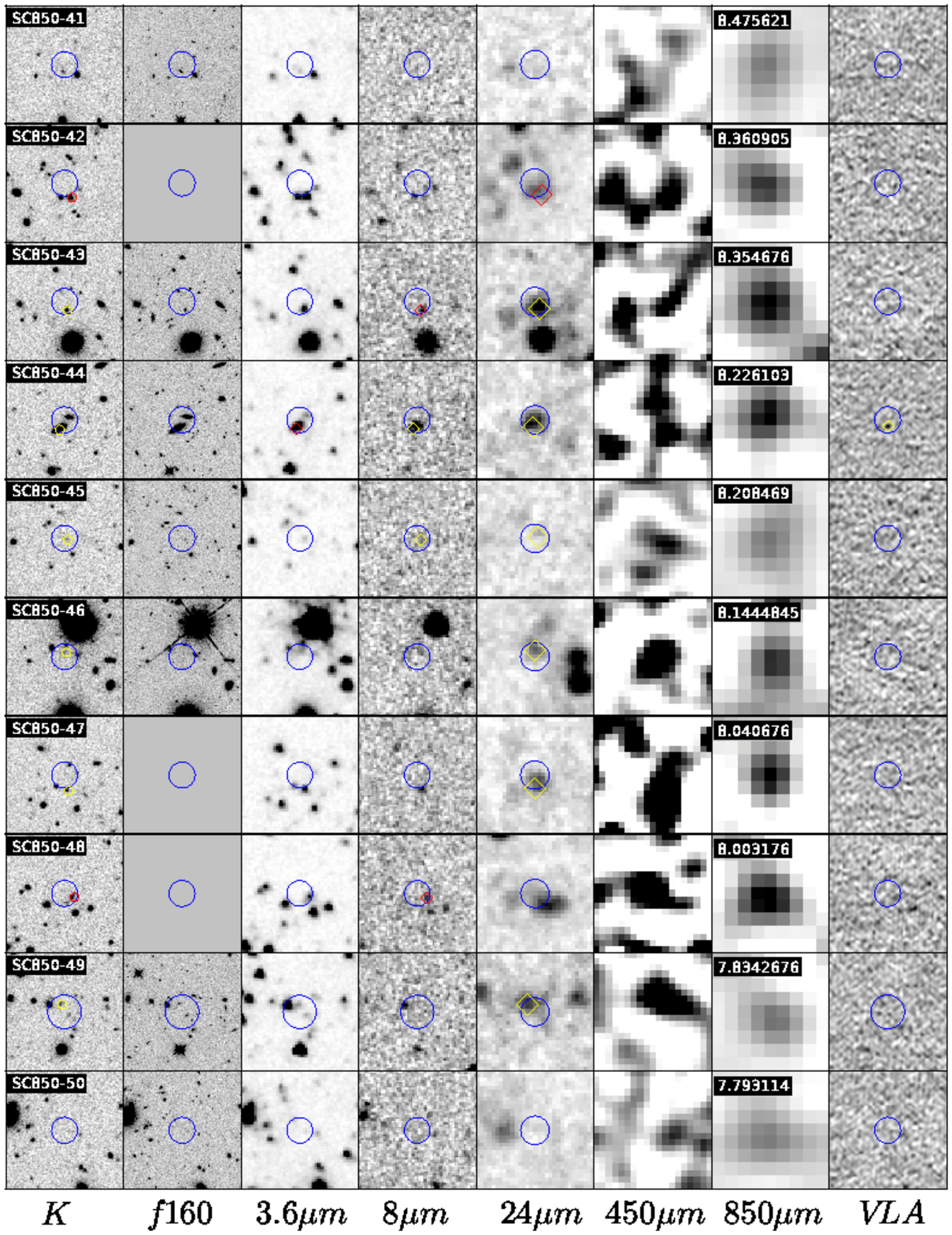}
\end{center}
\caption{(continued).}
\end{figure*}

\addtocounter{figure}{-1}
\begin{figure*}
\begin{center}
\includegraphics[scale=0.8,trim=5cm 2cm 5cm 2cm]{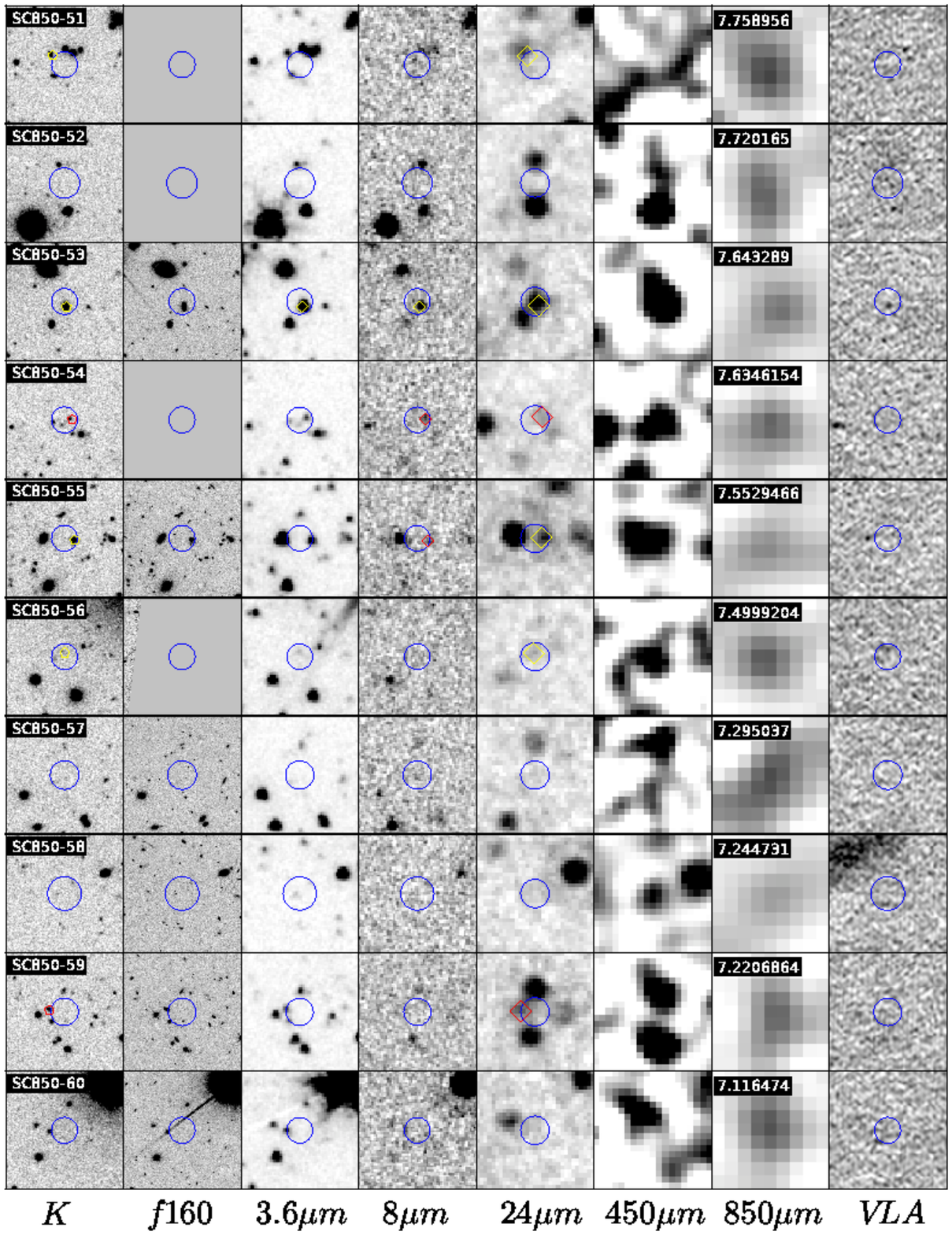}
\end{center}
\caption{(continued).}
\end{figure*}

\addtocounter{figure}{-1}
\begin{figure*}
\begin{center}
\includegraphics[scale=0.8,trim=5cm 2cm 5cm 2cm]{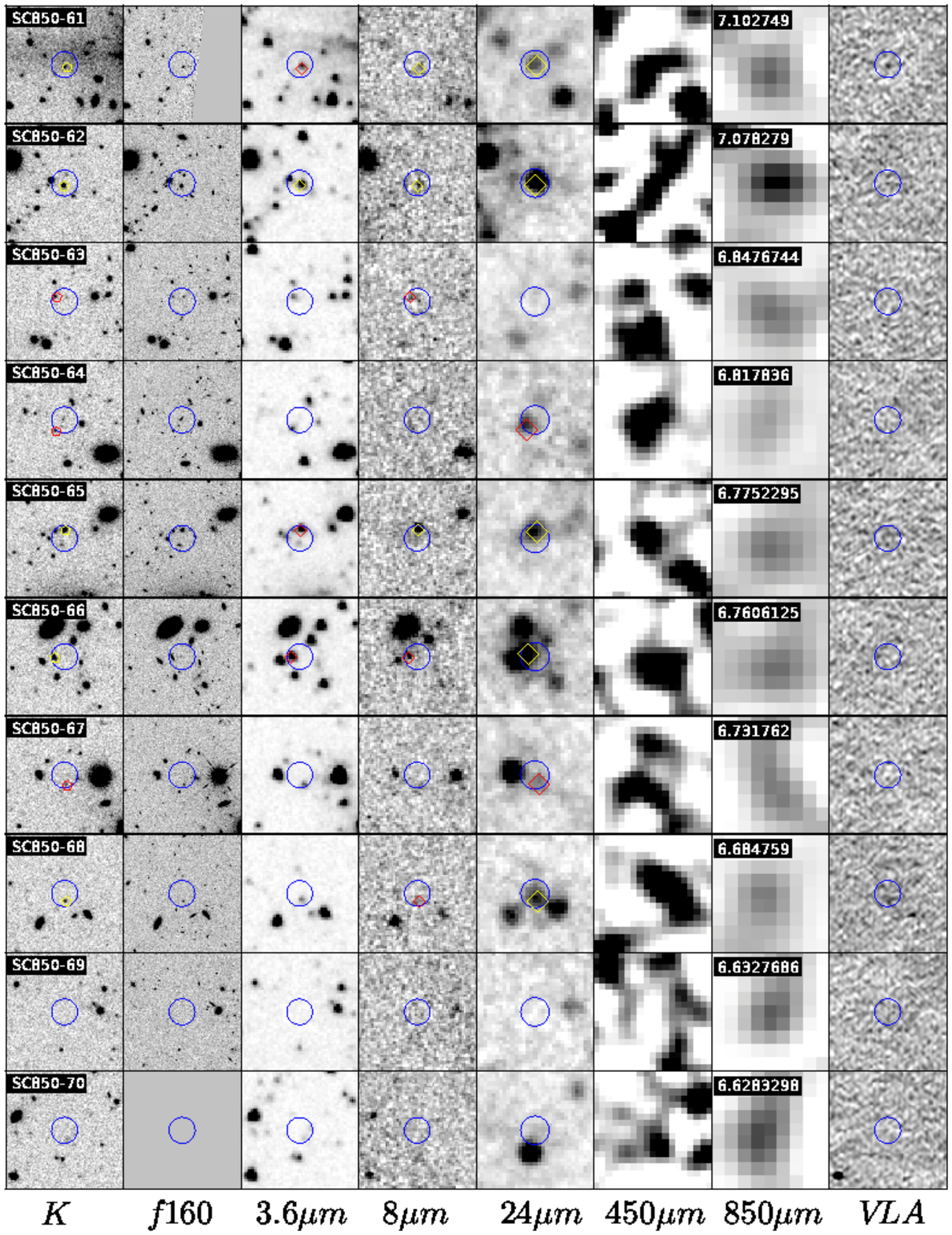}
\end{center}
\caption{(continued).}
\end{figure*}

\addtocounter{figure}{-1}
\begin{figure*}
\begin{center}
\includegraphics[scale=0.8,trim=5cm 2cm 5cm 2cm]{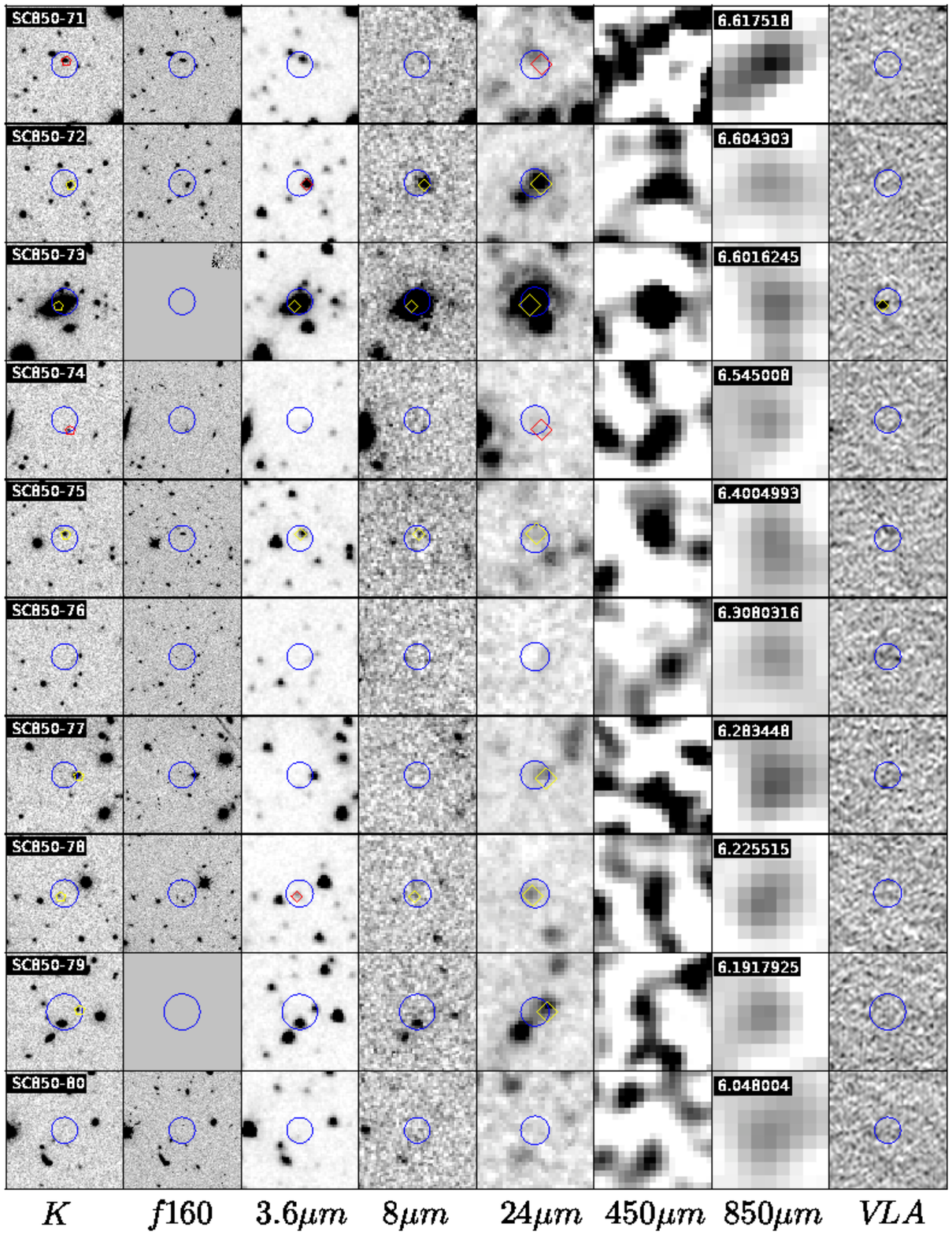}
\end{center}
\caption{(continued).}
\end{figure*}

\addtocounter{figure}{-1}
\begin{figure*}
\begin{center}
\includegraphics[scale=0.8,trim=5cm 2cm 5cm 2cm]{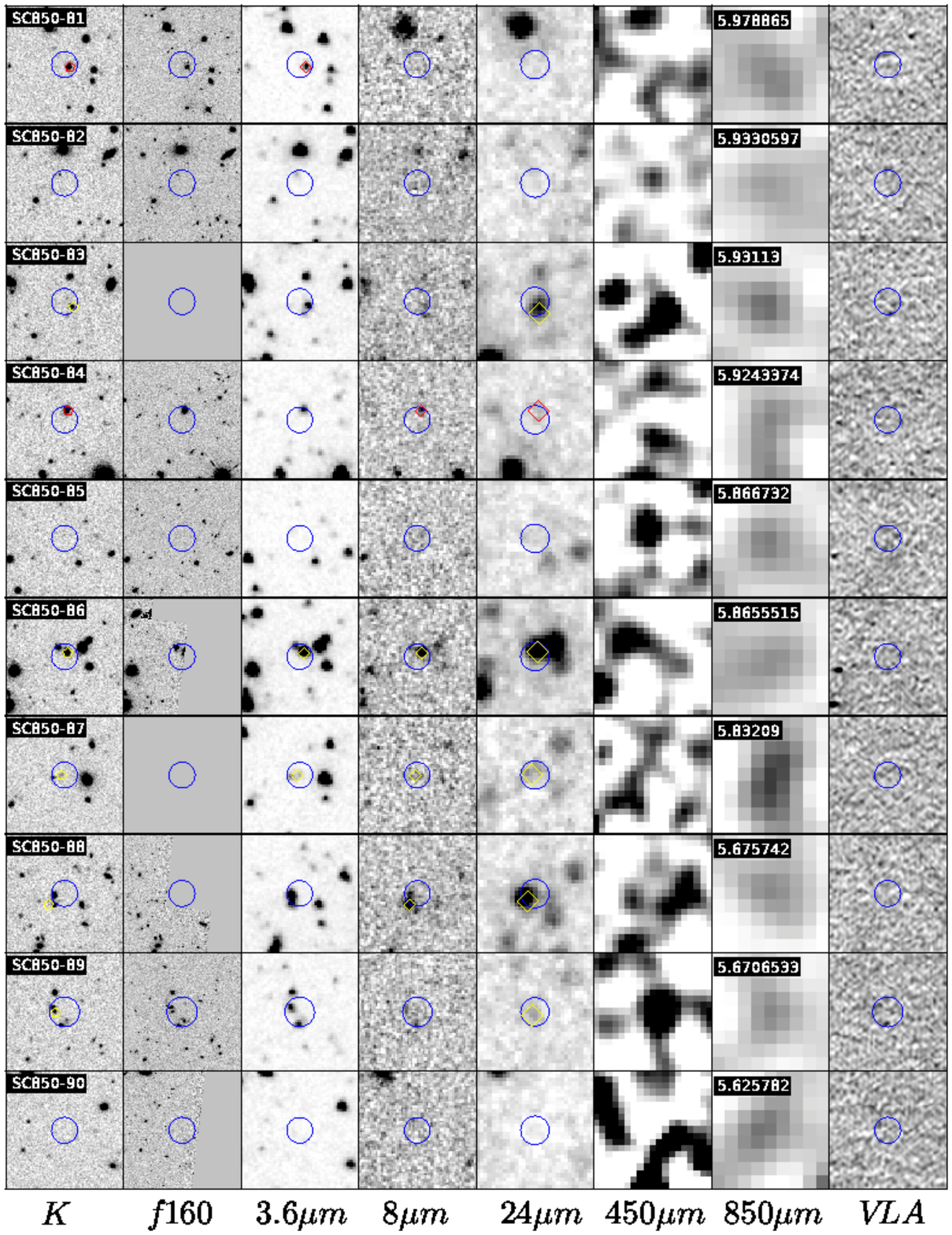}
\end{center}
\caption{(continued).}
\end{figure*}

\addtocounter{figure}{-1}
\begin{figure*}
\begin{center}
\includegraphics[scale=0.8,trim=5cm 2cm 5cm 2cm]{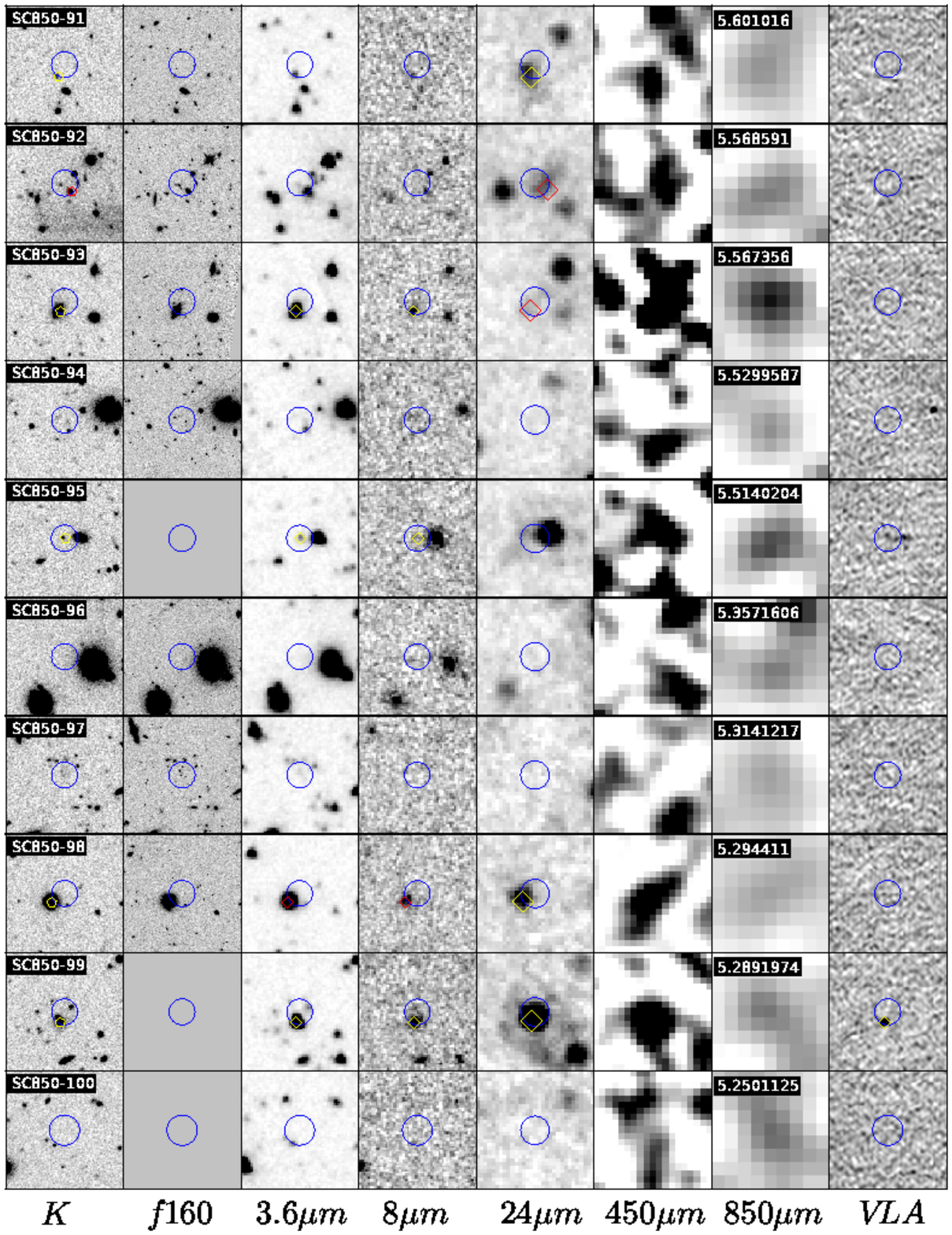}
\end{center}
\caption{(continued).}
\end{figure*}

\addtocounter{figure}{-1}
\begin{figure*}
\begin{center}
\includegraphics[scale=0.8,trim=5cm 12cm 5cm 2cm]{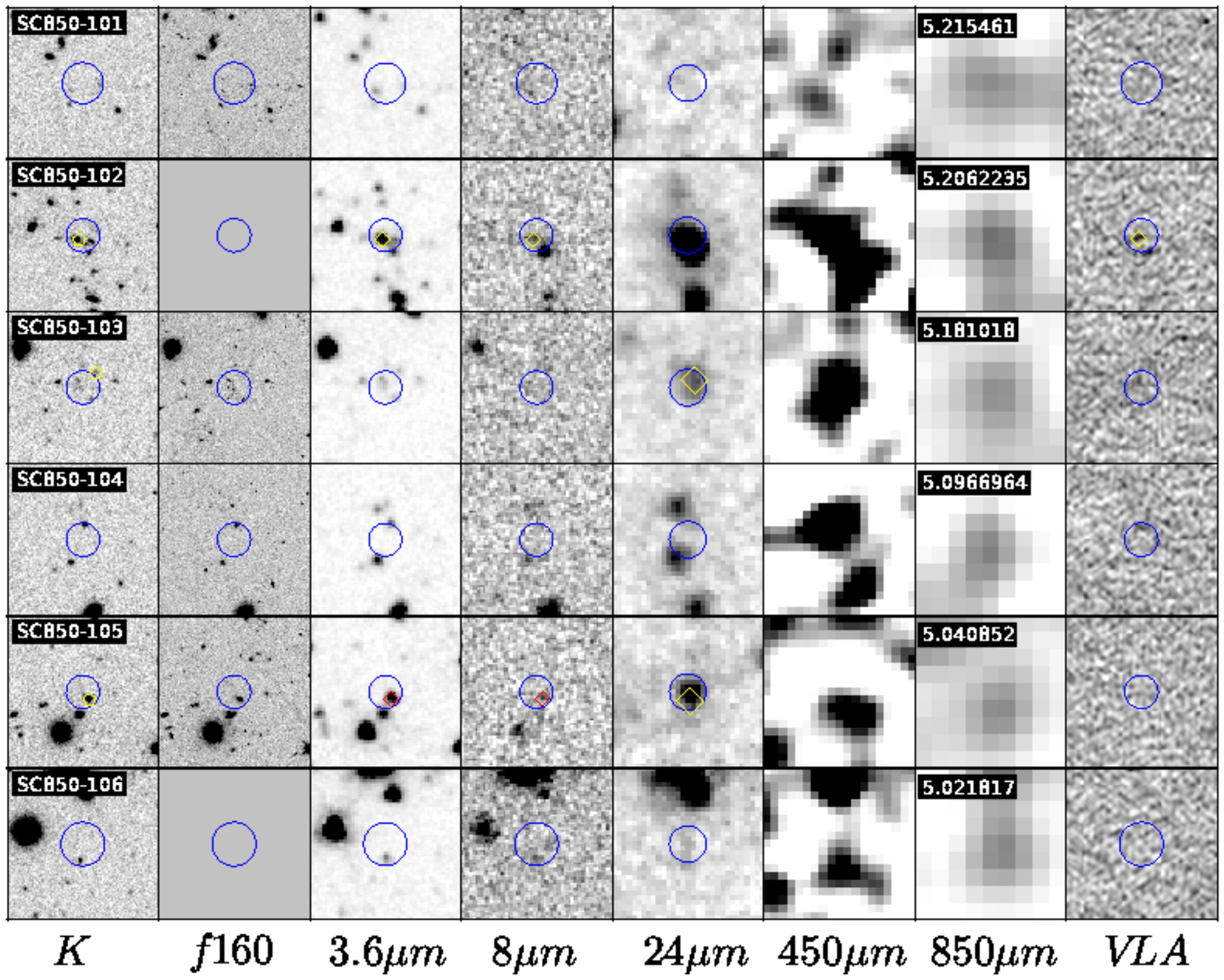}
\end{center}
\caption{(continued).}
\end{figure*}

\begin{figure*}
\begin{center}
\includegraphics[trim=5cm 2cm 5cm 2cm, scale=0.75]{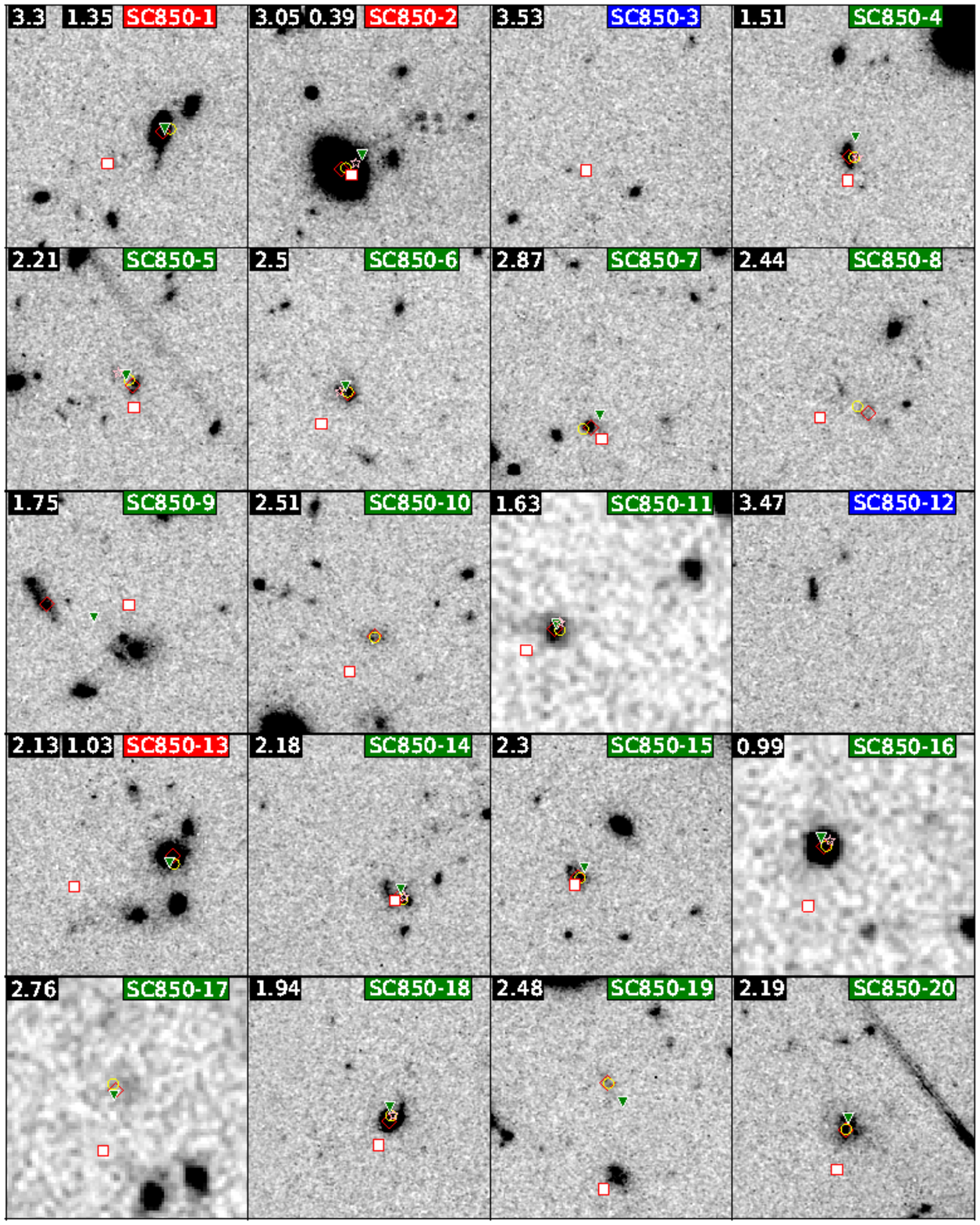}
\end{center}
\caption{Near-infrared ({\it HST} $H_{160}$ if available, otherwise UltraVISTA $K_S$-band) image stamps for all sources in the SCUBA-2 sample. Each stamp is centred on the 850\,${\rm \mu m}$ position,
and is 12 $\times$ 12\,arcsec, with North to the top and East to the left. 
The red squares indicate the positions of the 450\,${\rm \mu m}$ counterparts (when present). The green triangles show the 24\,${\rm \mu m}$ ID positions, 
the yellow circles the 8\,${\rm \mu m}$ ID positions, the pink stars the radio ID positions and the red diamonds the final optical ID positions. 
The colour-coding of the ID names in the upper-right corner is as follows. Green means that the adopted optical ID was deemed 
acceptable on the basis of agreement between $z_p$ and $z_{LW}$, and the final adopted value of $z_p$ is given in the top-left corner of the stamp. 
Red means that the proposed optical ID was rejected on the basis of inconsistency between $z_p$ and $z_{LW}$, 
and the first number in the upper-left corner of each stamp is then the adopted $z_{LW}$, with the second number being the rejected $z_p$. 
Where source names are given in blue, this means that no ID was secured for the SCUBA-2 source, and 
only the value of $z_{LW}$ was available (and is given in the upper-left corner). 
Seeveral of the rejected optical IDs appear to be rather obvious lenses (SC850-1, 2, 29, 44, 93).}
\label{fig:ID_stamps}
\end{figure*}

\addtocounter{figure}{-1}
\begin{figure*}
\begin{center}
\includegraphics[scale=0.75,trim=5cm 2cm 5cm 2cm]{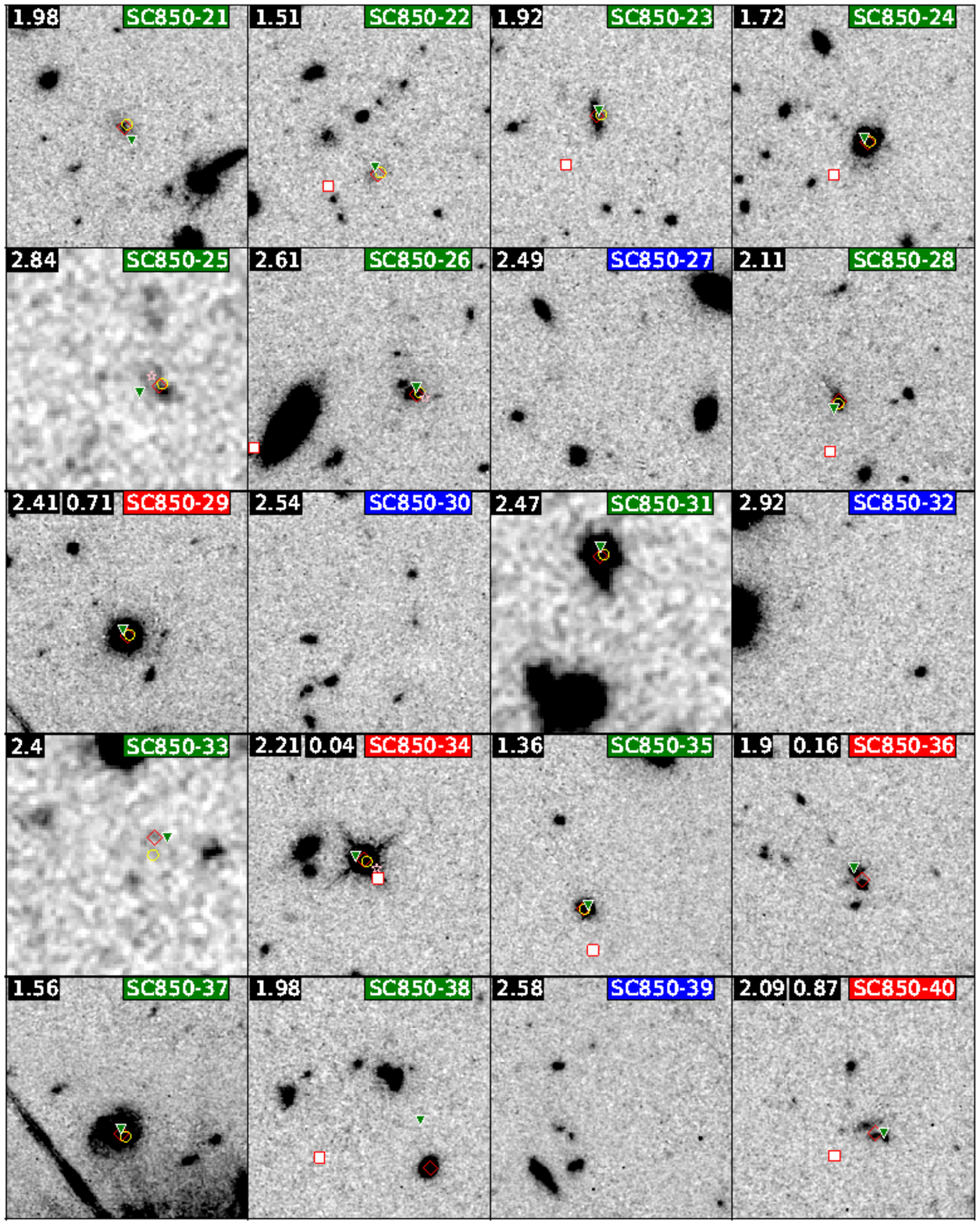}
\end{center}
\caption{(continued).}
\end{figure*}

\addtocounter{figure}{-1}
\begin{figure*}
\begin{center}
\includegraphics[scale=0.8,trim=5cm 2cm 5cm 2cm]{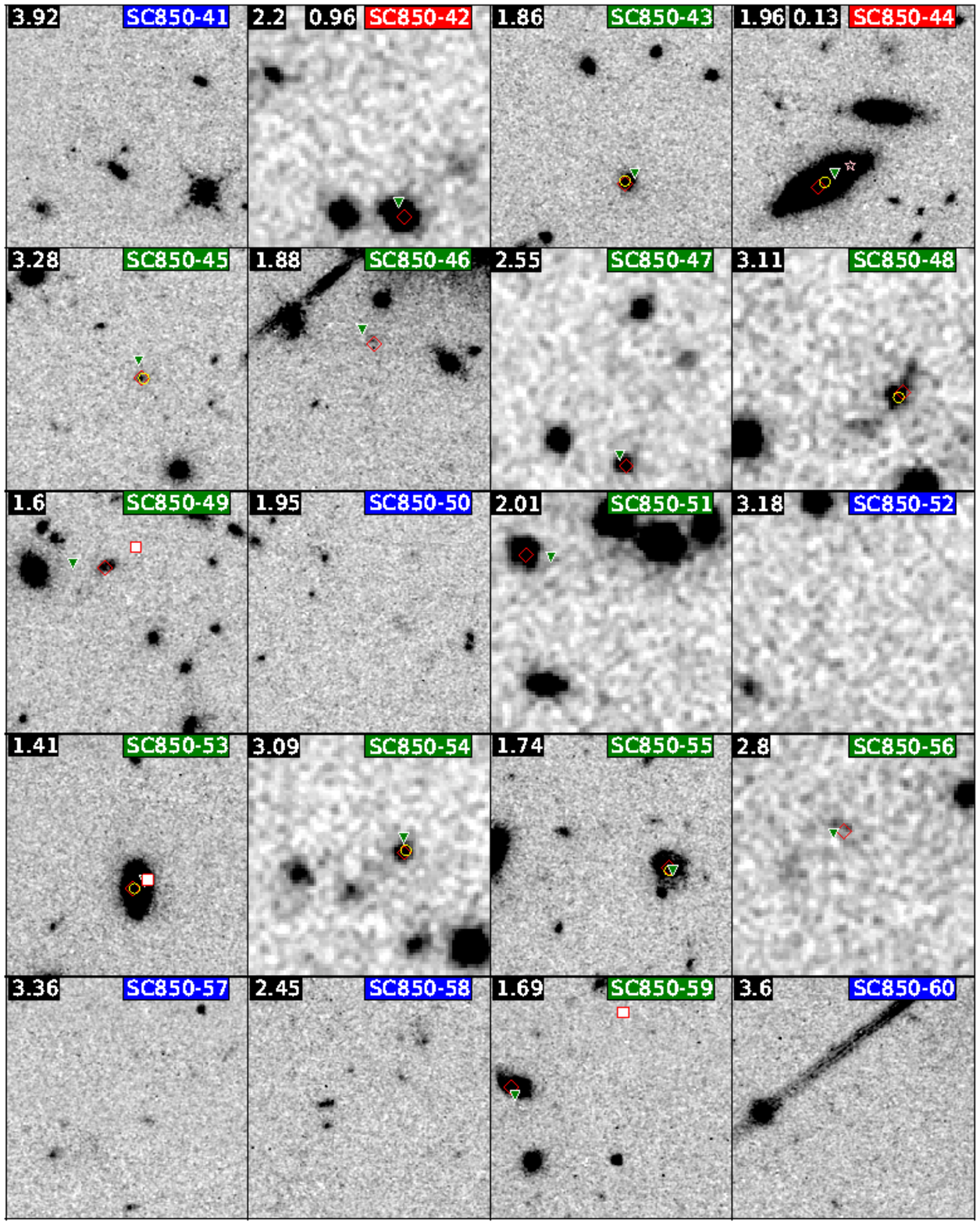}
\end{center}
\caption{(continued).}
\end{figure*}

\addtocounter{figure}{-1}
\begin{figure*}
\begin{center}
\includegraphics[scale=0.8,trim=5cm 2cm 5cm 2cm]{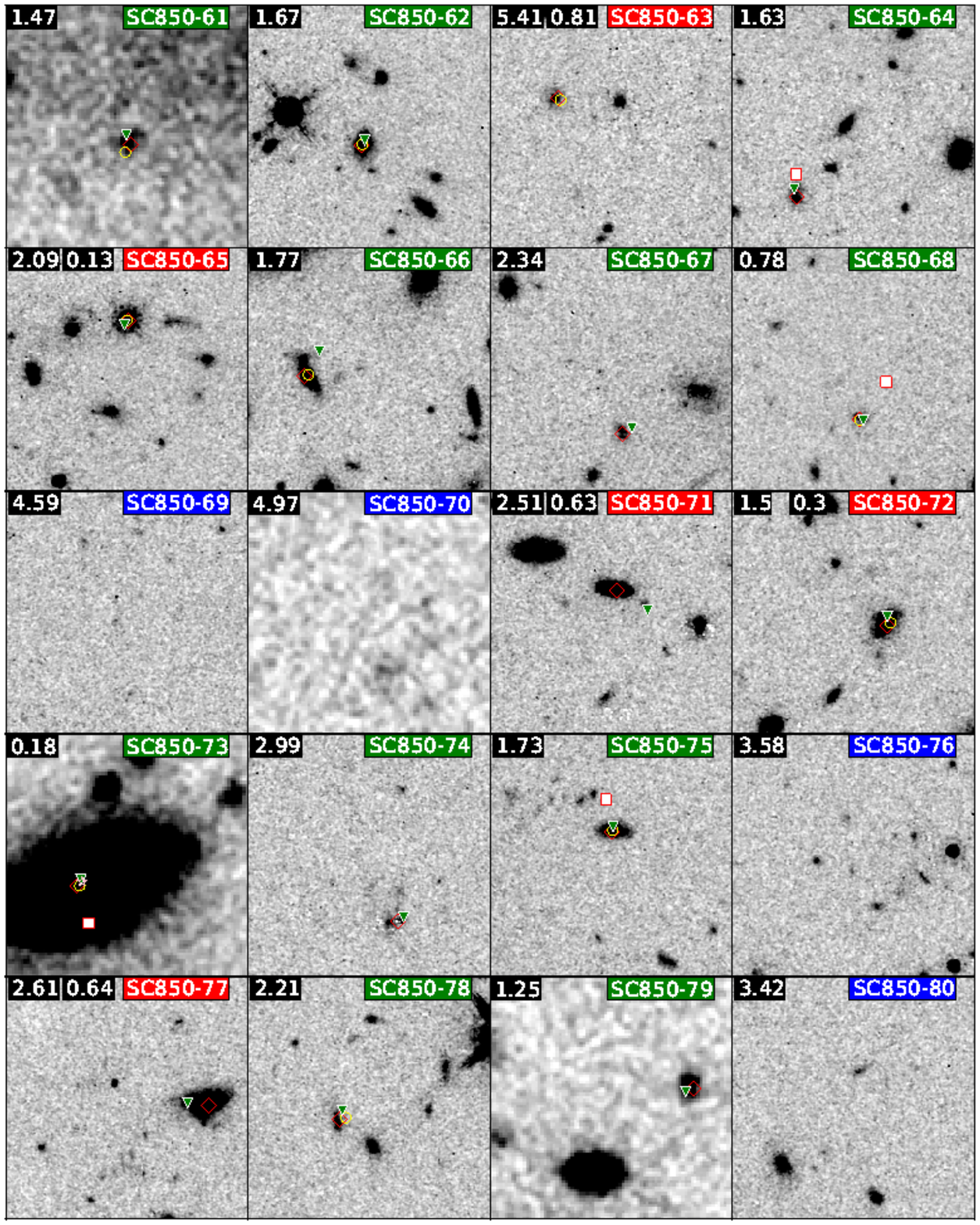}
\end{center}
\caption{(continued).}
\end{figure*}

\addtocounter{figure}{-1}
\begin{figure*}
\begin{center}
\includegraphics[scale=0.8,trim=5cm 2cm 5cm 2cm]{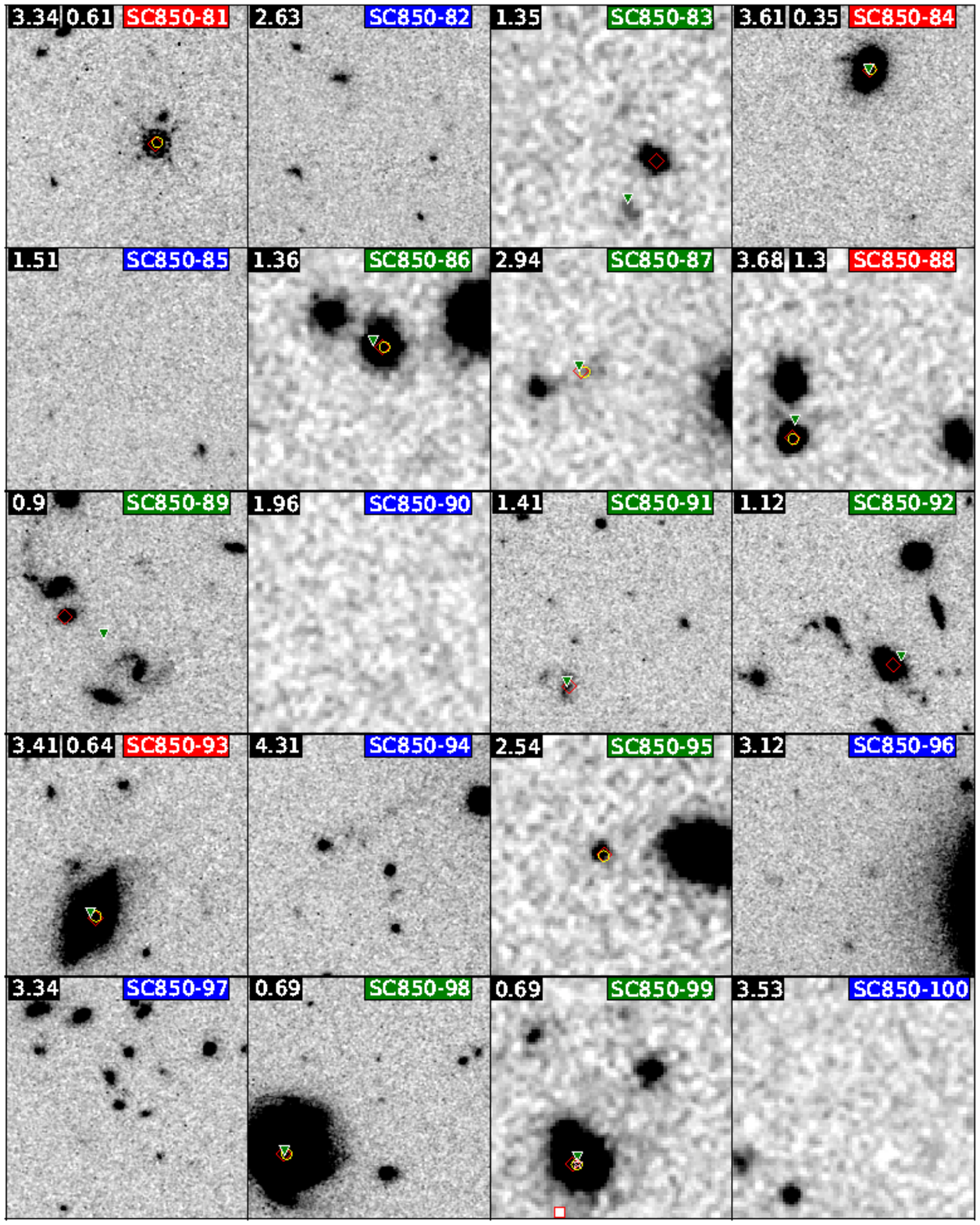}
\end{center}
\caption{(continued).}
\end{figure*}

\addtocounter{figure}{-1}
\begin{figure*}
\begin{center}
\includegraphics[scale=0.8,trim=5cm 15cm 5cm 2cm]{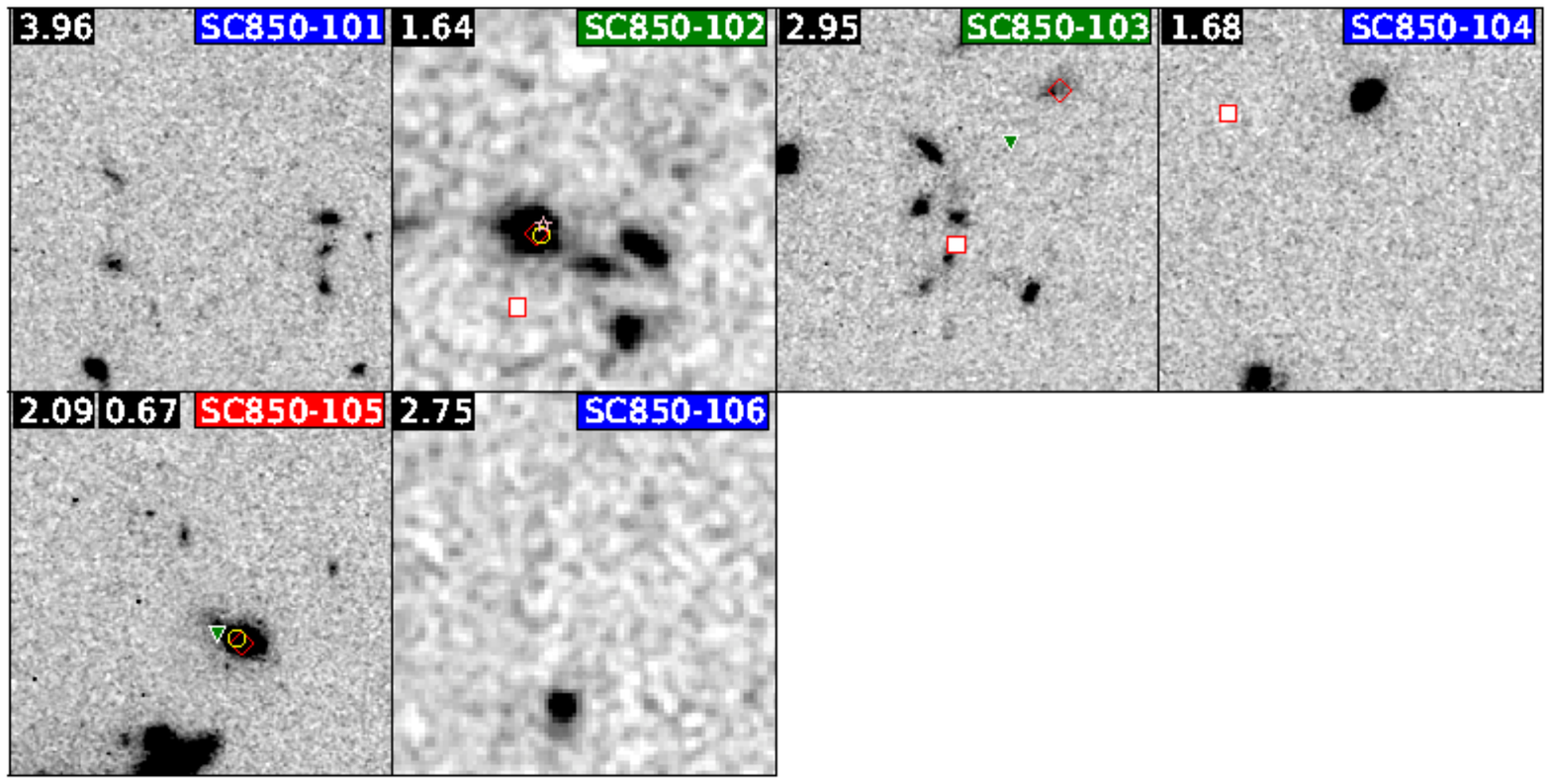}
\end{center}
\caption{(continued).}
\end{figure*}

\label{lastpage}

\end{document}

%% file: a1.tex
\begin{center}
\begin{footnotesize}
\begin{tabular}{lcccccccccc}
\hline
 ID & RA$_{850}$ & DEC$_{850}$ & $S_{850}$ & $\Delta S_{850}$ & SNR$_{850}$ & $S_{450}$ & $\Delta S_{450}$ & SNR$_{450}$ & $S_{850}/S_{450}$ & flag \\
    & /deg       & /deg        & /mJy      & /mJy             &             & /mJy      & /mJy             &             &                   &      \\
\hline
  1  &  150.06518  &  2.26412  &             15.64  &   0.38  &             41.69  &             26.99  &   2.38  &             11.34  &  $\phantom{>}0.58$  &  1  \\
  2  &  150.09985  &  2.29772  &             10.20  &   0.28  &             36.82  &             17.74  &   1.77  &             10.00  &  $\phantom{>}0.58$  &  1  \\
  3  &  150.10079  &  2.33499  &  \phantom{1}7.33  &   0.23  &             32.02  &             10.32  &   1.41  &  \phantom{1}7.34  &  $\phantom{>}0.71$  &  1  \\
  4  &  150.10549  &  2.31327  &  \phantom{1}7.79  &   0.24  &             31.96  &             23.42  &   1.53  &             15.35  &  $\phantom{>}0.33$  &  1  \\
  5  &  150.14320  &  2.35607  &  \phantom{1}7.88  &   0.26  &             29.98  &             19.71  &   1.54  &             12.78  &  $\phantom{>}0.40$  &  1  \\
  6  &  150.09833  &  2.36568  &  \phantom{1}8.20  &   0.28  &             29.20  &             22.81  &   1.80  &             12.71  &  $\phantom{>}0.36$  &  1  \\
  7  &  150.09847  &  2.32162  &  \phantom{1}7.04  &   0.25  &             28.44  &             16.66  &   1.53  &             10.88  &  $\phantom{>}0.42$  &  1  \\
  8  &  150.09820  &  2.26061  &  \phantom{1}6.44  &   0.33  &             19.32  &             14.89  &   2.13  &  \phantom{1}6.98  &  $\phantom{>}0.43$  &  1  \\
  9  &  150.07809  &  2.28168  &  \phantom{1}5.88  &   0.32  &             18.56  &             15.45  &   2.03  &  \phantom{1}7.62  &  $\phantom{>}0.38$  &  1  \\
 10  &  150.15390  &  2.32833  &  \phantom{1}4.75  &   0.26  &             18.17  &             11.15  &   1.55  &  \phantom{1}7.17  &  $\phantom{>}0.43$  &  1  \\
 11  &  150.04264  &  2.37371  &  \phantom{1}7.34  &   0.41  &             17.85  &             23.66  &   2.87  &  \phantom{1}8.23  &  $\phantom{>}0.31$  &  1  \\
 12  &  150.10996  &  2.25832  &  \phantom{1}5.54  &   0.34  &             16.55  &  \phantom{1}8.91  &   2.13  &  \phantom{1}4.18  &  $\phantom{>}0.62$  &  1  \\
 13  &  150.08512  &  2.29050  &  \phantom{1}4.87  &   0.30  &             16.25  &             12.79  &   1.97  &  \phantom{1}6.50  &  $\phantom{>}0.38$  &  1  \\
 14  &  150.10692  &  2.25218  &  \phantom{1}5.83  &   0.36  &             16.03  &             23.81  &   2.38  &  \phantom{1}9.99  &  $\phantom{>}0.24$  &  1  \\
 15  &  150.11717  &  2.33026  &  \phantom{1}3.41  &   0.21  &             15.95  &  \phantom{1}6.53  &   1.31  &  \phantom{1}4.99  &  $\phantom{>}0.52$  &  1  \\
 16  &  150.05633  &  2.37363  &  \phantom{1}5.42  &   0.35  &             15.53  &             24.31  &   2.39  &             10.17  &  $\phantom{>}0.22$  &  1  \\
 17  &  150.20799  &  2.38297  &  \phantom{1}7.34  &   0.47  &             15.50  &             15.66  &   2.68  &  \phantom{1}5.83  &  $\phantom{>}0.47$  &  1  \\
 18  &  150.16393  &  2.37274  &  \phantom{1}6.13  &   0.40  &             15.43  &             31.22  &   1.96  &             15.90  &  $\phantom{>}0.20$  &  1  \\
 19  &  150.11258  &  2.37633  &  \phantom{1}4.35  &   0.29  &             15.14  &  \phantom{1}9.91  &   1.84  &  \phantom{1}5.37  &  $\phantom{>}0.44$  &  1  \\
 20  &  150.15024  &  2.36457  &  \phantom{1}4.55  &   0.31  &             14.46  &             13.31  &   1.76  &  \phantom{1}7.56  &  $\phantom{>}0.34$  &  1  \\
 21  &  150.09873  &  2.31118  &  \phantom{1}3.68  &   0.26  &             13.89  &  \phantom{1}9.08  &   1.62  &  \phantom{1}5.61  &  $\phantom{>}0.41$  &  1  \\
 22  &  150.05727  &  2.29352  &  \phantom{1}4.60  &   0.33  &             13.88  &             12.83  &   2.19  &  \phantom{1}5.87  &  $\phantom{>}0.36$  &  1  \\
 23  &  150.12283  &  2.36081  &  \phantom{1}3.16  &   0.24  &             13.36  &             10.41  &   1.50  &  \phantom{1}6.94  &  $\phantom{>}0.30$  &  1  \\
 24  &  150.10937  &  2.29455  &  \phantom{1}3.43  &   0.27  &             12.58  &             11.88  &   1.74  &  \phantom{1}6.84  &  $\phantom{>}0.29$  &  1  \\
 25  &  150.03791  &  2.34079  &  \phantom{1}4.56  &   0.36  &             12.49  &  \phantom{1}9.20  &   2.48  &  \phantom{1}3.71  &  $\phantom{>}0.50$  &  0  \\
 26  &  150.08011  &  2.34091  &  \phantom{1}3.27  &   0.27  &             11.97  &             10.70  &   1.77  &  \phantom{1}6.06  &  $\phantom{>}0.31$  &  1  \\
 27  &  150.17416  &  2.35283  &  \phantom{1}4.07  &   0.34  &             11.83  &  \phantom{1}8.82  &   2.01  &  \phantom{1}4.38  &  $\phantom{>}0.46$  &  1  \\
 28  &  150.12169  &  2.34175  &  \phantom{1}2.48  &   0.21  &             11.59  &             10.56  &   1.30  &  \phantom{1}8.13  &  $\phantom{>}0.23$  &  1  \\
 29  &  150.10535  &  2.43531  &  \phantom{1}6.47  &   0.57  &             11.31  &             18.98  &   3.59  &  \phantom{1}5.28  &  $\phantom{>}0.34$  &  1  \\
 30  &  150.14489  &  2.37645  &  \phantom{1}3.37  &   0.32  &             10.54  &  \phantom{1}8.44  &   1.78  &  \phantom{1}4.75  &  $\phantom{>}0.40$  &  1  \\
 31  &  150.05250  &  2.24477  &  \phantom{1}7.85  &   0.76  &             10.40  &             14.21  &   5.63  &  \phantom{1}2.52  &  $\phantom{>}0.55$  &  0  \\
 32  &  150.06641  &  2.41264  &  \phantom{1}4.72  &   0.46  &             10.29  &  \phantom{1}6.49  &   3.18  &  \phantom{1}2.04  &  $\phantom{>}0.73$  &  0  \\
 33  &  150.04153  &  2.28039  &  \phantom{1}4.01  &   0.42  &  \phantom{1}9.53  &  \phantom{1}6.73  &   2.60  &  \phantom{1}2.59  &  $\phantom{>}0.60$  &  0  \\
 34  &  150.13514  &  2.39948  &  \phantom{1}3.03  &   0.32  &  \phantom{1}9.39  &             11.12  &   1.94  &  \phantom{1}5.73  &  $\phantom{>}0.27$  &  1  \\
 35  &  150.16742  &  2.29950  &  \phantom{1}3.29  &   0.35  &  \phantom{1}9.36  &             11.12  &   1.91  &  \phantom{1}5.84  &  $\phantom{>}0.30$  &  1  \\
 36  &  150.08208  &  2.41590  &  \phantom{1}3.95  &   0.43  &  \phantom{1}9.11  &             10.88  &   2.91  &  \phantom{1}3.74  &  $\phantom{>}0.36$  &  0  \\
 37  &  150.06812  &  2.27618  &  \phantom{1}3.06  &   0.34  &  \phantom{1}9.08  &             11.70  &   2.08  &  \phantom{1}5.62  &  $\phantom{>}0.26$  &  1  \\
 38  &  150.07620  &  2.38036  &  \phantom{1}3.14  &   0.35  &  \phantom{1}8.87  &             12.41  &   2.27  &  \phantom{1}5.46  &  $\phantom{>}0.25$  &  1  \\
 39  &  150.09322  &  2.24697  &  \phantom{1}3.69  &   0.43  &  \phantom{1}8.63  &             11.67  &   2.83  &  \phantom{1}4.12  &  $\phantom{>}0.32$  &  1  \\
 40  &  150.10570  &  2.32638  &  \phantom{1}1.94  &   0.23  &  \phantom{1}8.52  &  \phantom{1}7.87  &   1.38  &  \phantom{1}5.72  &  $\phantom{>}0.25$  &  1  \\
 41  &  150.12888  &  2.28474  &  \phantom{1}2.47  &   0.29  &  \phantom{1}8.48  &  \phantom{1}0.96  &   1.91  &  \phantom{1}0.50  &  $>          0.51$  &  0  \\
 42  &  150.02819  &  2.34702  &  \phantom{1}3.80  &   0.45  &  \phantom{1}8.36  &  \phantom{1}0.76  &   2.97  &  \phantom{1}0.26  &  $>          0.57$  &  0  \\
 43  &  150.17214  &  2.24149  &  \phantom{1}4.97  &   0.59  &  \phantom{1}8.35  &  \phantom{1}3.39  &   3.87  &  \phantom{1}0.88  &  $>          0.45$  &  0  \\
 44  &  150.13663  &  2.23305  &  \phantom{1}4.66  &   0.57  &  \phantom{1}8.23  &  \phantom{1}2.28  &   3.48  &  \phantom{1}0.66  &  $>          0.50$  &  0  \\
 45  &  150.12744  &  2.38798  &  \phantom{1}2.52  &   0.31  &  \phantom{1}8.21  &  \phantom{1}4.50  &   1.87  &  \phantom{1}2.41  &  $\phantom{>}0.56$  &  0  \\
 46  &  150.10606  &  2.42844  &  \phantom{1}3.94  &   0.48  &  \phantom{1}8.14  &             14.48  &   3.09  &  \phantom{1}4.69  &  $\phantom{>}0.27$  &  1  \\
 47  &  150.04863  &  2.25278  &  \phantom{1}4.95  &   0.62  &  \phantom{1}8.04  &  \phantom{1}3.14  &   4.64  &  \phantom{1}0.68  &  $>          0.40$  &  0  \\
 48  &  150.02227  &  2.28899  &  \phantom{1}5.05  &   0.63  &  \phantom{1}8.00  &             16.71  &   3.63  &  \phantom{1}4.60  &  $\phantom{>}0.30$  &  1  \\
 49  &  150.15725  &  2.35741  &  \phantom{1}2.58  &   0.33  &  \phantom{1}7.83  &             10.44  &   1.84  &  \phantom{1}5.67  &  $\phantom{>}0.25$  &  1  \\
 50  &  150.08103  &  2.36298  &  \phantom{1}2.39  &   0.31  &  \phantom{1}7.79  &  \phantom{1}0.75  &   1.95  &  \phantom{1}0.38  &  $>          0.51$  &  0  \\
 51  &  150.03551  &  2.28537  &  \phantom{1}3.56  &   0.46  &  \phantom{1}7.76  &  \phantom{1}1.63  &   2.76  &  \phantom{1}0.59  &  $>          0.50$  &  0  \\
 52  &  150.03693  &  2.31959  &  \phantom{1}2.85  &   0.37  &  \phantom{1}7.72  &  \phantom{1}1.68  &   2.47  &  \phantom{1}0.68  &  $>          0.43$  &  0  \\
 53  &  150.18780  &  2.32296  &  \phantom{1}2.54  &   0.33  &  \phantom{1}7.64  &             15.59  &   1.84  &  \phantom{1}8.46  &  $\phantom{>}0.16$  &  1  \\
\hline
\end{tabular}
\end{footnotesize}
\end{center}

%% file: a2.tex
\begin{center}
\begin{footnotesize}
\begin{tabular}{lcccccccccc}
\hline
 ID & RA$_{850}$ & DEC$_{850}$ & $S_{850}$ & $\Delta S_{850}$ & SNR$_{850}$ & $S_{450}$ & $\Delta S_{450}$ & SNR$_{450}$ & $S_{850}/S_{450}$ & flag \\
    & /deg       & /deg        & /mJy      & /mJy             &             & /mJy      & /mJy             &             &                   &      \\
\hline
 54  &  150.04307  &  2.29982  &  \phantom{1}2.87  &   0.36  &  \phantom{1}7.63  &             10.50  &   2.36  &  \phantom{1}4.45  &  $\phantom{>}0.27$  &  1  \\
 55  &  150.13442  &  2.37059  &  \phantom{1}2.06  &   0.27  &  \phantom{1}7.55  &             10.13  &   1.68  &  \phantom{1}6.04  &  $\phantom{>}0.20$  &  1  \\
 56  &  150.05005  &  2.38574  &  \phantom{1}3.09  &   0.41  &  \phantom{1}7.50  &  \phantom{1}7.69  &   2.95  &  \phantom{1}2.61  &  $\phantom{>}0.40$  &  0  \\
 57  &  150.15614  &  2.41984  &  \phantom{1}3.38  &   0.45  &  \phantom{1}7.30  &  \phantom{1}2.77  &   2.72  &  \phantom{1}1.02  &  $>          0.41$  &  0  \\
 58  &  150.10709  &  2.34444  &  \phantom{1}1.62  &   0.22  &  \phantom{1}7.24  &  \phantom{1}4.98  &   1.36  &  \phantom{1}3.66  &  $\phantom{>}0.32$  &  0  \\
 59  &  150.18368  &  2.38879  &  \phantom{1}2.83  &   0.39  &  \phantom{1}7.22  &             10.40  &   2.14  &  \phantom{1}4.86  &  $\phantom{>}0.27$  &  1  \\
 60  &  150.19199  &  2.27300  &  \phantom{1}3.25  &   0.46  &  \phantom{1}7.12  &  \phantom{1}0.68  &   2.95  &  \phantom{1}0.23  &  $>          0.49$  &  0  \\
 61  &  150.05419  &  2.39615  &  \phantom{1}3.06  &   0.43  &  \phantom{1}7.10  &  \phantom{1}8.03  &   3.08  &  \phantom{1}2.61  &  $\phantom{>}0.38$  &  0  \\
 62  &  150.16689  &  2.23608  &  \phantom{1}4.61  &   0.65  &  \phantom{1}7.08  &  \phantom{1}8.03  &   4.13  &  \phantom{1}1.95  &  $>          0.28$  &  0  \\
 63  &  150.07608  &  2.39821  &  \phantom{1}2.70  &   0.39  &  \phantom{1}6.85  &  \phantom{1}0.89  &   2.62  &  \phantom{1}0.34  &  $>          0.44$  &  0  \\
 64  &  150.13004  &  2.31505  &  \phantom{1}1.59  &   0.23  &  \phantom{1}6.82  &  \phantom{1}8.63  &   1.45  &  \phantom{1}5.95  &  $\phantom{>}0.18$  &  1  \\
 65  &  150.09167  &  2.39837  &  \phantom{1}2.71  &   0.40  &  \phantom{1}6.78  &             11.02  &   2.49  &  \phantom{1}4.42  &  $\phantom{>}0.25$  &  1  \\
 66  &  150.17480  &  2.40168  &  \phantom{1}2.70  &   0.40  &  \phantom{1}6.76  &  \phantom{1}9.29  &   2.18  &  \phantom{1}4.27  &  $\phantom{>}0.29$  &  1  \\
 67  &  150.11157  &  2.40409  &  \phantom{1}2.38  &   0.35  &  \phantom{1}6.73  &  \phantom{1}3.22  &   2.27  &  \phantom{1}1.42  &  $>          0.31$  &  0  \\
 68  &  150.13019  &  2.25338  &  \phantom{1}2.37  &   0.36  &  \phantom{1}6.68  &  \phantom{1}7.01  &   2.29  &  \phantom{1}3.06  &  $\phantom{>}0.34$  &  0  \\
 69  &  150.15507  &  2.24389  &  \phantom{1}3.10  &   0.47  &  \phantom{1}6.63  &  -2.98  &   3.02  &  -0.99  &  $>          0.51$  &  0  \\
 70  &  150.02490  &  2.29668  &  \phantom{1}3.43  &   0.52  &  \phantom{1}6.63  &  \phantom{1}3.49  &   3.16  &  \phantom{1}1.11  &  $>          0.35$  &  0  \\
 71  &  150.07211  &  2.23837  &  \phantom{1}4.44  &   0.67  &  \phantom{1}6.62  &  -6.58  &   4.78  &  -1.38  &  $>          0.46$  &  0  \\
 72  &  150.06512  &  2.32922  &  \phantom{1}1.93  &   0.29  &  \phantom{1}6.60  &  \phantom{1}8.04  &   2.09  &  \phantom{1}3.84  &  $\phantom{>}0.24$  &  0  \\
 73  &  150.20910  &  2.35567  &  \phantom{1}2.82  &   0.43  &  \phantom{1}6.60  &             18.21  &   2.40  &  \phantom{1}7.57  &  $\phantom{>}0.15$  &  1  \\
 74  &  150.07115  &  2.30605  &  \phantom{1}2.07  &   0.32  &  \phantom{1}6.55  &  \phantom{1}2.53  &   2.21  &  \phantom{1}1.15  &  $>          0.30$  &  0  \\
 75  &  150.15943  &  2.29648  &  \phantom{1}2.34  &   0.35  &  \phantom{1}6.40  &             10.40  &   1.96  &  \phantom{1}5.32  &  $\phantom{>}0.22$  &  1  \\
 76  &  150.18268  &  2.33601  &  \phantom{1}2.01  &   0.32  &  \phantom{1}6.31  &  -0.32  &   1.83  &  -0.17  &  $>          0.55$  &  0  \\
 77  &  150.07148  &  2.42307  &  \phantom{1}3.22  &   0.51  &  \phantom{1}6.28  &  \phantom{1}3.97  &   3.60  &  \phantom{1}1.10  &  $>          0.29$  &  0  \\
 78  &  150.09911  &  2.40516  &  \phantom{1}2.51  &   0.39  &  \phantom{1}6.23  &  \phantom{1}5.50  &   2.49  &  \phantom{1}2.21  &  $\phantom{>}0.46$  &  0  \\
 79  &  150.04249  &  2.32799  &  \phantom{1}2.20  &   0.35  &  \phantom{1}6.19  &  \phantom{1}3.49  &   2.36  &  \phantom{1}1.48  &  $>          0.27$  &  0  \\
 80  &  150.13624  &  2.26135  &  \phantom{1}2.06  &   0.34  &  \phantom{1}6.05  &  -0.66  &   2.16  &  -0.30  &  $>          0.48$  &  0  \\
 81  &  150.12630  &  2.41379  &  \phantom{1}2.21  &   0.37  &  \phantom{1}5.98  &  \phantom{1}1.27  &   2.35  &  \phantom{1}0.54  &  $>          0.37$  &  0  \\
 82  &  150.15286  &  2.32011  &  \phantom{1}1.59  &   0.27  &  \phantom{1}5.93  &  \phantom{1}3.65  &   1.60  &  \phantom{1}2.29  &  $\phantom{>}0.43$  &  0  \\
 83  &  150.02572  &  2.31335  &  \phantom{1}2.75  &   0.46  &  \phantom{1}5.93  &  \phantom{1}6.50  &   2.92  &  \phantom{1}2.22  &  $\phantom{>}0.42$  &  0  \\
 84  &  150.11186  &  2.40879  &  \phantom{1}2.18  &   0.37  &  \phantom{1}5.92  &  -1.58  &   2.36  &  -0.67  &  $>          0.46$  &  0  \\
 85  &  150.11984  &  2.41767  &  \phantom{1}2.32  &   0.39  &  \phantom{1}5.87  &  \phantom{1}7.06  &   2.52  &  \phantom{1}2.80  &  $\phantom{>}0.33$  &  0  \\
 86  &  150.05200  &  2.30554  &  \phantom{1}1.91  &   0.33  &  \phantom{1}5.87  &  \phantom{1}2.38  &   2.23  &  \phantom{1}1.07  &  $>          0.28$  &  0  \\
 87  &  150.22409  &  2.35646  &  \phantom{1}3.71  &   0.64  &  \phantom{1}5.83  &  -1.10  &   3.16  &  -0.35  &  $>          0.59$  &  0  \\
 88  &  150.05389  &  2.27630  &  \phantom{1}2.11  &   0.37  &  \phantom{1}5.68  &  \phantom{1}0.83  &   2.27  &  \phantom{1}0.37  &  $>          0.39$  &  0  \\
 89  &  150.16178  &  2.26814  &  \phantom{1}2.15  &   0.38  &  \phantom{1}5.67  &             13.81  &   2.32  &  \phantom{1}5.96  &  $\phantom{>}0.16$  &  1  \\
 90  &  150.05476  &  2.25801  &  \phantom{1}2.59  &   0.46  &  \phantom{1}5.63  &  -5.87  &   3.14  &  -1.87  &  $>          0.41$  &  0  \\
 91  &  150.07011  &  2.29022  &  \phantom{1}1.82  &   0.32  &  \phantom{1}5.60  &  \phantom{1}3.47  &   2.07  &  \phantom{1}1.67  &  $>          0.24$  &  0  \\
 92  &  150.05980  &  2.40055  &  \phantom{1}2.37  &   0.43  &  \phantom{1}5.57  &  \phantom{1}4.56  &   3.01  &  \phantom{1}1.52  &  $>          0.22$  &  0  \\
 93  &  150.05751  &  2.42810  &  \phantom{1}4.36  &   0.78  &  \phantom{1}5.57  &             14.01  &   5.47  &  \phantom{1}2.56  &  $\phantom{>}0.31$  &  0  \\
 94  &  150.06199  &  2.37970  &  \phantom{1}1.95  &   0.35  &  \phantom{1}5.53  &  -2.09  &   2.40  &  -0.87  &  $>          0.41$  &  0  \\
 95  &  150.01647  &  2.32095  &  \phantom{1}3.42  &   0.62  &  \phantom{1}5.51  &  \phantom{1}3.28  &   3.58  &  \phantom{1}0.92  &  $>          0.33$  &  0  \\
 96  &  150.10807  &  2.42369  &  \phantom{1}2.39  &   0.45  &  \phantom{1}5.36  &  \phantom{1}1.88  &   2.82  &  \phantom{1}0.67  &  $>          0.32$  &  0  \\
 97  &  150.09548  &  2.28661  &  \phantom{1}1.53  &   0.29  &  \phantom{1}5.31  &  \phantom{1}0.28  &   1.88  &  \phantom{1}0.15  &  $>          0.38$  &  0  \\
 98  &  150.16077  &  2.34168  &  \phantom{1}1.54  &   0.29  &  \phantom{1}5.29  &  \phantom{1}9.34  &   1.73  &  \phantom{1}5.40  &  $\phantom{>}0.17$  &  1  \\
 99  &  150.20984  &  2.31258  &  \phantom{1}2.53  &   0.48  &  \phantom{1}5.29  &             15.46  &   2.73  &  \phantom{1}5.67  &  $\phantom{>}0.16$  &  1  \\
100  &  150.21841  &  2.34489  &  \phantom{1}2.79  &   0.53  &  \phantom{1}5.25  &  \phantom{1}1.80  &   2.72  &  \phantom{1}0.66  &  $>          0.38$  &  0  \\
101  &  150.14854  &  2.25458  &  \phantom{1}2.01  &   0.38  &  \phantom{1}5.22  &  -2.33  &   2.44  &  -0.95  &  $>          0.41$  &  0  \\
102  &  150.03720  &  2.27215  &  \phantom{1}2.66  &   0.51  &  \phantom{1}5.21  &             14.78  &   3.25  &  \phantom{1}4.55  &  $\phantom{>}0.18$  &  1  \\
103  &  150.08604  &  2.38099  &  \phantom{1}1.94  &   0.36  &  \phantom{1}5.18  &             13.00  &   2.28  &  \phantom{1}5.70  &  $\phantom{>}0.15$  &  1  \\
104  &  150.14108  &  2.42386  &  \phantom{1}2.29  &   0.45  &  \phantom{1}5.10  &             15.70  &   2.83  &  \phantom{1}5.55  &  $\phantom{>}0.15$  &  1  \\
105  &  150.16471  &  2.40932  &  \phantom{1}2.04  &   0.40  &  \phantom{1}5.04  &  \phantom{1}4.68  &   2.27  &  \phantom{1}2.06  &  $\phantom{>}0.44$  &  0  \\
106  &  150.20893  &  2.35022  &  \phantom{1}2.12  &   0.42  &  \phantom{1}5.02  &  \phantom{1}3.52  &   2.40  &  \phantom{1}1.47  &  $>          0.25$  &  0  \\
\hline
\end{tabular}
\end{footnotesize}
\end{center}

%% file: p_stats.tex
\begin{tiny}
\begin{tabular}{lccccccccccccc}
\hline
 ID & RA$_{\rm opt}$ & DEC$_{\rm opt}$ & RA$_{\rm VLA}$ & DEC$_{\rm VLA}$ & $S_{8.0}$ & dist$_{8.0}$ & p$_{8.0}$ & $S_{24}$ & dist$_{24}$ & p$_{24}$ & $S_{\rm VLA}$ & dist$_{\rm VLA}$ & p$_{\rm VLA}$ \\
  SC850-  & deg            & deg     & deg            & deg     & $\mu$Jy   & "            &           & mJy      & "           &          & mJy       & "                &               \\
\hline
  1* &  150.06460  &  2.26405  &    ...      &  ...      &  \phantom{11}   18.88  &  2.47  &  ${\it0.062}$  &   0.13  &  2.18  &  ${\bf0.036}$  &  ...    &  ...   &  $    ...   $  \\
  2* &  150.10014  &  2.29713  &  150.09994  &  2.29721  &  \phantom{11}   35.00  &  2.23  &  ${\bf0.032}$  &   0.16  &  1.42  &  ${\bf0.016}$  &  0.187  &  1.85  &  ${\bf0.001}$  \\
  4  &  150.10546  &  2.31285  &  150.10535  &  2.31284  &  \phantom{11}   24.71  &  1.57  &  ${\bf0.026}$  &   0.23  &  0.65  &  ${\bf0.003}$  &  0.058  &  1.62  &  ${\bf0.002}$  \\
  5  &  150.14304  &  2.35585  &  150.14323  &  2.35602  &  \phantom{11}   14.21  &  0.66  &  ${\bf0.011}$  &   0.14  &  0.40  &  ${\bf0.002}$  &  0.517  &  0.20  &  ${\bf0.000}$  \\
  6  &  150.09854  &  2.36536  &  150.09865  &  2.36538  &  \phantom{11}   31.79  &  1.35  &  ${\bf0.016}$  &   0.24  &  1.20  &  ${\bf0.007}$  &  0.043  &  1.60  &  ${\bf0.002}$  \\
  7  &  150.09866  &  2.32081  &    ...      &  ...      &  \phantom{11}   15.93  &  3.16  &  ${\it0.092}$  &   0.12  &  2.30  &  ${\bf0.041}$  &  ...    &  ...   &  $    ...   $  \\
  8  &  150.09790  &  2.26001  &    ...      &  ...      &  \phantom{111}   9.47  &  1.91  &  ${\it0.070}$  &   ...   &  ...   &  $    ...   $  &  ...    &  ...   &  $    ...   $  \\
  9  &  150.07911  &  2.28180  &    ...      &  ...      &  ...                   &  ...   &  $    ...   $  &   0.33  &  1.35  &  ${\bf0.003}$  &  ...    &  ...   &  $    ...   $  \\
  9  &    ...      &  ...      &    ...      &  ...      &  \phantom{11}   27.09  &  2.21  &  ${\bf0.040}$  &   0.31  &  1.97  &  ${\bf0.012}$  &  ...    &  ...   &  $    ...   $  \\
  9  &    ...      &  ...      &    ...      &  ...      &  \phantom{11}   14.09  &  3.85  &  $    0.116 $  &   0.30  &  4.05  &  ${\bf0.035}$  &  ...    &  ...   &  $    ...   $  \\
 10  &  150.15374  &  2.32800  &    ...      &  ...      &  \phantom{11}   19.41  &  1.36  &  ${\bf0.026}$  &   ...   &  ...   &  $    ...   $  &  ...    &  ...   &  $    ...   $  \\
 11  &  150.04326  &  2.37348  &  150.04318  &  2.37357  &  \phantom{11}   20.86  &  2.15  &  ${\bf0.047}$  &   0.27  &  1.30  &  ${\bf0.004}$  &  0.100  &  2.03  &  ${\bf0.002}$  \\
 13* &  150.08440  &  2.29049  &    ...      &  ...      &  \phantom{11}   59.24  &  2.71  &  ${\bf0.027}$  &   0.42  &  2.43  &  ${\bf0.010}$  &  ...    &  ...   &  $    ...   $  \\
 14  &  150.10641  &  2.25161  &  150.10635  &  2.25161  &  \phantom{11}   26.43  &  2.94  &  ${\it0.059}$  &   0.58  &  2.53  &  ${\bf0.007}$  &  0.112  &  2.89  &  ${\bf0.003}$  \\
 15  &  150.11754  &  2.32996  &    ...      &  ...      &  \phantom{11}   14.97  &  1.69  &  ${\bf0.044}$  &   0.16  &  1.33  &  ${\bf0.009}$  &  ...    &  ...   &  $    ...   $  \\
 16  &  150.05657  &  2.37375  &  150.05649  &  2.37383  &  \phantom{1}   107.95  &  0.88  &  ${\bf0.003}$  &   0.46  &  0.78  &  ${\bf0.001}$  &  0.088  &  0.90  &  ${\bf0.001}$  \\
 17  &  150.20797  &  2.38308  &    ...      &  ...      &  \phantom{11}   21.93  &  0.71  &  ${\bf0.008}$  &   0.07  &  0.15  &  ${\bf0.001}$  &  ...    &  ...   &  $    ...   $  \\
 18  &  150.16357  &  2.37242  &  150.16351  &  2.37251  &  \phantom{11}   34.47  &  1.64  &  ${\bf0.020}$  &   0.56  &  1.41  &  ${\bf0.003}$  &  0.138  &  1.72  &  ${\bf0.001}$  \\
 19  &  150.11255  &  2.37654  &    ...      &  ...      &  \phantom{11}   10.47  &  0.76  &  ${\bf0.018}$  &   0.10  &  0.93  &  ${\bf0.013}$  &  ...    &  ...   &  $    ...   $  \\
 20  &  150.15026  &  2.36414  &    ...      &  ...      &  \phantom{11}   30.45  &  1.54  &  ${\bf0.021}$  &   0.21  &  1.00  &  ${\bf0.007}$  &  ...    &  ...   &  $    ...   $  \\
 21  &  150.09867  &  2.31118  &    ...      &  ...      &  \phantom{11}   11.97  &  0.37  &  ${\bf0.005}$  &   0.18  &  0.90  &  ${\bf0.007}$  &  ...    &  ...   &  $    ...   $  \\
 22  &  150.05706  &  2.29286  &    ...      &  ...      &  \phantom{11}   14.36  &  2.45  &  ${\it0.073}$  &   0.16  &  2.09  &  ${\bf0.028}$  &  ...    &  ...   &  $    ...   $  \\
 23  &  150.12294  &  2.36096  &    ...      &  ...      &  \phantom{11}   12.93  &  0.61  &  ${\bf0.011}$  &   0.23  &  0.65  &  ${\bf0.002}$  &  ...    &  ...   &  $    ...   $  \\
 24  &  150.10909  &  2.29433  &    ...      &  ...      &  \phantom{11}   21.57  &  1.37  &  ${\bf0.024}$  &   0.22  &  0.64  &  ${\bf0.002}$  &  ...    &  ...   &  $    ...   $  \\
 25  &  150.03729  &  2.34057  &  150.03740  &  2.34071  &  \phantom{111}   9.31  &  2.50  &  ${\it0.096}$  &   0.07  &  1.69  &  ${\bf0.040}$  &  0.062  &  1.86  &  ${\bf0.003}$  \\
 26  &  150.07937  &  2.34056  &  150.07925  &  2.34052  &  \phantom{11}   13.55  &  3.04  &  ${\it0.096}$  &   0.16  &  2.82  &  ${\bf0.042}$  &  0.061  &  3.38  &  ${\bf0.005}$  \\
 26  &    ...      &  ...      &    ...      &  ...      &  ...                   &  ...   &  $    ...   $  &   0.18  &  2.68  &  ${\bf0.022}$  &  ...    &  ...   &  $    ...   $  \\
 28  &  150.12181  &  2.34131  &    ...      &  ...      &  \phantom{11}   10.69  &  1.76  &  ${\it0.059}$  &   0.10  &  2.09  &  ${\bf0.042}$  &  ...    &  ...   &  $    ...   $  \\
 29* &  150.10525  &  2.43499  &    ...      &  ...      &  \phantom{11}   13.36  &  1.21  &  ${\bf0.030}$  &   0.08  &  0.89  &  ${\bf0.014}$  &  ...    &  ...   &  $    ...   $  \\
 31  &  150.05248  &  2.24555  &    ...      &  ...      &  \phantom{11}   37.59  &  2.89  &  ${\bf0.043}$  &   0.38  &  1.50  &  ${\bf0.003}$  &  ...    &  ...   &  $    ...   $  \\
 33  &  150.04098  &  2.28063  &    ...      &  ...      &  \phantom{11}   11.41  &  1.89  &  ${\it0.062}$  &   0.08  &  2.74  &  ${\it0.068}$  &  ...    &  ...   &  $    ...   $  \\
 34* &  150.13513  &  2.39942  &  150.13495  &  2.39930  &  \phantom{11}   14.60  &  0.38  &  ${\bf0.004}$  &   0.17  &  0.32  &  ${\bf0.001}$  &  0.056  &  0.96  &  ${\bf0.001}$  \\
 35  &  150.16771  &  2.29876  &    ...      &  ...      &  \phantom{11}   16.71  &  2.85  &  ${\it0.080}$  &   0.32  &  2.62  &  ${\bf0.018}$  &  ...    &  ...   &  $    ...   $  \\
 36* &  150.08187  &  2.41556  &    ...      &  ...      &  ...                   &  ...   &  $    ...   $  &   0.64  &  0.72  &  ${\bf0.001}$  &  ...    &  ...   &  $    ...   $  \\
 37  &  150.06811  &  2.27569  &    ...      &  ...      &  \phantom{11}   29.14  &  1.92  &  ${\bf0.030}$  &   0.45  &  1.53  &  ${\bf0.004}$  &  ...    &  ...   &  $    ...   $  \\
 38  &  150.07527  &  2.37940  &    ...      &  ...      &  ...                   &  ...   &  $    ...   $  &   0.13  &  1.28  &  ${\bf0.017}$  &  ...    &  ...   &  $    ...   $  \\
 40* &  150.10531  &  2.32590  &    ...      &  ...      &  ...                   &  ...   &  $    ...   $  &   0.06  &  3.12  &  ${\bf0.050}$  &  ...    &  ...   &  $    ...   $  \\
 42* &  150.02754  &  2.34577  &    ...      &  ...      &  ...                   &  ...   &  $    ...   $  &   0.16  &  4.32  &  ${\it0.069}$  &  ...    &  ...   &  $    ...   $  \\
 43  &  150.17186  &  2.24070  &    ...      &  ...      &  \phantom{11}   18.90  &  2.87  &  ${\it0.074}$  &   0.21  &  2.77  &  ${\bf0.032}$  &  ...    &  ...   &  $    ...   $  \\
 44* &  150.13702  &  2.23222  &  150.13658  &  2.23252  &  \phantom{11}   46.55  &  2.98  &  ${\bf0.038}$  &   0.22  &  2.39  &  ${\bf0.024}$  &  0.045  &  1.94  &  ${\bf0.003}$  \\
 45  &  150.12715  &  2.38786  &    ...      &  ...      &  \phantom{11}   12.64  &  1.21  &  ${\bf0.031}$  &   0.06  &  0.98  &  ${\bf0.020}$  &  ...    &  ...   &  $    ...   $  \\
 46  &  150.10590  &  2.42879  &    ...      &  ...      &  ...                   &  ...   &  $    ...   $  &   0.13  &  1.99  &  ${\bf0.033}$  &  ...    &  ...   &  $    ...   $  \\
 47  &  150.04825  &  2.25144  &    ...      &  ...      &  ...                   &  ...   &  $    ...   $  &   0.13  &  4.21  &  ${\bf0.047}$  &  ...    &  ...   &  $    ...   $  \\
 48  &  150.02141  &  2.28867  &    ...      &  ...      &  \phantom{11}   20.33  &  3.16  &  ${\it0.079}$  &   ...   &  ...   &  $    ...   $  &  ...    &  ...   &  $    ...   $  \\
 49  &  150.15747  &  2.35803  &    ...      &  ...      &  ...                   &  ...   &  $    ...   $  &   0.19  &  3.38  &  ${\bf0.047}$  &  ...    &  ...   &  $    ...   $  \\
 51  &  150.03652  &  2.28617  &    ...      &  ...      &  ...                   &  ...   &  $    ...   $  &   0.10  &  3.63  &  ${\bf0.049}$  &  ...    &  ...   &  $    ...   $  \\
 53  &  150.18763  &  2.32250  &    ...      &  ...      &  \phantom{11}   46.19  &  1.80  &  ${\bf0.018}$  &   0.24  &  1.73  &  ${\bf0.013}$  &  ...    &  ...   &  $    ...   $  \\
 54  &  150.04241  &  2.29985  &    ...      &  ...      &  \phantom{11}   19.09  &  2.46  &  ${\it0.061}$  &   0.09  &  2.47  &  ${\it0.056}$  &  ...    &  ...   &  $    ...   $  \\
 55  &  150.13354  &  2.37042  &    ...      &  ...      &  \phantom{11}   15.81  &  3.32  &  ${\it0.097}$  &   0.10  &  2.10  &  ${\bf0.027}$  &  ...    &  ...   &  $    ...   $  \\
 56  &  150.05002  &  2.38607  &    ...      &  ...      &  ...                   &  ...   &  $    ...   $  &   0.05  &  1.15  &  ${\bf0.026}$  &  ...    &  ...   &  $    ...   $  \\
 59  &  150.18497  &  2.38894  &    ...      &  ...      &  ...                   &  ...   &  $    ...   $  &   0.11  &  4.48  &  ${\it0.088}$  &  ...    &  ...   &  $    ...   $  \\
 61  &  150.05397  &  2.39590  &    ...      &  ...      &  \phantom{11}   11.07  &  1.39  &  ${\bf0.042}$  &   0.16  &  0.73  &  ${\bf0.006}$  &  ...    &  ...   &  $    ...   $  \\
 62  &  150.16691  &  2.23582  &    ...      &  ...      &  \phantom{11}   20.71  &  0.89  &  ${\bf0.013}$  &   0.37  &  0.69  &  ${\bf0.002}$  &  ...    &  ...   &  $    ...   $  \\
 63* &  150.07672  &  2.39860  &    ...      &  ...      &  \phantom{11}   12.09  &  2.56  &  ${\it0.086}$  &   ...   &  ...   &  $    ...   $  &  ...    &  ...   &  $    ...   $  \\
 64  &  150.13074  &  2.31408  &    ...      &  ...      &  ...                   &  ...   &  $    ...   $  &   0.18  &  4.05  &  ${\it0.059}$  &  ...    &  ...   &  $    ...   $  \\
 65* &  150.09156  &  2.39904  &    ...      &  ...      &  \phantom{11}   58.66  &  2.46  &  ${\bf0.024}$  &   0.16  &  2.19  &  ${\bf0.018}$  &  ...    &  ...   &  $    ...   $  \\
 66  &  150.17561  &  2.40159  &    ...      &  ...      &  \phantom{11}   27.26  &  2.72  &  ${\it0.052}$  &   0.16  &  2.34  &  ${\bf0.034}$  &  ...    &  ...   &  $    ...   $  \\
 67  &  150.11132  &  2.40320  &    ...      &  ...      &  \phantom{111}   9.76  &  3.36  &  $    0.121 $  &   0.11  &  3.22  &  ${\it0.067}$  &  ...    &  ...   &  $    ...   $  \\
 68  &  150.13001  &  2.25269  &    ...      &  ...      &  \phantom{11}   10.88  &  2.59  &  ${\it0.092}$  &   0.18  &  2.63  &  ${\bf0.034}$  &  ...    &  ...   &  $    ...   $  \\
 71* &  150.07194  &  2.23867  &    ...      &  ...      &  ...                   &  ...   &  $    ...   $  &   0.08  &  2.14  &  ${\it0.050}$  &  ...    &  ...   &  $    ...   $  \\
 72* &  150.06456  &  2.32903  &    ...      &  ...      &  \phantom{1}   111.50  &  2.24  &  ${\bf0.012}$  &   0.29  &  1.99  &  ${\bf0.013}$  &  ...    &  ...   &  $    ...   $  \\
 73  &  150.20962  &  2.35525  &  150.20955  &  2.35531  &               1446.07  &  2.32  &  ${\bf0.001}$  &   1.46  &  2.07  &  ${\bf0.001}$  &  0.273  &  2.09  &  ${\bf0.001}$  \\
 74  &  150.07066  &  2.30514  &    ...      &  ...      &  \phantom{11}   10.24  &  3.65  &  $    0.130 $  &   0.05  &  3.68  &  ${\it0.096}$  &  ...    &  ...   &  $    ...   $  \\
 75  &  150.15933  &  2.29680  &    ...      &  ...      &  \phantom{111}   8.45  &  1.30  &  ${\bf0.047}$  &   0.07  &  1.47  &  ${\bf0.034}$  &  ...    &  ...   &  $    ...   $  \\
 77* &  150.07027  &  2.42297  &    ...      &  ...      &  ...                   &  ...   &  $    ...   $  &   0.10  &  3.72  &  ${\bf0.050}$  &  ...    &  ...   &  $    ...   $  \\
 78  &  150.09944  &  2.40487  &    ...      &  ...      &  \phantom{11}   10.07  &  1.28  &  ${\bf0.042}$  &   0.10  &  1.08  &  ${\bf0.010}$  &  ...    &  ...   &  $    ...   $  \\
 79  &  150.04118  &  2.32813  &    ...      &  ...      &  ...                   &  ...   &  $    ...   $  &   0.13  &  3.85  &  ${\bf0.043}$  &  ...    &  ...   &  $    ...   $  \\
 79  &    ...      &  ...      &    ...      &  ...      &  \phantom{11}   43.14  &  3.81  &  ${\it0.059}$  &   ...   &  ...   &  $    ...   $  &  ...    &  ...   &  $    ...   $  \\
 81* &  150.12582  &  2.41354  &    ...      &  ...      &  ...                   &  ...   &  $    ...   $  &   ...   &  ...   &  $    ...   $  &  ...    &  ...   &  $    ...   $  \\
 83  &  150.02492  &  2.31287  &    ...      &  ...      &  \phantom{11}   12.00  &  3.40  &  $    0.127 $  &   0.16  &  4.09  &  ${\bf0.041}$  &  ...    &  ...   &  $    ...   $  \\
 84* &  150.11154  &  2.40957  &    ...      &  ...      &  \phantom{11}   19.59  &  3.09  &  ${\it0.086}$  &   0.04  &  3.05  &  ${\it0.086}$  &  ...    &  ...   &  $    ...   $  \\
 86  &  150.05166  &  2.30585  &    ...      &  ...      &  \phantom{11}   73.38  &  1.75  &  ${\bf0.012}$  &   0.34  &  1.67  &  ${\bf0.008}$  &  ...    &  ...   &  $    ...   $  \\
 87  &  150.22434  &  2.35644  &    ...      &  ...      &  \phantom{11}   11.10  &  0.74  &  ${\bf0.017}$  &   0.09  &  0.98  &  ${\bf0.016}$  &  ...    &  ...   &  $    ...   $  \\
 88* &  150.05456  &  2.27535  &    ...      &  ...      &  \phantom{11}   82.07  &  4.21  &  ${\bf0.044}$  &   0.27  &  3.35  &  ${\bf0.019}$  &  ...    &  ...   &  $    ...   $  \\
 88  &    ...      &  ...      &    ...      &  ...      &  \phantom{11}   31.33  &  2.41  &  ${\bf0.043}$  &   ...   &  ...   &  $    ...   $  &  ...    &  ...   &  $    ...   $  \\
 89  &  150.16255  &  2.26808  &    ...      &  ...      &  ...                   &  ...   &  $    ...   $  &   0.07  &  1.39  &  ${\bf0.032}$  &  ...    &  ...   &  $    ...   $  \\
 91  &  150.07060  &  2.28920  &    ...      &  ...      &  \phantom{11}   12.83  &  4.01  &  $    0.149 $  &   0.18  &  4.25  &  ${\bf0.040}$  &  ...    &  ...   &  $    ...   $  \\
 92  &  150.05916  &  2.39982  &    ...      &  ...      &  \phantom{11}   21.66  &  3.63  &  $    0.100 $  &   0.09  &  4.54  &  ${\it0.064}$  &  ...    &  ...   &  $    ...   $  \\
 93* &  150.05785  &  2.42723  &    ...      &  ...      &  \phantom{11}   44.59  &  3.25  &  ${\bf0.049}$  &   0.07  &  3.23  &  ${\it0.088}$  &  ...    &  ...   &  $    ...   $  \\
 95  &  150.01640  &  2.32096  &    ...      &  ...      &  \phantom{11}   14.17  &  0.27  &  ${\bf0.003}$  &   ...   &  ...   &  $    ...   $  &  ...    &  ...   &  $    ...   $  \\
 98  &  150.16186  &  2.34092  &    ...      &  ...      &  \phantom{11}   46.07  &  4.71  &  ${\it0.081}$  &   0.31  &  4.68  &  ${\bf0.043}$  &  ...    &  ...   &  $    ...   $  \\
 99  &  150.21020  &  2.31167  &  150.21013  &  2.31168  &  \phantom{11}   93.38  &  3.50  &  ${\bf0.031}$  &   0.91  &  3.11  &  ${\bf0.005}$  &  0.227  &  3.41  &  ${\bf0.003}$  \\
102  &  150.03745  &  2.27186  &  150.03670  &  2.27098  &  \phantom{11}   20.84  &  1.32  &  ${\bf0.026}$  &   ...   &  ...   &  $    ...   $  &  0.075  &  1.03  &  ${\bf0.001}$  \\
102  &    ...      &  ...      &  150.03738  &  2.27194  &  \phantom{11}   61.12  &  4.82  &  ${\it0.068}$  &   0.71  &  3.56  &  ${\bf0.009}$  &  0.080  &  4.57  &  ${\bf0.008}$  \\
103  &  150.08514  &  2.38195  &    ...      &  ...      &  ...                   &  ...   &  $    ...   $  &   0.39  &  2.45  &  ${\bf0.008}$  &  ...    &  ...   &  $    ...   $  \\
105* &  150.16426  &  2.40881  &    ...      &  ...      &  \phantom{11}   12.05  &  2.25  &  ${\it0.088}$  &   0.27  &  2.36  &  ${\bf0.012}$  &  ...    &  ...   &  $    ...   $  \\
\hline
\end{tabular}
\end{tiny}

%% file: photo1.tex
\begin{scriptsize}
\begin{tabular}{lcccccccccccccccc}
\hline
 ID & RA & DEC & $u$ & $g$ & $r$ & $i$ & $z$ & $Y$ & $J$ & $H$ & $Ks$ & $3.6\mu m$ &$4.5\mu m$ & F125W & F160W \\
\hline
  1  &  150.06460  &  2.26405  &  $25.02\pm0.10$  &  $24.60\pm0.10$  &  $24.17\pm0.10$  &  $23.65\pm0.10$  &  $23.09\pm0.10$  &  $22.67\pm0.10$  &  $22.30\pm0.10$  &  $21.85\pm0.10$  &  $21.51\pm0.10$  &  $20.97\pm0.10$  &  $20.77\pm0.10$  &  $22.24\pm0.10$  &  $21.80\pm0.10$  \\
  2  &  150.10014  &  2.29713  &  $24.51\pm0.10$  &  $22.47\pm0.10$  &  $20.96\pm0.10$  &  $20.31\pm0.10$  &  $20.05\pm0.10$  &  $19.81\pm0.10$  &  $19.51\pm0.10$  &  $19.18\pm0.10$  &  $18.87\pm0.10$  &  $19.62\pm0.10$  &  $19.77\pm0.10$  &  $19.49\pm0.10$  &  $19.19\pm0.10$  \\
  4  &  150.10546  &  2.31285  &  $-           $  &  $26.97\pm0.40$  &  $26.50\pm0.38$  &  $25.99\pm0.31$  &  $25.51\pm0.17$  &  $24.43\pm0.27$  &  $24.75\pm0.43$  &  $23.21\pm0.17$  &  $22.43\pm0.10$  &  $21.44\pm0.10$  &  $21.06\pm0.10$  &  $24.05\pm0.12$  &  $23.29\pm0.10$  \\
  5  &  150.14304  &  2.35585  &  $-           $  &  $26.78\pm0.34$  &  $26.04\pm0.25$  &  $25.50\pm0.20$  &  $25.63\pm0.19$  &  $24.90\pm0.41$  &  $24.38\pm0.31$  &  $24.08\pm0.37$  &  $22.97\pm0.16$  &  $22.24\pm0.10$  &  $21.83\pm0.10$  &  $24.39\pm0.16$  &  $23.72\pm0.10$  \\
  6  &  150.09854  &  2.36536  &  $-           $  &  $-           $  &  $26.46\pm0.36$  &  $26.25\pm0.40$  &  $26.05\pm0.27$  &  $-           $  &  $24.77\pm0.44$  &  $23.47\pm0.21$  &  $22.54\pm0.11$  &  $21.21\pm0.10$  &  $20.82\pm0.10$  &  $24.68\pm0.21$  &  $23.32\pm0.10$  \\
  7  &  150.09866  &  2.32081  &  $-           $  &  $26.09\pm0.18$  &  $25.39\pm0.14$  &  $25.22\pm0.15$  &  $25.25\pm0.13$  &  $-           $  &  $24.70\pm0.41$  &  $23.56\pm0.23$  &  $23.09\pm0.18$  &  $22.36\pm0.10$  &  $21.86\pm0.10$  &  $24.39\pm0.16$  &  $23.75\pm0.10$  \\
  8  &  150.09790  &  2.26001  &  $26.61\pm0.29$  &  $25.85\pm0.14$  &  $25.32\pm0.13$  &  $24.93\pm0.12$  &  $24.91\pm0.10$  &  $-           $  &  $24.54\pm0.36$  &  $24.10\pm0.38$  &  $23.73\pm0.32$  &  $23.37\pm0.11$  &  $23.21\pm0.14$  &  $24.52\pm0.18$  &  $24.18\pm0.15$  \\
  9  &  150.07911  &  2.28180  &  $27.29\pm0.54$  &  $26.35\pm0.23$  &  $26.41\pm0.35$  &  $25.31\pm0.17$  &  $24.98\pm0.10$  &  $24.66\pm0.33$  &  $23.87\pm0.19$  &  $23.01\pm0.14$  &  $22.59\pm0.11$  &  $21.59\pm0.10$  &  $21.35\pm0.10$  &  $23.68\pm0.10$  &  $23.20\pm0.10$  \\
 10  &  150.15374  &  2.32800  &  $-           $  &  $-           $  &  $-           $  &  $-           $  &  $-           $  &  $-           $  &  $-           $  &  $-           $  &  $23.49\pm0.26$  &  $22.29\pm0.10$  &  $21.77\pm0.10$  &  $-           $  &  $24.25\pm0.16$  \\
 13  &  150.08440  &  2.29049  &  $23.55\pm0.10$  &  $23.11\pm0.10$  &  $22.91\pm0.10$  &  $22.47\pm0.10$  &  $22.11\pm0.10$  &  $22.14\pm0.10$  &  $21.69\pm0.10$  &  $21.23\pm0.10$  &  $20.70\pm0.10$  &  $20.10\pm0.10$  &  $19.89\pm0.10$  &  $21.55\pm0.10$  &  $21.15\pm0.10$  \\
 14  &  150.10641  &  2.25161  &  $-           $  &  $26.67\pm0.30$  &  $26.27\pm0.30$  &  $25.50\pm0.20$  &  $25.11\pm0.12$  &  $-           $  &  $23.69\pm0.16$  &  $22.86\pm0.12$  &  $21.95\pm0.10$  &  $20.70\pm0.10$  &  $20.31\pm0.10$  &  $23.84\pm0.10$  &  $22.92\pm0.10$  \\
 15  &  150.11754  &  2.32996  &  $-           $  &  $-           $  &  $26.35\pm0.33$  &  $26.09\pm0.34$  &  $25.77\pm0.21$  &  $-           $  &  $24.65\pm0.39$  &  $23.36\pm0.19$  &  $22.55\pm0.11$  &  $21.62\pm0.10$  &  $21.29\pm0.10$  &  $24.34\pm0.16$  &  $23.44\pm0.10$  \\
 16  &  150.05657  &  2.37375  &  $25.14\pm0.10$  &  $24.66\pm0.10$  &  $24.01\pm0.10$  &  $23.22\pm0.10$  &  $22.57\pm0.10$  &  $22.25\pm0.10$  &  $21.78\pm0.10$  &  $21.33\pm0.10$  &  $20.71\pm0.10$  &  $20.16\pm0.10$  &  $19.82\pm0.10$  &  $21.74\pm0.10$  &  $21.24\pm0.10$  \\
 18  &  150.16357  &  2.37242  &  $25.12\pm0.10$  &  $24.32\pm0.10$  &  $23.98\pm0.10$  &  $23.65\pm0.10$  &  $23.45\pm0.10$  &  $23.29\pm0.10$  &  $22.57\pm0.10$  &  $22.06\pm0.10$  &  $21.48\pm0.10$  &  $20.75\pm0.10$  &  $20.48\pm0.10$  &  $22.49\pm0.10$  &  $22.11\pm0.10$  \\
 19  &  150.11255  &  2.37654  &  $-           $  &  $-           $  &  $-           $  &  $-           $  &  $-           $  &  $-           $  &  $-           $  &  $-           $  &  $24.00\pm0.41$  &  $22.46\pm0.10$  &  $21.93\pm0.10$  &  $-           $  &  $24.98\pm0.31$  \\
 20  &  150.15026  &  2.36414  &  $26.01\pm0.17$  &  $25.19\pm0.10$  &  $24.71\pm0.10$  &  $24.71\pm0.10$  &  $24.32\pm0.10$  &  $23.53\pm0.12$  &  $23.38\pm0.12$  &  $22.65\pm0.10$  &  $21.86\pm0.10$  &  $20.95\pm0.10$  &  $20.67\pm0.10$  &  $23.45\pm0.10$  &  $22.69\pm0.10$  \\
 21  &  150.09867  &  2.31118  &  $-           $  &  $-           $  &  $-           $  &  $-           $  &  $26.59\pm0.45$  &  $-           $  &  $-           $  &  $-           $  &  $23.20\pm0.20$  &  $22.08\pm0.10$  &  $21.72\pm0.10$  &  $24.89\pm0.26$  &  $23.95\pm0.12$  \\
 22  &  150.05706  &  2.29286  &  $-           $  &  $-           $  &  $-           $  &  $-           $  &  $-           $  &  $-           $  &  $-           $  &  $23.97\pm0.33$  &  $22.91\pm0.15$  &  $21.72\pm0.10$  &  $21.24\pm0.10$  &  $25.42\pm0.42$  &  $24.43\pm0.19$  \\
 23  &  150.12294  &  2.36096  &  $-           $  &  $27.04\pm0.43$  &  $26.32\pm0.32$  &  $25.66\pm0.23$  &  $25.07\pm0.11$  &  $-           $  &  $23.76\pm0.17$  &  $23.51\pm0.22$  &  $22.27\pm0.10$  &  $21.35\pm0.10$  &  $21.04\pm0.10$  &  $23.60\pm0.10$  &  $23.15\pm0.10$  \\
 24  &  150.10909  &  2.29433  &  $26.25\pm0.21$  &  $25.22\pm0.10$  &  $25.02\pm0.10$  &  $24.17\pm0.10$  &  $23.73\pm0.10$  &  $23.32\pm0.10$  &  $22.46\pm0.10$  &  $21.92\pm0.10$  &  $21.47\pm0.10$  &  $20.77\pm0.10$  &  $20.52\pm0.10$  &  $22.50\pm0.10$  &  $22.00\pm0.10$  \\
 26  &  150.07937  &  2.34056  &  $25.53\pm0.11$  &  $24.29\pm0.10$  &  $23.76\pm0.10$  &  $23.45\pm0.10$  &  $23.28\pm0.10$  &  $23.15\pm0.10$  &  $22.78\pm0.10$  &  $22.27\pm0.10$  &  $21.89\pm0.10$  &  $21.59\pm0.10$  &  $21.37\pm0.10$  &  $22.72\pm0.10$  &  $22.23\pm0.10$  \\
 28  &  150.12181  &  2.34131  &  $26.87\pm0.37$  &  $25.47\pm0.10$  &  $25.23\pm0.12$  &  $25.35\pm0.17$  &  $24.83\pm0.10$  &  $25.10\pm0.50$  &  $23.74\pm0.17$  &  $23.20\pm0.17$  &  $22.64\pm0.12$  &  $22.29\pm0.10$  &  $22.22\pm0.10$  &  $23.82\pm0.10$  &  $23.31\pm0.10$  \\
 29  &  150.10525  &  2.43499  &  $24.44\pm0.10$  &  $23.59\pm0.10$  &  $22.75\pm0.10$  &  $21.82\pm0.10$  &  $21.33\pm0.10$  &  $21.30\pm0.10$  &  $21.07\pm0.10$  &  $20.81\pm0.10$  &  $20.49\pm0.10$  &  $20.42\pm0.10$  &  $20.92\pm0.10$  &  $21.07\pm0.10$  &  $20.81\pm0.10$  \\
 34  &  150.13513  &  2.39942  &  $21.72\pm0.10$  &  $20.55\pm0.10$  &  $20.07\pm0.10$  &  $19.80\pm0.10$  &  $19.83\pm0.10$  &  $19.76\pm0.10$  &  $19.74\pm0.10$  &  $19.78\pm0.10$  &  $20.04\pm0.10$  &  $20.70\pm0.10$  &  $20.81\pm0.10$  &  $19.74\pm0.10$  &  $19.69\pm0.10$  \\
 35  &  150.16771  &  2.29876  &  $-           $  &  $-           $  &  $-           $  &  $-           $  &  $25.79\pm0.21$  &  $25.17\pm0.53$  &  $24.02\pm0.22$  &  $23.19\pm0.16$  &  $22.04\pm0.10$  &  $20.96\pm0.10$  &  $20.70\pm0.10$  &  $23.91\pm0.10$  &  $23.25\pm0.10$  \\
 36  &  150.08187  &  2.41556  &  $26.65\pm0.30$  &  $24.45\pm0.10$  &  $23.73\pm0.10$  &  $23.43\pm0.10$  &  $23.39\pm0.10$  &  $23.45\pm0.11$  &  $23.38\pm0.12$  &  $22.78\pm0.11$  &  $22.82\pm0.14$  &  $23.01\pm0.10$  &  $22.98\pm0.11$  &  $23.12\pm0.10$  &  $22.79\pm0.10$  \\
 37  &  150.06811  &  2.27569  &  $23.96\pm0.10$  &  $22.51\pm0.10$  &  $22.82\pm0.10$  &  $22.82\pm0.10$  &  $22.32\pm0.10$  &  $22.01\pm0.10$  &  $21.54\pm0.10$  &  $21.11\pm0.10$  &  $20.81\pm0.10$  &  $-           $  &  $-           $  &  $21.79\pm0.10$  &  $21.35\pm0.10$  \\
 38  &  150.07527  &  2.37940  &  $26.25\pm0.21$  &  $25.24\pm0.10$  &  $24.77\pm0.10$  &  $24.33\pm0.10$  &  $23.94\pm0.10$  &  $23.80\pm0.15$  &  $22.95\pm0.10$  &  $22.54\pm0.10$  &  $21.65\pm0.10$  &  $21.37\pm0.10$  &  $21.22\pm0.10$  &  $22.98\pm0.10$  &  $22.42\pm0.10$  \\
 40  &  150.10531  &  2.32590  &  $26.12\pm0.18$  &  $25.78\pm0.13$  &  $25.33\pm0.13$  &  $24.58\pm0.10$  &  $24.39\pm0.10$  &  $24.22\pm0.22$  &  $24.42\pm0.32$  &  $23.17\pm0.16$  &  $22.93\pm0.15$  &  $22.39\pm0.10$  &  $22.07\pm0.10$  &  $23.85\pm0.10$  &  $23.35\pm0.10$  \\
 43  &  150.17186  &  2.24070  &  $-           $  &  $-           $  &  $-           $  &  $25.98\pm0.31$  &  $25.86\pm0.23$  &  $-           $  &  $24.08\pm0.23$  &  $22.94\pm0.13$  &  $22.15\pm0.10$  &  $21.02\pm0.10$  &  $20.84\pm0.10$  &  $23.87\pm0.10$  &  $23.01\pm0.10$  \\
 44  &  150.13702  &  2.23222  &  $21.67\pm0.10$  &  $20.68\pm0.10$  &  $20.43\pm0.10$  &  $20.14\pm0.10$  &  $20.20\pm0.10$  &  $20.08\pm0.10$  &  $20.01\pm0.10$  &  $19.95\pm0.10$  &  $19.93\pm0.10$  &  $20.60\pm0.10$  &  $20.83\pm0.10$  &  $20.01\pm0.10$  &  $19.89\pm0.10$  \\
 45  &  150.12715  &  2.38786  &  $-           $  &  $-           $  &  $26.18\pm0.28$  &  $26.13\pm0.35$  &  $25.69\pm0.20$  &  $-           $  &  $24.96\pm0.53$  &  $-           $  &  $23.24\pm0.21$  &  $23.35\pm0.11$  &  $22.86\pm0.10$  &  $24.99\pm0.28$  &  $24.38\pm0.18$  \\
 46  &  150.10590  &  2.42879  &  $24.71\pm0.10$  &  $23.68\pm0.10$  &  $23.45\pm0.10$  &  $23.94\pm0.10$  &  $23.45\pm0.10$  &  $22.96\pm0.10$  &  $23.05\pm0.10$  &  $22.73\pm0.11$  &  $22.91\pm0.15$  &  $-           $  &  $-           $  &  $23.96\pm0.11$  &  $23.66\pm0.10$  \\
 49  &  150.15747  &  2.35803  &  $26.82\pm0.35$  &  $25.97\pm0.16$  &  $25.58\pm0.16$  &  $25.53\pm0.20$  &  $25.23\pm0.13$  &  $24.84\pm0.39$  &  $24.91\pm0.50$  &  $23.94\pm0.33$  &  $23.09\pm0.18$  &  $22.22\pm0.10$  &  $21.89\pm0.10$  &  $24.21\pm0.14$  &  $23.87\pm0.11$  \\
 53  &  150.18763  &  2.32250  &  $25.71\pm0.13$  &  $25.13\pm0.10$  &  $24.50\pm0.10$  &  $23.71\pm0.10$  &  $22.95\pm0.10$  &  $22.22\pm0.10$  &  $21.57\pm0.10$  &  $20.90\pm0.10$  &  $20.38\pm0.10$  &  $19.61\pm0.10$  &  $19.48\pm0.10$  &  $21.55\pm0.10$  &  $20.95\pm0.10$  \\
 55  &  150.13354  &  2.37042  &  $24.99\pm0.10$  &  $24.51\pm0.10$  &  $24.25\pm0.10$  &  $23.94\pm0.10$  &  $23.54\pm0.10$  &  $23.29\pm0.10$  &  $22.35\pm0.10$  &  $21.86\pm0.10$  &  $21.30\pm0.10$  &  $20.77\pm0.10$  &  $20.59\pm0.10$  &  $22.37\pm0.10$  &  $21.85\pm0.10$  \\
 59  &  150.18497  &  2.38894  &  $24.79\pm0.10$  &  $24.14\pm0.10$  &  $23.99\pm0.10$  &  $23.63\pm0.10$  &  $23.39\pm0.10$  &  $23.11\pm0.10$  &  $22.49\pm0.10$  &  $22.25\pm0.10$  &  $22.06\pm0.10$  &  $21.67\pm0.10$  &  $21.53\pm0.10$  &  $22.60\pm0.10$  &  $22.30\pm0.10$  \\
 61  &  150.05397  &  2.39590  &  $-           $  &  $25.38\pm0.10$  &  $24.49\pm0.10$  &  $25.12\pm0.14$  &  $24.14\pm0.10$  &  $24.28\pm0.23$  &  $23.93\pm0.20$  &  $23.63\pm0.24$  &  $22.99\pm0.16$  &  $21.48\pm0.10$  &  $21.34\pm0.10$  &  $24.91\pm0.26$  &  $24.31\pm0.17$  \\
 62  &  150.16691  &  2.23582  &  $26.48\pm0.26$  &  $25.74\pm0.13$  &  $25.51\pm0.15$  &  $24.76\pm0.10$  &  $24.47\pm0.10$  &  $23.85\pm0.16$  &  $23.02\pm0.10$  &  $22.33\pm0.10$  &  $21.67\pm0.10$  &  $20.71\pm0.10$  &  $20.46\pm0.10$  &  $22.97\pm0.10$  &  $22.44\pm0.10$  \\
 63  &  150.07672  &  2.39860  &  $-           $  &  $-           $  &  $-           $  &  $26.44\pm0.47$  &  $25.79\pm0.22$  &  $-           $  &  $-           $  &  $23.42\pm0.20$  &  $22.75\pm0.13$  &  $22.00\pm0.10$  &  $21.61\pm0.10$  &  $24.57\pm0.19$  &  $23.84\pm0.11$  \\
 64  &  150.13074  &  2.31408  &  $-           $  &  $-           $  &  $-           $  &  $26.45\pm0.48$  &  $26.01\pm0.26$  &  $-           $  &  $24.55\pm0.36$  &  $23.37\pm0.19$  &  $23.05\pm0.17$  &  $21.75\pm0.10$  &  $21.36\pm0.10$  &  $24.48\pm0.18$  &  $23.77\pm0.10$  \\
 65  &  150.09156  &  2.39904  &  $23.49\pm0.10$  &  $22.57\pm0.10$  &  $22.42\pm0.10$  &  $22.04\pm0.10$  &  $21.52\pm0.10$  &  $21.35\pm0.10$  &  $21.16\pm0.10$  &  $20.92\pm0.10$  &  $20.51\pm0.10$  &  $20.77\pm0.10$  &  $20.73\pm0.10$  &  $21.14\pm0.10$  &  $21.04\pm0.10$  \\
 66  &  150.17561  &  2.40159  &  $26.00\pm0.16$  &  $25.41\pm0.10$  &  $25.08\pm0.10$  &  $24.45\pm0.10$  &  $24.13\pm0.10$  &  $23.93\pm0.17$  &  $22.92\pm0.10$  &  $22.40\pm0.10$  &  $21.68\pm0.10$  &  $20.71\pm0.10$  &  $20.43\pm0.10$  &  $22.84\pm0.10$  &  $22.30\pm0.10$  \\
 67  &  150.11132  &  2.40320  &  $-           $  &  $26.34\pm0.23$  &  $26.22\pm0.29$  &  $25.87\pm0.28$  &  $26.72\pm0.50$  &  $25.12\pm0.51$  &  $24.68\pm0.41$  &  $23.92\pm0.32$  &  $23.31\pm0.22$  &  $-           $  &  $-           $  &  $24.64\pm0.20$  &  $23.94\pm0.12$  \\
 68  &  150.13001  &  2.25269  &  $-           $  &  $-           $  &  $-           $  &  $26.29\pm0.41$  &  $26.44\pm0.39$  &  $-           $  &  $-           $  &  $24.34\pm0.47$  &  $23.04\pm0.17$  &  $22.26\pm0.10$  &  $21.78\pm0.10$  &  $24.78\pm0.23$  &  $24.04\pm0.13$  \\
 71  &  150.07194  &  2.23867  &  $25.67\pm0.12$  &  $24.99\pm0.10$  &  $24.18\pm0.10$  &  $23.16\pm0.10$  &  $23.28\pm0.10$  &  $22.56\pm0.10$  &  $22.23\pm0.10$  &  $21.88\pm0.10$  &  $21.38\pm0.10$  &  $21.12\pm0.10$  &  $21.45\pm0.10$  &  $22.20\pm0.10$  &  $21.88\pm0.10$  \\
 72  &  150.06456  &  2.32903  &  $23.81\pm0.10$  &  $22.98\pm0.10$  &  $22.28\pm0.10$  &  $21.96\pm0.10$  &  $21.66\pm0.10$  &  $21.54\pm0.10$  &  $21.56\pm0.10$  &  $21.15\pm0.10$  &  $20.66\pm0.10$  &  $20.35\pm0.10$  &  $20.00\pm0.10$  &  $21.46\pm0.10$  &  $21.23\pm0.10$  \\
 74  &  150.07066  &  2.30514  &  $-           $  &  $25.51\pm0.11$  &  $24.75\pm0.10$  &  $24.48\pm0.10$  &  $24.49\pm0.10$  &  $24.40\pm0.26$  &  $24.03\pm0.22$  &  $23.46\pm0.21$  &  $22.99\pm0.16$  &  $22.84\pm0.10$  &  $22.40\pm0.10$  &  $24.52\pm0.18$  &  $23.62\pm0.10$  \\
 75  &  150.15933  &  2.29680  &  $-           $  &  $-           $  &  $26.54\pm0.39$  &  $25.83\pm0.27$  &  $25.17\pm0.12$  &  $24.71\pm0.35$  &  $23.80\pm0.18$  &  $22.82\pm0.12$  &  $22.13\pm0.10$  &  $21.24\pm0.10$  &  $20.95\pm0.10$  &  $23.40\pm0.10$  &  $22.84\pm0.10$  \\
 77  &  150.07027  &  2.42297  &  $23.42\pm0.10$  &  $22.90\pm0.10$  &  $22.40\pm0.10$  &  $21.75\pm0.10$  &  $21.66\pm0.10$  &  $21.60\pm0.10$  &  $21.44\pm0.10$  &  $21.26\pm0.10$  &  $20.98\pm0.10$  &  $20.96\pm0.10$  &  $21.39\pm0.10$  &  $21.30\pm0.10$  &  $21.17\pm0.10$  \\
 78  &  150.09944  &  2.40487  &  $-           $  &  $26.80\pm0.34$  &  $25.83\pm0.20$  &  $25.92\pm0.29$  &  $25.47\pm0.16$  &  $24.99\pm0.45$  &  $24.50\pm0.35$  &  $23.64\pm0.25$  &  $23.14\pm0.19$  &  $22.38\pm0.10$  &  $21.88\pm0.10$  &  $24.30\pm0.15$  &  $23.77\pm0.10$  \\
 81  &  150.12582  &  2.41354  &  $25.95\pm0.16$  &  $23.98\pm0.10$  &  $22.83\pm0.10$  &  $21.50\pm0.10$  &  $21.00\pm0.10$  &  $20.76\pm0.10$  &  $20.55\pm0.10$  &  $20.50\pm0.10$  &  $20.65\pm0.10$  &  $21.50\pm0.10$  &  $21.95\pm0.10$  &  $20.53\pm0.10$  &  $20.45\pm0.10$  \\
 84  &  150.11154  &  2.40957  &  $23.05\pm0.10$  &  $22.29\pm0.10$  &  $21.57\pm0.10$  &  $21.28\pm0.10$  &  $21.10\pm0.10$  &  $21.09\pm0.10$  &  $20.91\pm0.10$  &  $20.78\pm0.10$  &  $20.57\pm0.10$  &  $21.34\pm0.10$  &  $21.46\pm0.10$  &  $20.90\pm0.10$  &  $20.73\pm0.10$  \\
 89  &  150.16255  &  2.26808  &  $24.39\pm0.10$  &  $24.16\pm0.10$  &  $23.88\pm0.10$  &  $23.14\pm0.10$  &  $22.88\pm0.10$  &  $22.81\pm0.10$  &  $22.59\pm0.10$  &  $22.50\pm0.10$  &  $22.29\pm0.10$  &  $22.27\pm0.10$  &  $22.33\pm0.10$  &  $22.54\pm0.10$  &  $22.43\pm0.10$  \\
 91  &  150.07060  &  2.28920  &  $-           $  &  $-           $  &  $-           $  &  $26.52\pm0.51$  &  $25.87\pm0.23$  &  $-           $  &  $24.91\pm0.50$  &  $23.76\pm0.28$  &  $22.53\pm0.11$  &  $21.69\pm0.10$  &  $21.30\pm0.10$  &  $24.19\pm0.14$  &  $23.58\pm0.10$  \\
 92  &  150.05916  &  2.39982  &  $26.21\pm0.20$  &  $25.31\pm0.10$  &  $24.41\pm0.10$  &  $23.67\pm0.10$  &  $22.92\pm0.10$  &  $22.59\pm0.10$  &  $22.18\pm0.10$  &  $21.49\pm0.10$  &  $21.08\pm0.10$  &  $20.32\pm0.10$  &  $20.43\pm0.10$  &  $21.97\pm0.10$  &  $21.46\pm0.10$  \\
 93  &  150.05785  &  2.42723  &  $25.93\pm0.16$  &  $24.16\pm0.10$  &  $22.58\pm0.10$  &  $21.25\pm0.10$  &  $20.73\pm0.10$  &  $20.39\pm0.10$  &  $19.91\pm0.10$  &  $19.33\pm0.10$  &  $18.84\pm0.10$  &  $18.85\pm0.10$  &  $19.34\pm0.10$  &  $19.89\pm0.10$  &  $19.38\pm0.10$  \\
 98  &  150.16186  &  2.34092  &  $23.10\pm0.10$  &  $22.38\pm0.10$  &  $21.54\pm0.10$  &  $20.61\pm0.10$  &  $20.28\pm0.10$  &  $20.03\pm0.10$  &  $19.73\pm0.10$  &  $19.34\pm0.10$  &  $18.98\pm0.10$  &  $19.11\pm0.10$  &  $19.60\pm0.10$  &  $19.73\pm0.10$  &  $19.40\pm0.10$  \\
103  &  150.08514  &  2.38195  &  $-           $  &  $26.55\pm0.27$  &  $26.02\pm0.24$  &  $26.25\pm0.40$  &  $25.46\pm0.16$  &  $-           $  &  $-           $  &  $24.03\pm0.35$  &  $23.48\pm0.26$  &  $23.02\pm0.10$  &  $22.59\pm0.10$  &  $24.92\pm0.27$  &  $24.23\pm0.15$  \\
105  &  150.16426  &  2.40881  &  $-           $  &  $25.57\pm0.11$  &  $24.23\pm0.10$  &  $23.13\pm0.10$  &  $22.60\pm0.10$  &  $22.27\pm0.10$  &  $21.96\pm0.10$  &  $21.57\pm0.10$  &  $21.21\pm0.10$  &  $20.92\pm0.10$  &  $20.95\pm0.10$  &  $21.91\pm0.10$  &  $21.58\pm0.10$  \\
\hline
\end{tabular}
\end{scriptsize}

%% file: photo2.tex
\begin{scriptsize}
\begin{tabular}{lcccccccccccccccc}
\hline
 ID & RA & DEC & $u$ & $g$ & $r$ & $i$ & $z$ & $Y$ & $J$ & $H$ & $Ks$ & $3.6\mu m$ &$4.5\mu m$ & $5.8\mu m$ & $8.0\mu m$ \\
\hline
 11  &  150.04326  &  2.37348  &  $-           $  &  $-           $  &  $-           $  &  $26.85\pm0.48$  &  $25.86\pm0.20$  &  $-           $  &  $24.12\pm0.22$  &  $23.29\pm0.15$  &  $22.45\pm0.10$  &  $21.30\pm0.10$  &  $20.92\pm0.10$  &  $20.72\pm0.10$  &  $21.18\pm0.11$  \\
 17  &  150.20797  &  2.38308  &  $27.27\pm0.45$  &  $26.36\pm0.16$  &  $26.11\pm0.18$  &  $25.91\pm0.18$  &  $25.91\pm0.21$  &  $-           $  &  $24.92\pm0.52$  &  $24.31\pm0.44$  &  $23.65\pm0.26$  &  $22.66\pm0.10$  &  $22.15\pm0.10$  &  $21.65\pm0.10$  &  $21.12\pm0.11$  \\
 25  &  150.03729  &  2.34057  &  $-           $  &  $27.76\pm0.72$  &  $26.31\pm0.22$  &  $25.73\pm0.15$  &  $25.87\pm0.20$  &  $-           $  &  $25.08\pm0.63$  &  $23.41\pm0.17$  &  $23.10\pm0.15$  &  $22.48\pm0.10$  &  $22.10\pm0.10$  &  $21.49\pm0.10$  &  $22.08\pm0.22$  \\
 31  &  150.05248  &  2.24555  &  $27.07\pm0.36$  &  $25.62\pm0.10$  &  $24.77\pm0.10$  &  $24.27\pm0.10$  &  $23.89\pm0.10$  &  $23.54\pm0.11$  &  $22.91\pm0.10$  &  $21.85\pm0.10$  &  $21.21\pm0.10$  &  $20.61\pm0.10$  &  $20.37\pm0.10$  &  $20.21\pm0.10$  &  $20.52\pm0.10$  \\
 33  &  150.04098  &  2.28063  &  $-           $  &  $-           $  &  $-           $  &  $-           $  &  $-           $  &  $-           $  &  $-           $  &  $-           $  &  $-           $  &  $22.83\pm0.10$  &  $22.46\pm0.10$  &  $21.97\pm0.10$  &  $21.85\pm0.20$  \\
 42  &  150.02754  &  2.34577  &  $25.86\pm0.10$  &  $25.18\pm0.10$  &  $24.28\pm0.10$  &  $23.33\pm0.10$  &  $22.84\pm0.10$  &  $22.59\pm0.10$  &  $22.05\pm0.10$  &  $21.66\pm0.10$  &  $21.16\pm0.10$  &  $21.04\pm0.10$  &  $21.36\pm0.10$  &  $21.03\pm0.10$  &  $21.79\pm0.18$  \\
 47  &  150.04825  &  2.25144  &  $-           $  &  $-           $  &  $-           $  &  $-           $  &  $26.80\pm0.54$  &  $-           $  &  $-           $  &  $23.78\pm0.25$  &  $22.88\pm0.12$  &  $21.81\pm0.10$  &  $21.40\pm0.10$  &  $20.72\pm0.10$  &  $21.07\pm0.10$  \\
 48  &  150.02141  &  2.28867  &  $-           $  &  $26.31\pm0.15$  &  $25.30\pm0.10$  &  $24.87\pm0.10$  &  $24.80\pm0.10$  &  $24.73\pm0.36$  &  $24.28\pm0.26$  &  $23.22\pm0.14$  &  $22.46\pm0.10$  &  $21.81\pm0.10$  &  $21.57\pm0.10$  &  $21.27\pm0.10$  &  $21.21\pm0.11$  \\
 51  &  150.03652  &  2.28617  &  $26.01\pm0.12$  &  $24.99\pm0.10$  &  $24.59\pm0.10$  &  $24.28\pm0.10$  &  $23.97\pm0.10$  &  $23.77\pm0.14$  &  $23.04\pm0.10$  &  $22.68\pm0.10$  &  $21.90\pm0.10$  &  $21.65\pm0.10$  &  $21.49\pm0.10$  &  $21.73\pm0.10$  &  $22.50\pm0.36$  \\
 54  &  150.04241  &  2.29985  &  $-           $  &  $-           $  &  $27.33\pm0.68$  &  $-           $  &  $26.42\pm0.35$  &  $-           $  &  $24.86\pm0.49$  &  $23.72\pm0.23$  &  $22.85\pm0.12$  &  $22.12\pm0.10$  &  $21.85\pm0.10$  &  $21.78\pm0.10$  &  $21.28\pm0.13$  \\
 56  &  150.05002  &  2.38607  &  $-           $  &  $27.23\pm0.38$  &  $26.84\pm0.38$  &  $26.38\pm0.29$  &  $26.63\pm0.44$  &  $-           $  &  $-           $  &  $24.52\pm0.56$  &  $23.98\pm0.37$  &  $23.11\pm0.10$  &  $22.89\pm0.10$  &  $-           $  &  $-           $  \\
 73  &  150.20962  &  2.35525  &  $23.48\pm0.10$  &  $21.60\pm0.10$  &  $20.41\pm0.10$  &  $19.63\pm0.10$  &  $19.24\pm0.10$  &  $18.95\pm0.10$  &  $18.43\pm0.10$  &  $17.98\pm0.10$  &  $17.65\pm0.10$  &  $18.07\pm0.10$  &  $18.24\pm0.10$  &  $-           $  &  $-           $  \\
 79  &  150.04118  &  2.32813  &  $-           $  &  $-           $  &  $27.37\pm0.71$  &  $26.42\pm0.30$  &  $26.00\pm0.23$  &  $-           $  &  $24.38\pm0.29$  &  $23.81\pm0.26$  &  $22.59\pm0.10$  &  $21.80\pm0.10$  &  $21.55\pm0.10$  &  $21.23\pm0.10$  &  $21.89\pm0.21$  \\
 83  &  150.02492  &  2.31287  &  $26.26\pm0.16$  &  $25.74\pm0.10$  &  $25.22\pm0.10$  &  $24.59\pm0.10$  &  $24.01\pm0.10$  &  $23.48\pm0.10$  &  $23.07\pm0.10$  &  $22.37\pm0.10$  &  $21.94\pm0.10$  &  $21.55\pm0.10$  &  $21.31\pm0.10$  &  $21.58\pm0.10$  &  $21.80\pm0.18$  \\
 86  &  150.05166  &  2.30585  &  $25.09\pm0.10$  &  $24.76\pm0.10$  &  $24.35\pm0.10$  &  $23.61\pm0.10$  &  $23.03\pm0.10$  &  $22.41\pm0.10$  &  $21.80\pm0.10$  &  $21.29\pm0.10$  &  $20.86\pm0.10$  &  $20.35\pm0.10$  &  $20.22\pm0.10$  &  $-           $  &  $-           $  \\
 87  &  150.22434  &  2.35644  &  $-           $  &  $27.18\pm0.36$  &  $26.52\pm0.27$  &  $26.51\pm0.33$  &  $26.37\pm0.33$  &  $-           $  &  $-           $  &  $24.34\pm0.45$  &  $23.62\pm0.25$  &  $23.04\pm0.10$  &  $22.73\pm0.10$  &  $-           $  &  $21.88\pm0.23$  \\
 88  &  150.05456  &  2.27535  &  $26.77\pm0.18$  &  $26.06\pm0.10$  &  $25.52\pm0.10$  &  $24.82\pm0.10$  &  $24.02\pm0.10$  &  $23.67\pm0.10$  &  $22.87\pm0.10$  &  $22.12\pm0.10$  &  $21.26\pm0.10$  &  $21.24\pm0.10$  &  $20.85\pm0.10$  &  $-           $  &  $-           $  \\
 95  &  150.01640  &  2.32096  &  $-           $  &  $27.66\pm0.63$  &  $26.83\pm0.38$  &  $26.35\pm0.28$  &  $25.89\pm0.20$  &  $-           $  &  $24.97\pm0.55$  &  $23.50\pm0.19$  &  $22.91\pm0.12$  &  $22.41\pm0.10$  &  $22.13\pm0.10$  &  $21.79\pm0.10$  &  $21.61\pm0.16$  \\
 99  &  150.21020  &  2.31167  &  $24.05\pm0.10$  &  $23.12\pm0.10$  &  $22.14\pm0.10$  &  $21.16\pm0.10$  &  $20.88\pm0.10$  &  $20.70\pm0.10$  &  $20.34\pm0.10$  &  $19.86\pm0.10$  &  $19.37\pm0.10$  &  $19.23\pm0.10$  &  $19.59\pm0.10$  &  $19.49\pm0.10$  &  $19.51\pm0.10$  \\
102  &  150.03745  &  2.27186  &  $25.07\pm0.10$  &  $24.42\pm0.10$  &  $24.09\pm0.10$  &  $23.66\pm0.10$  &  $23.50\pm0.10$  &  $22.97\pm0.10$  &  $22.48\pm0.10$  &  $22.08\pm0.10$  &  $21.68\pm0.10$  &  $21.07\pm0.10$  &  $20.82\pm0.10$  &  $20.81\pm0.10$  &  $21.18\pm0.11$  \\
\hline
\end{tabular}
\end{scriptsize}

%% file: z1.tex
\begin{center}
\begin{footnotesize}
\begin{tabular}{lcccrccccc}
\hline
 ID     & $z_{spec}$ & $z_{\rm p}$ & $z_{\rm LW}$ & r & flag & $z$ & SFR                       & ${\rm log(M_{\star})}$ & source \\
 SC850- &            &             &              &   &      &     & $/{\rm M_{\odot}yr^{-1}}$ & $/{\rm M_{\odot}}$     & \\
\hline
  1  &  ...       &  $1.35^{+0.10}_{-0.10}$  &  $ 3.30^{+0.22}_{-0.14}$  &  \phantom{-}  0.83  &     0  &  $ 3.30^{+0.22}_{-0.14}$  &  $            282.8\pm             32.8$  &  ...                      & ...\\
  2  &  $0.3600$  &  $0.39^{+0.11}_{-0.09}$  &  $ 3.05^{+0.19}_{-0.19}$  &  \phantom{-}  1.98  &     0  &  $ 3.05^{+0.19}_{-0.19}$  &  $            171.2\pm             11.2$  &  ...                      & VIMOS \\
  3  &  ...       &  ...                     &  $ 3.53^{+0.30}_{-0.30}$  &  ...                &     2  &  $ 3.53^{+0.30}_{-0.30}$  &  $            447.6\pm             14.5$  &  ...                      & ...\\
  4  &  ...       &  $1.51^{+0.19}_{-0.11}$  &  $ 2.09^{+0.08}_{-0.10}$  &  \phantom{-}  0.23  &     1  &  $ 1.51^{+0.19}_{-0.11}$  &  $            325.4\pm \phantom{1}  9.6$  &  $10.21^{+0.77}_{-0.45}$  & ...\\
  5  &  ...       &  $2.21^{+0.24}_{-0.21}$  &  $ 2.03^{+0.11}_{-0.10}$  &              -0.06  &     1  &  $ 2.21^{+0.24}_{-0.21}$  &  $            444.3\pm             13.1$  &  $10.36^{+0.77}_{-0.68}$  & ...\\
  6  &  ...       &  $2.50^{+0.20}_{-0.15}$  &  $ 2.28^{+0.16}_{-0.11}$  &              -0.06  &     1  &  $ 2.50^{+0.20}_{-0.15}$  &  $            490.2\pm             15.1$  &  $11.35^{+0.65}_{-0.49}$  & ...\\
  7  &  ...       &  $2.87^{+0.18}_{-0.17}$  &  $ 2.50^{+0.15}_{-0.16}$  &              -0.10  &     1  &  $ 2.87^{+0.18}_{-0.17}$  &  $            447.6\pm             14.5$  &  $11.01^{+0.51}_{-0.48}$  & ...\\
  8  &  ...       &  $2.44^{+0.36}_{-0.24}$  &  $ 2.87^{+0.28}_{-0.28}$  &  \phantom{-}  0.12  &     1  &  $ 2.44^{+0.36}_{-0.24}$  &  $            356.2\pm             17.5$  &  $ 9.54^{+1.00}_{-0.67}$  & ...\\
  9  &  ...       &  $1.75^{+0.15}_{-0.40}$  &  $ 2.20^{+0.20}_{-0.15}$  &  \phantom{-}  0.16  &     1  &  $ 1.75^{+0.15}_{-0.40}$  &  $            261.5\pm             13.3$  &  $10.45^{+0.57}_{-1.52}$  & ...\\
 10  &  ...       &  $2.51^{+0.29}_{-0.26}$  &  $ 2.28^{+0.31}_{-0.22}$  &              -0.07  &     1  &  $ 2.51^{+0.29}_{-0.26}$  &  $            279.8\pm             14.2$  &  $10.94^{+0.90}_{-0.81}$  & ...\\
 11  &  ...       &  $1.63^{+0.42}_{-0.13}$  &  $ 2.13^{+0.10}_{-0.11}$  &  \phantom{-}  0.19  &     1  &  $ 1.63^{+0.42}_{-0.13}$  &  $            261.1\pm             13.9$  &  $10.82^{+1.73}_{-0.53}$  & ...\\
 12  &  ...       &  ...                     &  $ 3.47^{+0.65}_{-0.50}$  &  ...                &     2  &  $ 3.47^{+0.65}_{-0.50}$  &  $            409.6\pm             17.8$  &  ...                      & ...\\
 13  &  ...       &  $1.03^{+0.12}_{-0.13}$  &  $ 2.13^{+0.17}_{-0.16}$  &  \phantom{-}  0.54  &     0  &  $ 2.13^{+0.17}_{-0.16}$  &  $            261.5\pm             13.3$  &  ...                      & ...\\
 14  &  ...       &  $2.18^{+0.17}_{-0.13}$  &  $ 1.44^{+0.05}_{-0.05}$  &              -0.23  &     1  &  $ 2.18^{+0.17}_{-0.13}$  &  $            409.6\pm             17.8$  &  $11.04^{+0.59}_{-0.45}$  & ...\\
 15  &  ...       &  $2.30^{+0.20}_{-0.20}$  &  $ 2.64^{+0.41}_{-0.36}$  &  \phantom{-}  0.10  &     1  &  $ 2.30^{+0.20}_{-0.20}$  &  $            186.9\pm             11.1$  &  $10.93^{+0.66}_{-0.66}$  & ...\\
 16  &  $0.6670$  &  $0.99^{+0.11}_{-0.09}$  &  $ 1.77^{+0.08}_{-0.07}$  &  \phantom{-}  0.66  &     0  &  $ 1.77^{+0.08}_{-0.07}$  &  $\phantom{1}  69.5\pm \phantom{1}  5.0$  &  $10.64^{+0.59}_{-0.48}$  & VIMOS \\
 17  &  ...       &  $2.76^{+0.29}_{-0.51}$  &  $ 2.66^{+0.26}_{-0.17}$  &              -0.03  &     1  &  $ 2.76^{+0.29}_{-0.51}$  &  $            442.9\pm             25.7$  &  $10.79^{+0.83}_{-1.46}$  & ...\\
 18  &  ...       &  $1.94^{+0.21}_{-0.19}$  &  $ 1.45^{+0.05}_{-0.03}$  &              -0.17  &     1  &  $ 1.94^{+0.21}_{-0.19}$  &  $            429.9\pm             16.2$  &  $10.68^{+0.76}_{-0.69}$  & ...\\
 19  &  ...       &  $2.48^{+0.52}_{-1.18}$  &  $ 2.29^{+0.31}_{-0.22}$  &              -0.05  &     1  &  $ 2.48^{+0.52}_{-1.18}$  &  $            257.0\pm             15.5$  &  $10.95^{+1.64}_{-3.71}$  & ...\\
 20  &  ...       &  $2.19^{+0.11}_{-0.09}$  &  $ 2.29^{+0.27}_{-0.17}$  &  \phantom{-}  0.03  &     1  &  $ 2.19^{+0.11}_{-0.09}$  &  $            255.5\pm             15.4$  &  $11.17^{+0.39}_{-0.32}$  & ...\\
 21  &  ...       &  $1.98^{+0.42}_{-0.63}$  &  $ 1.70^{+0.13}_{-0.13}$  &              -0.09  &     1  &  $ 1.98^{+0.42}_{-0.63}$  &  $            203.6\pm             12.4$  &  $10.89^{+1.53}_{-2.30}$  & ...\\
 22  &  ...       &  $1.51^{+0.94}_{-0.76}$  &  $ 2.17^{+0.35}_{-0.25}$  &  \phantom{-}  0.26  &     1  &  $ 1.51^{+0.94}_{-0.76}$  &  $            194.3\pm             13.5$  &  $11.19^{+4.19}_{-3.39}$  & ...\\
 23  &  ...       &  $1.92^{+0.08}_{-0.17}$  &  $ 1.77^{+0.17}_{-0.13}$  &              -0.05  &     1  &  $ 1.92^{+0.08}_{-0.17}$  &  $            172.5\pm             11.1$  &  $10.60^{+0.29}_{-0.62}$  & ...\\
 24  &  ...       &  $1.72^{+0.03}_{-0.12}$  &  $ 1.70^{+0.11}_{-0.13}$  &              -0.01  &     1  &  $ 1.72^{+0.03}_{-0.12}$  &  $            171.2\pm             11.2$  &  $10.92^{+0.12}_{-0.48}$  & ...\\
 25  &  ...       &  $2.84^{+0.21}_{-0.24}$  &  $ 2.09^{+0.21}_{-0.22}$  &              -0.20  &     1  &  $ 2.84^{+0.21}_{-0.24}$  &  $            299.8\pm             21.2$  &  $10.93^{+0.60}_{-0.68}$  & ...\\
 26  &  $2.6760$  &  $2.61^{+0.09}_{-0.26}$  &  $ 1.44^{+0.10}_{-0.10}$  &              -0.34  &     1  &  $ 2.68$                  &  $            217.9\pm             15.1$  &  $10.48^{+0.26}_{-0.75}$  & VIMOS \\
 27  &  ...       &  ...                     &  $ 2.49^{+0.44}_{-0.34}$  &  ...                &     2  &  $ 2.49^{+0.44}_{-0.34}$  &  $\phantom{1}  17.3\pm \phantom{1}  2.0$  &  ...                      & ...\\
 28  &  ...       &  $2.11^{+0.09}_{-0.16}$  &  $ 1.53^{+0.17}_{-0.11}$  &              -0.19  &     1  &  $ 2.11^{+0.09}_{-0.16}$  &  $            157.8\pm             10.7$  &  $10.85^{+0.31}_{-0.56}$  & ...\\
 29  &  $0.7270$  &  $0.71^{+0.14}_{-0.11}$  &  $ 2.41^{+0.24}_{-0.26}$  &  \phantom{-}  0.97  &     0  &  $ 2.41^{+0.24}_{-0.26}$  &  $            132.5\pm             19.5$  &  ...                      & VIMOS\\
 30  &  ...       &  ...                     &  $ 2.54^{+0.48}_{-0.41}$  &  ...                &     2  &  $ 2.54^{+0.48}_{-0.41}$  &  $            128.9\pm             11.2$  &  ...                      & ...\\
 31  &  ...       &  $2.47^{+0.08}_{-0.12}$  &  $ 2.29^{+0.22}_{-0.15}$  &              -0.05  &     1  &  $ 2.47^{+0.08}_{-0.12}$  &  $            450.5\pm             36.5$  &  $11.23^{+0.26}_{-0.39}$  & ...\\
 32  &  ...       &  ...                     &  $ 2.92^{+0.64}_{-0.45}$  &  ...                &     2  &  $ 2.92^{+0.64}_{-0.45}$  &  $\phantom{1}  57.8\pm \phantom{1}  7.9$  &  ...                      & ...\\
 33  &  ...       &  $2.40^{+1.40}_{-0.65}$  &  $ 3.26^{+0.97}_{-0.56}$  &  \phantom{-}  0.25  &     1  &  $ 2.40^{+1.40}_{-0.65}$  &  $            206.6\pm             21.4$  &  $11.66^{+4.80}_{-2.23}$  & ...\\
 34  &  $0.0010$  &  $0.04^{+0.06}_{-0.04}$  &  $ 2.21^{+0.41}_{-0.21}$  &  \phantom{-}  2.21  &     0  &  $ 2.21^{+0.41}_{-0.21}$  &  $            164.2\pm             19.1$  &  ...                      & 3D-{\it HST} \\
 35  &  ...       &  $1.36^{+0.24}_{-0.16}$  &  $ 2.06^{+0.23}_{-0.27}$  &  \phantom{-}  0.30  &     1  &  $ 1.36^{+0.24}_{-0.16}$  &  $            102.3\pm             10.5$  &  $10.90^{+1.11}_{-0.74}$  & ...\\
 36  &  ...       &  $0.16^{+0.14}_{-0.11}$  &  $ 1.90^{+0.20}_{-0.14}$  &  \phantom{-}  1.50  &     0  &  $ 1.90^{+0.20}_{-0.14}$  &  $            132.5\pm             19.5$  &  ...                      & ...\\
 37  &  ...       &  ...                     &  $ 1.56^{+0.20}_{-0.12}$  &  ...                &     2  &  $ 1.56^{+0.20}_{-0.12}$  &  $            261.5\pm             13.3$  &  ...                      & ...\\
 38  &  ...       &  $1.98^{+0.12}_{-0.28}$  &  $ 2.03^{+0.28}_{-0.39}$  &  \phantom{-}  0.02  &     1  &  $ 1.98^{+0.12}_{-0.28}$  &  $            174.5\pm             16.8$  &  $10.71^{+0.43}_{-1.01}$  & ...\\
 39  &  ...       &  ...                     &  $ 2.58^{+0.63}_{-0.42}$  &  ...                &     2  &  $ 2.58^{+0.63}_{-0.42}$  &  $            356.2\pm             17.5$  &  ...                      & ...\\
 40  &  ...       &  $0.87^{+0.08}_{-0.17}$  &  $ 2.09^{+0.26}_{-0.32}$  &  \phantom{-}  0.65  &     0  &  $ 2.09^{+0.26}_{-0.32}$  &  $            447.6\pm             14.5$  &  ...                      & ...\\
 41  &  ...       &  ...                     &  $>3.92$                  &  ...                &     2  &  $>3.92$                  &  $            171.2\pm             11.2$  &  ...                      & ...\\
 42  &  $0.9370$  &  $0.96^{+0.09}_{-0.11}$  &  $ 2.20^{+0.36}_{-0.34}$  &  \phantom{-}  0.65  &     0  &  $ 2.20^{+0.36}_{-0.34}$  &  $            299.8\pm             21.2$  &  ...                      & DEIMOS \\
 43  &  ...       &  $1.86^{+0.44}_{-0.26}$  &  $ 2.83^{+0.79}_{-0.49}$  &  \phantom{-}  0.34  &     1  &  $ 1.86^{+0.44}_{-0.26}$  &  $            176.2\pm             22.1$  &  $10.90^{+1.68}_{-0.99}$  & ...\\
 44  &  $0.1220$  &  $0.13^{+0.12}_{-0.13}$  &  $ 1.96^{+0.58}_{-0.12}$  &  \phantom{-}  1.64  &     0  &  $ 1.96^{+0.58}_{-0.12}$  &  $\phantom{1}  16.8\pm \phantom{1}  2.9$  &  ...                      & VIMOS \\
 45  &  ...       &  $3.28^{+0.47}_{-0.13}$  &  $ 3.06^{+1.21}_{-0.70}$  &              -0.05  &     1  &  $ 3.28^{+0.47}_{-0.13}$  &  $            164.2\pm             19.1$  &  $10.58^{+1.16}_{-0.32}$  & ...\\
 46  &  ...       &  ...                     &  $ 1.88^{+0.21}_{-0.14}$  &  ...                &     2  &  $ 1.88^{+0.21}_{-0.14}$  &  $            132.5\pm             19.5$  &  ...                      & ...\\
 47  &  ...       &  $2.55^{+0.75}_{-0.45}$  &  $ 2.63^{+0.63}_{-0.48}$  &  \phantom{-}  0.02  &     1  &  $ 2.55^{+0.75}_{-0.45}$  &  $            282.8\pm             32.8$  &  $11.12^{+2.35}_{-1.41}$  & ...\\
\hline
\end{tabular}
\end{footnotesize}
\end{center}

%% file: z2.tex
\begin{center}
\begin{footnotesize}
\begin{tabular}{lcccrccccc}
\hline
 ID     & $z_{spec}$ & $z_{\rm p}$ & $z_{\rm LW}$ & r & flag & $z$ & SFR                       & ${\rm log(M_{\star})}$ \\
 SC850- &            &             &              &   &      &     & $/{\rm M_{\odot}yr^{-1}}$ & $/{\rm M_{\odot}}$     \\
\hline
 48  &  ...       &  $3.11^{+0.09}_{-0.16}$  &  $ 2.06^{+0.24}_{-0.25}$  &              -0.26  &     1  &  $ 3.11^{+0.09}_{-0.16}$  &  $            369.5\pm             37.6$  &  $11.32^{+0.25}_{-0.44}$  & ...\\
 49  &  ...       &  $1.60^{+0.25}_{-0.20}$  &  $ 2.05^{+0.22}_{-0.30}$  &  \phantom{-}  0.17  &     1  &  $ 1.60^{+0.25}_{-0.20}$  &  $            108.5\pm             11.8$  &  $ 9.48^{+0.91}_{-0.73}$  & ...\\
 50  &  ...       &  ...                     &  $ 1.95^{+0.87}_{-0.27}$  &  ...                &     2  &  $ 1.95^{+0.87}_{-0.27}$  &  $            490.2\pm             15.1$  &  ...                      & ...\\ 
 51  &  ...       &  $2.01^{+0.09}_{-0.11}$  &  $ 3.96^{+2.04}_{-1.12}$  &  \phantom{-}  0.65  &     1  &  $ 2.01^{+0.09}_{-0.11}$  &  $            152.4\pm             20.8$  &  $10.82^{+0.32}_{-0.40}$  & ...\\
 52  &  ...       &  ...                     &  $ 3.18^{+2.82}_{-1.28}$  &  ...                &     2  &  $ 3.18^{+2.82}_{-1.28}$  &  $\phantom{1}  48.4\pm \phantom{1}  8.2$  &  ...                      & ...\\
 53  &  ...       &  $1.41^{+0.14}_{-0.11}$  &  $ 1.47^{+0.09}_{-0.08}$  &  \phantom{-}  0.02  &     1  &  $ 1.41^{+0.14}_{-0.11}$  &  $            120.6\pm \phantom{1}  9.8$  &  $11.51^{+0.67}_{-0.53}$  & ...\\
 54  &  ...       &  $3.09^{+0.26}_{-0.44}$  &  $ 2.09^{+0.34}_{-0.30}$  &              -0.24  &     1  &  $ 3.09^{+0.26}_{-0.44}$  &  $            204.5\pm             22.4$  &  $11.02^{+0.70}_{-1.19}$  & ...\\
 55  &  ...       &  $1.74^{+0.11}_{-0.04}$  &  $ 1.25^{+0.13}_{-0.08}$  &              -0.18  &     1  &  $ 1.74^{+0.11}_{-0.04}$  &  $            128.9\pm             11.2$  &  $11.27^{+0.45}_{-0.16}$  & ...\\
 56  &  ...       &  $2.80^{+0.35}_{-0.40}$  &  $ 2.13^{+0.48}_{-0.31}$  &              -0.18  &     1  &  $ 2.80^{+0.35}_{-0.40}$  &  $            194.9\pm             23.6$  &  $10.76^{+0.99}_{-1.13}$  & ...\\
 57  &  ...       &  ...                     &  $ 3.36^{+2.47}_{-0.72}$  &  ...                &     2  &  $ 3.36^{+2.47}_{-0.72}$  &  $            143.9\pm             17.0$  &  ...                      & ...\\
 58  &  ...       &  ...                     &  $ 2.45^{+0.60}_{-0.49}$  &  ...                &     2  &  $ 2.45^{+0.60}_{-0.49}$  &  $            157.8\pm             10.7$  &  ...                      & ...\\
 59  &  ...       &  $1.69^{+0.11}_{-0.04}$  &  $ 2.04^{+0.22}_{-0.26}$  &  \phantom{-}  0.13  &     1  &  $ 1.69^{+0.11}_{-0.04}$  &  $            120.9\pm             13.8$  &  $10.30^{+0.42}_{-0.15}$  & ...\\
 60  &  ...       &  ...                     &  $>3.60$                  &  ...                &     2  &  $>3.60$                  &  $\phantom{1}  25.6\pm \phantom{1}  3.9$  &  ...                      & ...\\
 61  &  ...       &  ...                     &  $ 1.47^{+0.16}_{-0.14}$  &  ...                &     2  &  $ 1.47^{+0.16}_{-0.14}$  &  $\phantom{1}  57.8\pm \phantom{1}  7.9$  &  ...                      & ...\\
 62  &  ...       &  $1.67^{+0.08}_{-0.17}$  &  $ 2.24^{+0.33}_{-0.26}$  &  \phantom{-}  0.21  &     1  &  $ 1.67^{+0.08}_{-0.17}$  &  $            129.6\pm             17.3$  &  $10.96^{+0.33}_{-0.70}$  & ...\\
 63  &  ...       &  $0.81^{+0.29}_{-0.16}$  &  $ 5.41^{+0.59}_{-2.13}$  &  \phantom{-}  2.54  &     0  &  $ 5.41^{+0.59}_{-2.13}$  &  $\phantom{1}  57.8\pm \phantom{1}  7.9$  &  ...                      & ...\\
 64  &  ...       &  $1.63^{+0.47}_{-0.28}$  &  $ 1.53^{+0.24}_{-0.15}$  &              -0.04  &     1  &  $ 1.63^{+0.47}_{-0.28}$  &  $\phantom{1}  85.5\pm \phantom{1}  9.4$  &  $10.69^{+1.91}_{-1.14}$  & ...\\
 65  &  $2.4750$  &  $0.13^{+0.07}_{-0.03}$  &  $ 2.09^{+0.33}_{-0.29}$  &              -0.11  &     1  &  $ 2.48$                  &  $            132.5\pm             19.5$  &  ...                      & VIMOS\\
 66  &  ...       &  $1.77^{+0.08}_{-0.12}$  &  $ 1.57^{+0.27}_{-0.22}$  &              -0.07  &     1  &  $ 1.77^{+0.08}_{-0.12}$  &  $            143.9\pm             17.0$  &  $10.84^{+0.31}_{-0.47}$  & ...\\
 67  &  ...       &  $2.34^{+0.11}_{-0.19}$  &  $ 1.96^{+0.50}_{-0.23}$  &              -0.11  &     1  &  $ 2.34^{+0.11}_{-0.19}$  &  $            134.3\pm             18.0$  &  $10.09^{+0.33}_{-0.57}$  & ...\\
 68  &  ...       &  $0.78^{+0.22}_{-0.18}$  &  $ 1.60^{+0.23}_{-0.16}$  &  \phantom{-}  0.46  &     1  &  $ 0.78^{+0.22}_{-0.18}$  &  $\phantom{1}  16.8\pm \phantom{1}  2.9$  &  $ 9.71^{+1.20}_{-0.98}$  & ...\\
 69  &  ...       &  ...                     &  $>4.59$                  &  ...                &     2  &  $>4.59$                  &  $            129.6\pm             17.3$  &  ...                      & ...\\
 70  &  ...       &  ...                     &  $ 4.97^{+1.03}_{-1.79}$  &  ...                &     2  &  $ 4.97^{+1.03}_{-1.79}$  &  $            369.5\pm             37.6$  &  ...                      & ...\\
 71  &  ...       &  $0.63^{+0.12}_{-0.03}$  &  $ 2.51^{+0.80}_{-0.44}$  &  \phantom{-}  1.15  &     0  &  $ 2.51^{+0.80}_{-0.44}$  &  $            450.5\pm             36.5$  &  ...                      & ...\\
 72  &  $2.4460$  &  $0.30^{+0.10}_{-0.05}$  &  $ 1.50^{+0.22}_{-0.15}$  &              -0.27  &     1  &  $ 2.45$                  &  $            217.9\pm             15.1$  &  ...                      & VIMOS\\
 73  &  $0.1660$  &  $0.18^{+0.07}_{-0.08}$  &  $ 0.42^{+0.02}_{-0.03}$  &  \phantom{-}  0.22  &     1  &  $ 0.17$                  &  $\phantom{11}  6.2\pm \phantom{1}  0.0$  &  $10.40^{+0.62}_{-0.71}$  & VIMOS\\
 74  &  ...       &  $2.99^{+0.16}_{-0.09}$  &  $ 1.94^{+0.46}_{-0.31}$  &              -0.26  &     1  &  $ 2.99^{+0.16}_{-0.09}$  &  $            133.2\pm             19.2$  &  $10.89^{+0.44}_{-0.25}$  & ...\\
 75  &  ...       &  $1.73^{+0.37}_{-0.13}$  &  $ 2.07^{+0.23}_{-0.31}$  &  \phantom{-}  0.12  &     1  &  $ 1.73^{+0.37}_{-0.13}$  &  $            109.6\pm             13.2$  &  $10.62^{+1.44}_{-0.51}$  & ...\\
 76  &  ...       &  ...                     &  $>3.58$                  &  ...                &     2  &  $>3.58$                  &  $            120.6\pm \phantom{1}  9.8$  &  ...                      & ...\\
 77  &  ...       &  $0.64^{+0.11}_{-0.09}$  &  $ 2.61^{+1.11}_{-0.65}$  &  \phantom{-}  1.20  &     0  &  $ 2.61^{+1.11}_{-0.65}$  &  $\phantom{1}  57.8\pm \phantom{1}  7.9$  &  ...                      & ...\\
 78  &  ...       &  $2.21^{+0.19}_{-0.46}$  &  $ 2.61^{+1.09}_{-0.64}$  &  \phantom{-}  0.12  &     1  &  $ 2.21^{+0.19}_{-0.46}$  &  $            132.5\pm             19.5$  &  $10.29^{+0.61}_{-1.47}$  & ...\\
 79  &  ...       &  $1.25^{+0.65}_{-0.35}$  &  $ 1.92^{+0.40}_{-0.24}$  &  \phantom{-}  0.30  &     1  &  $ 1.25^{+0.65}_{-0.35}$  &  $\phantom{1}  48.4\pm \phantom{1}  8.2$  &  $ 9.63^{+2.78}_{-1.50}$  & ...\\
 80  &  ...       &  ...                     &  $>3.42$                  &  ...                &     2  &  $>3.42$                  &  $\phantom{1}  16.8\pm \phantom{1}  2.9$  &  ...                      & ...\\
 81  &  ...       &  $0.61^{+0.04}_{-0.11}$  &  $>3.34$                  &  \phantom{-}  3.35  &     0  &  $>3.34$                  &  $            134.3\pm             18.0$  &  ...                      & ...\\
 82  &  ...       &  ...                     &  $ 2.63^{+1.38}_{-0.75}$  &  ...                &     2  &  $ 2.63^{+1.38}_{-0.75}$  &  $            279.8\pm             14.2$  &  ...                      & ...\\
 83  &  ...       &  $1.35^{+0.05}_{-0.15}$  &  $ 2.04^{+0.39}_{-0.30}$  &  \phantom{-}  0.29  &     1  &  $ 1.35^{+0.05}_{-0.15}$  &  $\phantom{1}  70.5\pm             12.2$  &  $10.31^{+0.22}_{-0.66}$  & ...\\
 84  &  $0.3500$  &  $0.35^{+0.15}_{-0.10}$  &  $>3.61$                  &  \phantom{-}  4.19  &     0  &  $>3.61$                  &  $            134.3\pm             18.0$  &  ...                      & VIMOS\\
 85  &  ...       &  ...                     &  $ 1.51^{+0.26}_{-0.24}$  &  ...                &     2  &  $ 1.51^{+0.26}_{-0.24}$  &  $            134.3\pm             18.0$  &  ...                      & ...\\
 86  &  $1.4530$  &  $1.36^{+0.19}_{-0.11}$  &  $ 1.38^{+0.22}_{-0.15}$  &              -0.03  &     1  &  $ 1.45$                  &  $\phantom{1}  75.8\pm             10.2$  &  $11.11^{+0.89}_{-0.52}$  & VIMOS\\
 87  &  ...       &  $2.94^{+0.31}_{-0.34}$  &  $ 4.82^{+1.18}_{-1.70}$  &  \phantom{-}  0.48  &     1  &  $ 2.94^{+0.31}_{-0.34}$  &  $            200.9\pm             35.4$  &  $10.77^{+0.85}_{-0.93}$  & ...\\
 88  &  ...       &  $1.30^{+0.10}_{-0.10}$  &  $ 3.68^{+2.32}_{-1.28}$  &  \phantom{-}  1.03  &     0  &  $ 3.68^{+2.32}_{-1.28}$  &  $            206.6\pm             21.4$  &  ...                      & ...\\
 89  &  $0.9050$  &  $0.90^{+0.15}_{-0.10}$  &  $ 1.65^{+0.12}_{-0.20}$  &  \phantom{-}  0.39  &     1  &  $ 0.91$                  &  $\phantom{1}  25.6\pm \phantom{1}  3.9$  &  $ 9.78^{+0.77}_{-0.51}$  & 3D-{\it HST}\\
 90  &  ...       &  ...                     &  $ 1.96^{+1.14}_{-0.31}$  &  ...                &     2  &  $ 1.96^{+1.14}_{-0.31}$  &  $            282.8\pm             32.8$  &  ...                      & ...\\
 91  &  ...       &  $1.41^{+0.14}_{-0.21}$  &  $ 1.90^{+0.67}_{-0.36}$  &  \phantom{-}  0.20  &     1  &  $ 1.41^{+0.14}_{-0.21}$  &  $\phantom{1}  64.6\pm             11.1$  &  $10.53^{+0.61}_{-0.92}$  & ...\\
 92  &  ...       &  $1.12^{+0.03}_{-0.12}$  &  $ 1.37^{+0.18}_{-0.14}$  &  \phantom{-}  0.12  &     1  &  $ 1.12^{+0.03}_{-0.12}$  &  $\phantom{1}  57.8\pm \phantom{1}  7.9$  &  $10.51^{+0.15}_{-0.59}$  & ...\\
 93  &  $0.6550$  &  $0.64^{+0.06}_{-0.09}$  &  $ 3.41^{+1.67}_{-0.85}$  &  \phantom{-}  1.66  &     0  &  $ 3.41^{+1.67}_{-0.85}$  &  $\phantom{1}  57.8\pm \phantom{1}  7.9$  &  ...                      & DEIMOS\\
 94  &  ...       &  ...                     &  $>4.31$                  &  ...                &     2  &  $>4.31$                  &  $\phantom{1}  69.5\pm \phantom{1}  5.0$  &  ...                      & ...\\
 95  &  ...       &  $2.54^{+0.61}_{-0.29}$  &  $ 1.96^{+0.55}_{-0.24}$  &              -0.16  &     1  &  $ 2.54^{+0.61}_{-0.29}$  &  $            192.0\pm             31.9$  &  $10.67^{+1.84}_{-0.87}$  & ...\\
 96  &  ...       &  ...                     &  $>3.12$                  &  ...                &     2  &  $>3.12$                  &  $            134.3\pm             18.0$  &  ...                      & ...\\
 97  &  ...       &  ...                     &  $>3.34$                  &  ...                &     2  &  $>3.34$                  &  $            171.2\pm             11.2$  &  ...                      & ...\\
 98  &  ...       &  $0.69^{+0.16}_{-0.14}$  &  $ 1.13^{+0.09}_{-0.08}$  &  \phantom{-}  0.26  &     1  &  $ 0.69^{+0.16}_{-0.14}$  &  $\phantom{1}  17.3\pm \phantom{1}  2.0$  &  $11.03^{+1.04}_{-0.91}$  & ...\\
 99  &  ...       &  $0.69^{+0.06}_{-0.09}$  &  $ 0.62^{+0.03}_{-0.04}$  &              -0.07  &     1  &  $ 0.75$                  &  $\phantom{1}  91.0\pm \phantom{1}  2.1$  &  $10.54^{+0.37}_{-0.56}$  & ...\\
100  &  ...       &  ...                     &  $>3.53$                  &  ...                &     2  &  $>3.53$                  &  $            200.9\pm             35.4$  &  ...                      & ...\\
101  &  ...       &  ...                     &  $>3.96$                  &  ...                &     2  &  $>3.96$                  &  $\phantom{1}  16.8\pm \phantom{1}  2.9$  &  ...                      & ...\\
102  &  $1.7410$  &  $1.64^{+0.06}_{-0.14}$  &  $ 1.46^{+0.21}_{-0.16}$  &              -0.10  &     1  &  $ 1.74$                  &  $            133.5\pm             17.5$  &  $10.40^{+0.24}_{-0.55}$  & VIMOS\\
103  &  ...       &  $2.95^{+0.20}_{-0.55}$  &  $ 1.68^{+0.22}_{-0.28}$  &              -0.32  &     1  &  $ 2.95^{+0.20}_{-0.55}$  &  $            151.8\pm             21.6$  &  $10.54^{+0.53}_{-1.47}$  & ...\\
104  &  ...       &  ...                     &  $ 1.68^{+0.18}_{-0.29}$  &  ...                &     2  &  $ 1.68^{+0.18}_{-0.29}$  &  $            134.3\pm             18.0$  &  ...                      & ...\\
105  &  ...       &  $0.67^{+0.08}_{-0.12}$  &  $ 2.09^{+0.72}_{-0.40}$  &  \phantom{-}  0.85  &     0  &  $ 2.09^{+0.72}_{-0.40}$  &  $            143.9\pm             17.0$  &  ...                      & ...\\
106  &  ...       &  ...                     &  $ 2.75^{+1.66}_{-0.83}$  &  ...                &     2  &  $ 2.75^{+1.66}_{-0.83}$  &  $\phantom{11}  6.2\pm \phantom{1}  0.0$  &  ...                      & ...\\
\hline
\end{tabular}
\end{footnotesize}
\end{center}